\documentclass{aa}  

\usepackage{natbib}
\usepackage{graphicx}
\usepackage{txfonts}
\usepackage{placeins}

\begin{document}

   \title{Black holes in the low-mass galaxy regime:}

   \subtitle{Imprint of AGN feedback on the circumgalactic medium of central dwarf galaxies}

   \author{R. Flores-Freitas\inst{1,2}
          \and
          D. Wylezalek\inst{1}
          \and
          M. Trevisan\inst{2}
          \and
          M. Albán\inst{1}
          \and
          R. A. Riffel\inst{3}
          \and \\
          C. Bertemes\inst{1}
          \and
          A. Schnorr-Müller\inst{2} 
          \and 
          R. Riffel\inst{2} 
          \and          
          B. Dall'Agnol de Oliveira\inst{1}
          \and
          P. Kukreti\inst{1}
          }

   \institute{Zentrum für Astronomie der Universität Heidelberg, Astronomisches Rechen-Institut, Mönchhofstr, 12-14, 69120 Heidelberg, Germany
              \and
              Departamento de Astronomia, Instituto de Física, Universidade Federal do Rio Grande do Sul, 91501-970, Av. Bento Gonçalves, 9500 Porto Alegre, RS, Brazil
              \and
              Departamento de Física, Centro de Ciências Naturais e Exatas, Universidade Federal de Santa Maria, 97105-900 Santa Maria, RS, Brazil
             }

   \date{Received July 17, 2025; accepted November 02, 2025}
 
  \abstract
   {Active galactic nuclei (AGN) have been observed in dwarf galaxies, yet the impact of black hole feedback in these low-mass systems remains unclear.}
   {To uncover the potential effects of AGN in the low-mass galaxy regime, we study the properties and demographics of active dwarf galaxies at $z=0$, using the IllustrisTNG simulations.}
   {We use data from the TNG50-1 simulation, selecting central galaxies with stellar masses in the range $8 \leq \log(M_\ast/{\rm M_\odot}) \leq 9.5$, and selecting AGN based on their Eddington ratio ($\lambda_{\rm Edd}$). We analyzed the properties and environment of AGN host galaxies and compared them with inactive control galaxies.}
   {The AGN fractions found in the simulation depend strongly on the threshold for $\lambda_{\rm Edd}$ in the AGN selection, ranging from $\sim$~1\% ($\lambda_{\rm Edd} \geq 0.05$) to $\sim$~24\% ($\lambda_{\rm Edd} \geq 0.01$). In comparison with non-AGN galaxies of similar stellar and halo mass, dwarf AGN hosts are deficient in neutral gas, having $\sim$~3.9 times less neutral mass, in qualitative agreement with observations. The dearth in neutral gas is stronger beyond two stellar half-mass radii ($r \gtrsim 3$~kpc), and AGN hosts have more extended gas components than non-AGN galaxies, with a gas half-mass radius, on average, $\gtrsim$~10 kpc larger. AGN hosts are also slightly less star-forming, but have no differences in local environment.}
   {We found that AGN can significantly decrease the neutral gas component of dwarf galaxies, a direct effect of the high-accretion feedback mode employed in IllustrisTNG. However, it is important to test our findings with observations to unveil the complete role of AGN in dwarf galaxies. In TNG50, dwarf AGN fractions are an order of magnitude larger than those observed, motivating a detailed investigation to precisely quantify the mismatch between simulations and observations.}


   \maketitle

\section{Introduction}

Dwarf galaxies are usually defined as having stellar masses ($M_\ast$) below $10^{9.5} \ \rm M_\odot$, a limit close to the stellar mass of the Large Magellanic Cloud \citep{Marel2002}. They are the most numerous type of galaxies at all epochs \citep{Fontana2006, Grazian2015, Davidzon2017}, forming the basis for the galaxy hierarchical formation scenario \citep{White1991, DeLucia2007}. Their evolution is impacted by environmental effects and internal processes \citep{Peng2010,Woo2013,Liu2019}, which can regulate, enhance, or halt their stellar mass growth. The increase of stellar mass due to star formation in galaxies at the local Universe is thought to be regulated by feedback processes, most commonly either from active galactic nuclei (AGN) \citep{DiMatteo2005, Bower2006,Sijacki2007, Cattaneo2009,Silk2013, Somerville2015, Weinberger2017} or stellar feedback \citep{Dekel1986, Hopkins2011}. At low redshifts ($z \lesssim 0.1$), feedback on dwarf galaxies is expected to be dominated by supernovae, with AGN usually assuming a more dominant role in massive galaxies \citep{SilkMamon2012}. However, a few recent studies have been challenging the idea that AGN feedback is important only for the high-mass end \citep{Silk2017, Dashyan2018, Kaviraj2019}, proposing that the presence of accreting black holes in dwarf galaxies may have non-negligible effects (e.g., gas heating and star formation suppression) on their evolution \citep{Arjona-Galvez2024}.

As new facilities become operational and galaxy observations are pushed to fainter luminosities, the importance of accurately quantifying the effect of AGN feedback in the shallower gravitational potentials of dwarf galaxies increases. Throughout the years, dwarf galaxies hosting AGN were found using multiple methods applied across the different regions of the electromagnetic spectrum \citep{Greene2004, Greene2007, Mezcua2016, Reines2020, Birchall2020,Mezcua2020, Latimer2021b, Davis2022,Aravindan2024, Mezcua2024,Eberhard2025}. In the population of galaxies with $\log (M_\ast/{\rm M_\odot}) \lesssim 9.5$ at $z<0.1$, it was found that the fraction of galaxies hosting AGN is usually below 1\% \citep{Reines2013, Sartori2015,Mezcua2020, Birchall2020}. For example, based on spectroscopic optical data from the Sloan Digital Sky Survey (SDSS), \cite{Reines2013} found a dwarf AGN fraction of $\sim 0.5$~\%. Also, using optical data, a recent analysis of the Dark Energy Spectroscopic Instrument (DESI) early data release \citep{Pucha2025} identified thousands of active dwarf galaxies, and found an AGN fraction in emission-line galaxies of $\sim 2$~\%, 4 times higher than prior estimates from SDSS. The recent increase in the number of reported dwarf AGN candidates motivates the investigation of the possible imprints of accreting black holes on the properties of low-mass galaxies. Furthermore, since the fraction of AGN in low-mass galaxies is not used to calibrate cosmological simulations, this quantity may be useful to differentiate galaxy formation models \citep{Haidar2022}. Moreover, the fraction of AGN in dwarf galaxies can provide lower limits to the black hole (BH) occupation fraction in the low-mass regime. This occupation fraction at $M_\ast \lesssim 10^{10}$ is an important diagnostic for understanding the BH seeding process \citep{Volonteri2008, Greene2020} and it has been explored in the context of simulations and semi-analytical models \citep{Habouzit2017, Ricarte-Natarajan2018, Bellovary2019, Haidar2022}.

On the theoretical side, different works suggest that AGN feedback may have a relevant impact on dwarf galaxies, decreasing their gas mass fractions, suppressing star formation, or contributing to the hindering of cosmic gas inflow \citep{Koudmani2019, Koudmani2021,Sharma2023, Arjona-Galvez2024}. Additionally, recent simulations by \cite{Koudmani2024} highlight that the complex interplay between stellar feedback, BH feedback, and cosmic rays can significantly impact the density profiles of dwarf galaxies. Another recent work using the CAMELS simulations \citep{Medlock2025} found that AGN and supernova feedback can be connected in a complex manner, resulting in halo property effects that might seem counterintuitive. This suggests that to reproduce observations, it is important to accurately model the full range of baryonic processes in dwarf galaxy simulations. These results motivate a detailed analysis of widely used cosmological simulations and their predictions for the effect of AGN feedback in the low-mass regime of galaxies. 

In this work, we analyze how the presence of accreting black holes impacts the properties of dwarf galaxies within the IllustrisTNG simulations framework \citep{Pillepich2019, Nelson2019b}. We present the predictions from the TNG50-1 simulation and compare them with trends seen in observational works. The sections of this paper are organized as follows. We present the simulation data and our methods in Section 2, we describe our results for dwarf AGN fractions and properties compared to control galaxies in Section 3, we discuss our findings in Section 4, and present our conclusions in Section 5.  
To be consistent with the IllustrisTNG simulations, in this work we adopt the standard $\Lambda$CDM cosmology with $\Omega_{\rm m,0} = 0.3089$, $\Omega_{\rm \Lambda,0} = 0.6911$, $\Omega_{\rm b,0} = 0.0486$ and $H_0 = 67.74$ km\,s$^{-1}$\, Mpc$^{-1}$, from \cite{Planck2016}.

\section{Data and methods}
\label{sec:data_methods}

\subsection{Simulation details}
\label{methods:TNG}

The IllustrisTNG simulation suite \citep{Nelson2019, Pillepich2019} is a series of magneto-hydrodynamical simulations developed with the \textsc{arepo} code \citep{Weinberger2020}. The TNG subgrid model implements different AGN feedback modes, individual chemical element tracing, stellar evolution, and other relevant baryonic processes \citep{Weinberger2017, Pillepich2018}. 
The simulation self-consistently evolves the dynamics of dark matter, gas, stars, and magnetic fields within uniform periodic-boundary cubes (in the case of the TNG50-1 run, a cube of side $\sim 51.7$ cMpc). TNG50-1 is one of the few hydrodynamic cosmological simulations able to resolve dwarf galaxies while simulating them in a larger cosmological context. Regarding the simulation resolution, the average initial stellar particle mass is $8.5 \times 10^4 \ \mathrm{M_\odot}$, the dark matter particle mass is $4.5 \times 10^5 \ \mathrm{M_\odot}$, and the gravitational softening length ($\epsilon$) for both particles is 288 pc at $z = 0$. Thus, we choose TNG50-1 because of its spatial and mass resolution while providing a reasonable number of dwarf galaxies for statistical comparisons.  

Given the vast range of scales that would need to be resolved to fully simulate black hole formation and accretion within the entire cosmological volume of TNG50-1, it is necessary to employ an effective subgrid model to represent processes occurring below the simulation’s resolution limit. Since this work focuses on the effect of accreting black holes on dwarf galaxies, it is relevant to briefly describe the black hole model employed in IllustrisTNG. Black holes initially have a seed mass of $\log (M_{\rm seed} / {\rm M_\odot}) = 6.072$ and are seeded on halos of the simulation\footnote{Halos are identified with the friends-of-friends (FOF) algorithm \citep{Springel2001}.} that reach a minimum mass. In case of TNG50-1, this minimum mass is $\log (M_{\rm FOF}/{\rm M_{\odot}}) = 10.868$, where $M_{\rm FOF}$ is the total mass of the halo. Black holes are then accreting according to an Eddington-limited Bondi–Hoyle–Lyttleton \citep{Hoyle1939, Bondi1944, Bondi1952} accretion rate  ($\dot{M}_{\rm BH}$) with radiative efficiency, $\epsilon_{r}$, set to 0.2 \citep{Weinberger2017}. Regarding the feedback from BH, it is implemented as a two-mode accretion-rate-dependent model. High-accretion-rate feedback from black holes operates through continuous thermal energy injection into the gas surrounding the black hole. The energy per unit of time injected by this mode is equal to $\epsilon_f \epsilon_r\dot{M}_{\rm BH} c$, where $\epsilon_f=0.1$ is the fraction of energy that couples to the gas, and $c$ is the speed of light. In contrast, low-accretion-rate feedback manifests as a black hole-driven kinetic wind, where the energy is injected into the surroundings as pure kinetic feedback in randomly directed pulses. Both feedback modes will affect the gaseous component differently, either heating or expelling the gas from the galaxies. The transition between the two modes depends on a threshold $\chi$ for the Eddington ratio\footnote{Defined as the ratio between the black hole accretion rate and the Eddington accretion rate, $\lambda_{\rm Edd} = \dot{M}_{\rm BH} / \dot{M}_{\rm Edd}$.} ($\lambda_{\rm Edd}$), where $\chi$ scales with the black hole mass. Because of the way this transition is implemented, lower mass black holes will likely stay in the high-accretion mode until they reach high enough masses. For more details on the TNG seeding and feedback models, we refer the reader to \cite{Weinberger2017} and \cite{Pillepich2018}. This work uses subhalo/halo catalogs, supplementary catalogs, merger trees, and direct snapshot data from TNG50-1. The following subsections describe the quantities used in our analysis and the relevant details.

\subsection{Galaxy properties and environment}
Throughout this work, we refer to many physical quantities of individual galaxies used for sample selection or analysis (e.g., stellar mass, halo mass, star formation rate), and we define them in detail in Table \ref{tab:quantities}. Most of the physical quantities can be extracted directly from the subhalo catalogs of IllustrisTNG, but some had to be explicitly computed for this work (e.g., neutral gas mass, gas temperature, bolometric luminosity). One important quantity for this work is the neutral gas mass. To compute it, we simply use $M_{\rm neutral} = \sum_i f_i X_i m_i$, where $m_i$, $X_i$, and $f_i$ are the mass, the hydrogen abundance, and the neutral hydrogen fraction of individual gas cells, respectively. For gas below the star-formation threshold density, we use the exact values of $f_i$ computed self-consistently in IllustrisTNG \citep{Vogelsberger2013}, but we assume $f_i = 1$ for star-forming cells. In the star-forming cells, the cold fraction is predicted to be between 0.9 and 1 \citep{Diemer2018}, and although the radiation of young stars could partially ionize the interstellar medium, we neglect this contribution and assume that all gas in a star-forming cell is neutral. Here we refer to the neutral atomic component as \ion{H}{i}, and neutral molecular component as $\rm H_2$, with the whole neutral component being the sum of \ion{H}{i} and $\rm H_2$. We obtain the masses of \ion{H}{i} and $\rm H_2$ phases from the supplementary catalog of \cite{Diemer2019}, which post-processed IllustrisTNG simulations using four different $\rm \ion{H}{i} /H_2$ partition models. We use the "volumetric" masses available in their catalog and adopt an agnostic approach regarding the models. Thus, to obtain either $M_{\rm \ion{H}{i}}$ or $M_{\rm H_2}$ for a galaxy in the simulation, we compute an average over the four values available for each of these quantities for the given galaxy, giving equal weight to each model. By tracing the individual neutral phases, we can study how the AGN feedback can impact the short ($\rm H_2$) and long (\ion{H}{i}) term gas reservoirs for star formation in the dwarf galaxies.

We also trace the thermodynamics of the gas in galaxies using direct snapshot data, with gas temperature computed as $T = (\gamma - 1) (u /k_B) \mu $, where $\gamma = 5/3$ is the adiabatic index, $u$ is the internal energy of the gas cell, $\mu$ is the mean molecular weight and $k_B$ is the Boltzmann constant. However, due to resolution limits in TNG50-1, star formation and pressurization of the multiphase interstellar medium are treated following an effective model \citep{Pillepich2018}. Gas cells with hydrogen number density above $\sim$0.1~$\rm cm^{-3}$ become star-forming and no longer follow the ideal-gas equation of state. Instead, these cells are governed by a two-phase effective equation of state, accounting for pressurization from unresolved SNe \citep{Springel&Hernquist2003}. To avoid unrealistic temperature values, when averaging over many cells to compute a mean temperature, we will ignore cells with a star formation rate (SFR) above zero.

To measure the environment of galaxies in the simulation, we used the Euclidean distance to the neighbors of galaxies, assuming different minimum thresholds for the stellar mass of neighbors. These quantities were computed in a previous work by \cite{Flores-Freitas2024} and are publicly available on the IllustrisTNG supplementary catalogs\footnote{\url{http://www.tng-project.org/floresfreitas24}}.

\begin{table*}
\caption{Definition of simulated galaxy quantities used in this work.}             
\label{tab:quantities}   
\centering
\begin{tabular}{p{4.3cm} p{11cm} c}
\hline\hline       

Quantity & Description & References  \\ \hline
Stellar half-mass radius ($R_{\rm e,\ast}$) & Radius containing half of the total stellar mass of a galaxy. & 1 \\
Stellar mass ($M_\ast$) & Stellar mass within a spherical aperture of 2$R_{\rm e,\ast}$.  &  1 \\ 
Host halo mass ($M_{\rm 200c}$) & Total mass of the galaxy host halo enclosed in a sphere whose mean density is 200 times the critical density of the Universe, at the time the halo is considered. & 1 \\ 
Star formation rate (SFR) & Instantaneous star formation rate within a spherical aperture of 2$R_{\rm e,\ast}$. & 1 \\ 
Eddington ratio ($\lambda_{\rm Edd}$) & Ratio between the black hole accretion rate and the Eddington accretion rate. & 1 \\
Gas half-mass radius ($R_{\rm e,gas}$) & Radius containing half of the total gas mass of a galaxy. & 1 \\
Gas mass ($M_{\rm gas}$) & Total gas mass gravitationally bound. & 1 \\
Neutral gas mass ($M_{\rm neutral}$) & Total neutral gas mass gravitationally bound. & This work \\
Neutral atomic gas mass ($M_{\rm \ion{H}{I}}$) & Total gas mass in the form of \ion{H}{i} gravitationally bound. & 2 \\
Molecular gas mass ($M_{\rm H_2}$) &  Total gas mass in the form of $\rm H_2$ gravitationally bound. & 2 \\
Stellar metallicity ($Z_{\rm \ast}$) & Mass-weighted average stellar metallicity within a spherical aperture of 2$R_{\rm e,\ast}$. & 1 \\
Stellar formation time ($t_{\rm 50}$) & The age of the Universe when 50\% of the current stellar mass within a spherical aperture of 2$R_{\rm e,\ast}$ was already formed. & This work \\
Distance to neighbor ($d_{i,{\rm M}x}$) & Euclidean distance in real space between a given galaxy and its $i$-th nearest neighbor with $\log (M_\ast/{\rm M_\odot}) \geq x$. & 3\\ 
\hline                
\end{tabular}
\tablefoot{
The columns are, from left to right: the name of the quantity and the adopted symbol, the description of the quantity, and the corresponding reference for the catalog from which it was retrieved.
}
\tablebib{
(1)~\citet{Nelson2019}; (2) \citet{Diemer2019}; (3) \citet{Flores-Freitas2024}.
}
\end{table*}

In order to discuss the detectability of the dwarf AGN discussed in this work, we also present values for their luminosities. We compute bolometric luminosities for AGN using $L_{\rm bol} = \epsilon_{r} \dot{M}_{BH} c^2$, where $\epsilon_c = 0.2$ is the radiative efficiency, $\dot{M}_{BH}$ is the black hole instantaneous accretion rate, and $c$ is the speed of light. To obtain their X-ray luminosities ($L_{\rm X,AGN}$), we apply the bolometric corrections from \cite{Shen2020}. The contributions from X-ray binaries (XRBs) are estimated using the following relation from \cite{Lehmer2016}, for $z=0$:
\begin{equation}
    \frac{L_{\rm X,XRB}}{\rm erg \ s^{-1}} = \alpha_0 \bigg ( \frac{M_\ast}{\rm M_\odot} \bigg ) + \beta_0 \bigg ( \frac{\rm SFR}{\rm M_\odot \ yr^{-1}} \bigg ) \ , 
\end{equation}
with the parameters $\alpha_0$ and $\beta_0$ having the following values for soft and hard X-ray respectively: ($\alpha_0$, $\beta_0$) = ($10^{29.04}$, $10^{39.38}$), ($\alpha_0$, $\beta_0$) = ($10^{29.37}$, $10^{39.28}$). The total X-ray luminosities shown in Section \ref{sec:agn_fraction_obs} refer to the sum of soft (0.5-2 keV) and hard (2-10 keV) X-ray luminosities. For all the luminosities presented here, extinction is not accounted for.

\subsection{Cosmic evolution}
\label{sec:method_evolution}
To trace the evolution of galaxies and their halos in the simulation, we follow the main progenitor branch of their merger trees, including all available snapshots in the analysis. The snapshots are divided into "full" and "mini" encompassing the entire simulation volume, however, "mini" snapshots only have a subset of gas particle information available. Due to this limitation, we can only compute neutral hydrogen gas masses in "full" snapshots. Thus, when analyzing neutral gas properties quantities across cosmic time, the time interval between snapshots will be $\sim 1$~Gyr. When examining the past evolution of AGN, we use the black hole seeding time ($t_{\rm seed}$), which we define as the cosmic time of the earliest snapshot in which a galaxy has a black hole mass above zero. This quantity is defined using the highest time resolution available - both "full" and "mini" snapshots, which are separated by $\sim 170 \ \rm Myr$ - and is used to re-scale cosmic time in the analysis of the evolution of AGN hosts. When comparing the evolution of AGN and non-AGN samples in Section \ref{sec:evolution}, not all non-AGN control galaxies host a BH. Thus, for these samples, $t_{\rm seed}$ always refers to the seeding time of AGN hosts.

\subsection{Selection of dwarf AGN hosts at $z=0$}
To select dwarf galaxies in the simulation, we use their stellar masses measured in an aperture with a radius of two stellar half-mass radii. We adopt the $M_\ast = 10^{9.5} \ \rm M_\odot$ upper limit and, to avoid including the more poorly resolved dwarf galaxies in our analysis, here we select only the subhalos\footnote{In the context of the simulation data, the terms subhalo and galaxy are used here interchangeably.} with $M_\ast \geq 10^8 {\rm \ M_\odot}$. In addition, due to the characteristics of the black hole seeding method employed in TNG, only central galaxies are selected for the main analysis of this work. 
In TNG catalogs, central galaxies are usually the most massive subhalos in their host halos. More specifically, we consider central galaxies to be the ones that are the first in the halo list of subhalos, ordered in descending total number of bound particles. We also impose \texttt{subhalo\_flag=1} and a minimum of 100 dark matter particles for all the selected galaxies, excluding possible tidal dwarf galaxies. For this work, we restrict our sample selection to the last snapshot of the TNG50-1 simulation, corresponding to $z=0$.

To classify galaxies as AGN hosts, we use the straightforward criteria of the Eddington ratio. To be considered an AGN, the galaxy must have a black hole accreting with $\lambda_{\rm Edd} \geq \lambda_{\rm Edd,min}$, where we adopt $\lambda_{\rm Edd,min} = 0.01$ as a fiducial value for our selection. The choice of the specific value of 0.01 for the minimum Eddington ratio is motivated by two factors. First, it is similar to thresholds adopted in previous studies \citep{McAlpine2020, Kristensen2021, Ward2022}, and thus, facilitates comparison with the other works. Second, with our adopted $\lambda_{\rm Edd,min}$, we assume that the nuclear sources are radiatively efficient, thin discs, with high X-ray luminosities \citep{Rosas-Guevara2016, Ward2022}, a type of nuclear source that may be more easily detected in observations of dwarf galaxies.

The adopted Eddington ratio used in the selection and analysis is not instantaneous but a time average over the last three snapshots, corresponding to an average over the timescale of $\sim 340$ Myr. By using the time average, we can more robustly create a sample of control galaxies that is guaranteed to have null or weak black hole accretion in the very recent past. Our main results do not change significantly if the selection of AGN hosts and control galaxies is done using the instantaneous value of $\lambda_{\rm Edd}$ at $z=0$ (see Appendix \ref{app:selection}). Additionally, as an exercise to understand the effects of our selection criteria, in Section \ref{sec:results_agn_fraction} we vary the minimum threshold ($\lambda_{\rm Edd,min}$) for AGN selection. Thus, except for the dwarf AGN fractions, the main conclusions of this work are related to our fiducial choice, where $\lambda_{\rm Edd,min} = 0.01$.

\begin{figure}
    \centering
    \includegraphics[width=0.49\textwidth]{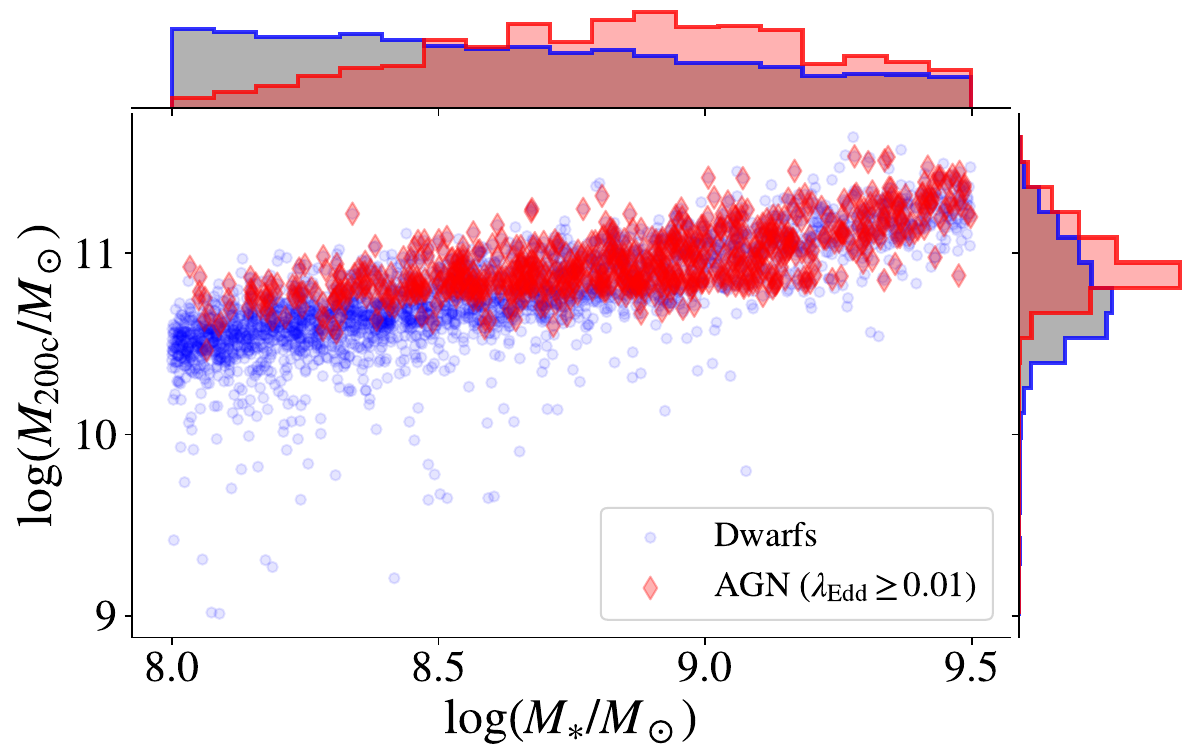}
    \caption{Joint distribution of stellar ($M_{\ast}$) and halo masses ($M_{\rm 200c}$) for TNG50-1 central dwarf galaxies at $z=0$. All 3297 dwarf galaxies are shown as blue squares, while the 789 dwarf galaxies hosting AGN are shown as red diamonds. Density histograms represent the marginal distributions of samples in each axis.}
    \label{fig:whole_sample_Mstar_M200}
\end{figure}

After applying all the criteria described above, we select a total of 3297 central dwarf galaxies, with 789 of them being AGN hosts by our criteria. We present their stellar and halo mass distributions in Figure \ref{fig:whole_sample_Mstar_M200}. It is evident that the distribution of AGN hosts is skewed towards higher masses, thus, we must build control samples of dwarf galaxies with similar $M_{\ast}$ and $M_{\rm 200c}$ to isolate the effect of AGN.

\subsection{Control samples}
\label{sec:control_sample}

\begin{figure}
    \centering
    \includegraphics[width=\linewidth]{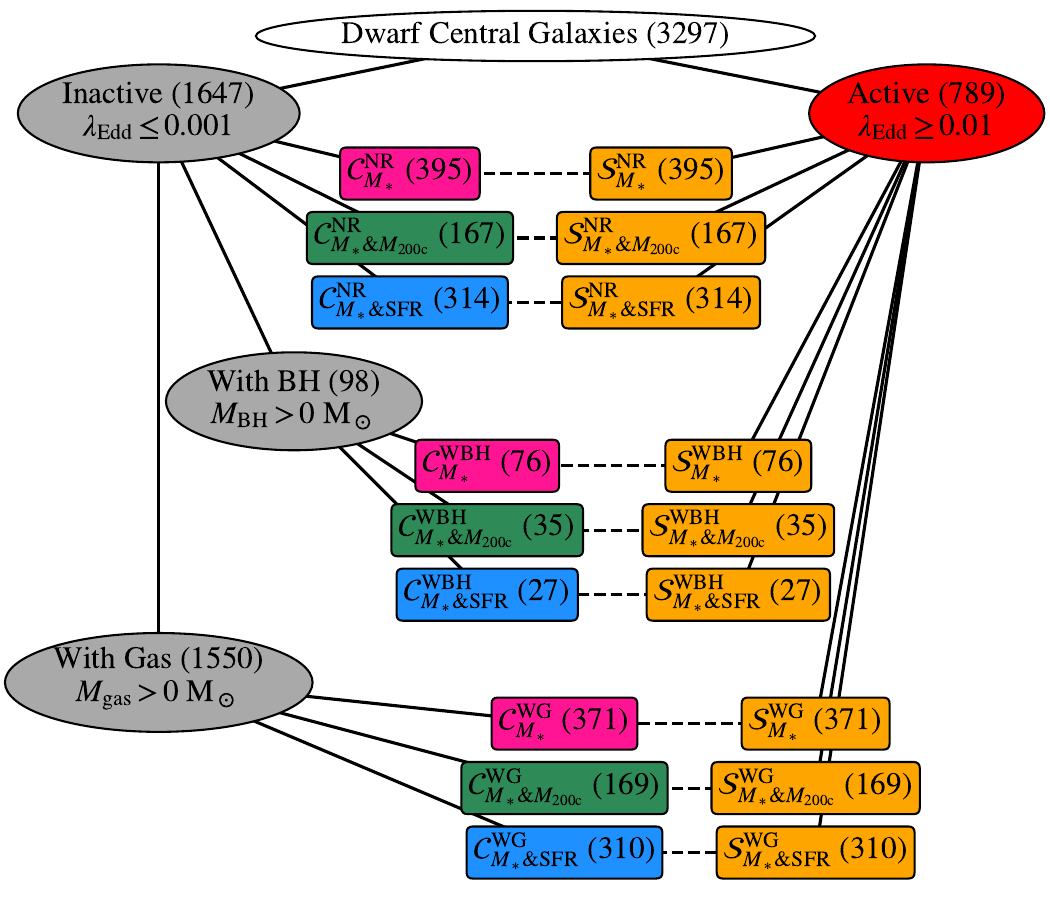}
    \caption{Scheme to illustrate the samples referred to in this work. Numbers in parentheses indicate the number of galaxies. From the main sample of dwarf central galaxies in TNG50-1 (white ellipse), we create a sample of AGN hosts (red ellipse). We also create three parent samples of non-AGN galaxies (grey ellipses), one sample of galaxies with no restriction (NR) on their properties, another of galaxies with the restriction of having gas (WG), and another of galaxies with the restriction of having black holes (WBH). Specific non-AGN control samples (pink, green, and blue boxes) are thus drawn from these parent samples, while the corresponding AGN samples (orange boxes) are drawn from the sample of active dwarf galaxies (red ellipse). Dashed lines indicate the pairing of a specific non-AGN control sample and its respective AGN samples, following the method described in Section \ref{sec:control_sample}. The colors of the boxes of control samples indicate the variables used as control, and correspond to the same colors used in Figure \ref{fig:boxplots1}. }
    \label{fig:sample_scheme}
\end{figure}

To understand the effect of AGN, we want to compare the AGN hosts with non-AGN galaxies. The latter are defined as subhalos where $\lambda_{\rm Edd} \leq 0.1 \lambda_{\rm Edd, min}$, and we use this threshold to ensure that the non-AGN sample is sufficiently different from the AGN sample in terms of black hole accretion rate. All the control samples described below - and represented in Figure \ref{fig:sample_scheme} - are constructed from this sample of non-AGN galaxies ($\lambda_{\rm Edd} \leq 0.001$). For clarity, it is relevant to define here the terms used throughout the text.  When referring to a dwarf galaxy hosting an AGN, we use the terms "active galaxy" or "AGN host", similarly, when referring to galaxies not hosting an AGN, we use the terms "inactive galaxy" or "non-AGN galaxy". The non-AGN galaxies are always part of control samples in our analysis, thus they can also be referred to as "control galaxies".

Since many galaxy properties scale with their stellar and halo masses, the non-AGN control samples must have a similar distribution of these variables to isolate the effect of AGN. Motivated by previous works \citep{Elisson2019}, which show the importance of considering star formation rate when analyzing AGN, we also build samples paired by SFR. To evaluate different biases, we create multiple control samples drawn from three different parent samples of non-AGN central galaxies: the set of all non-AGN (with no restriction, abbreviated as NR), the set of non-AGN containing gas (WG), and the set of non-AGN containing black holes (WBH). Galaxies are assigned to these parent samples based on their gas mass ($M_{\rm gas}$) and black hole mass ($M_{\rm BH}$):
\begin{itemize}
    \item No restriction (NR): any value of $M_{\rm gas}$ and $M_{\rm BH}$;
    \item With gas (WG): $M_{\rm gas} > 0 \ \rm M_\odot$, any value of $M_{\rm BH}$;
    \item With black holes (WBH): $M_{\rm BH} > 0 \ \rm M_\odot$, any value of $M_{\rm gas}$.
\end{itemize}
Considering these three sets of non-AGN galaxies, we can independently check if significant differences found in the AGN population still hold if we enforce control galaxies to have gas or black holes. This is important because the simulation can produce dwarf galaxies with no black holes and also dwarf galaxies with no gas reservoir, both of which would be considered non-AGN by our criteria. On the other hand, by definition, our AGN selection in the simulation will always select dwarf galaxies with $M_{\rm gas} > 0 \ \rm M_\odot$ and $M_{\rm BH} > 0 \ \rm M_\odot$.  When comparing the dwarf AGN with control galaxies in the NR control samples, we are simply comparing galaxies with and without recent black hole accretion, regardless of their properties. When comparing dwarf AGN with non-AGN galaxies in the WG control samples, we ensure that any differences found will not simply result from control galaxies lacking gas reservoirs. Similarly, when comparing dwarf AGN with non-AGN galaxies in the WBH control samples, we ensure any differences found will not simply result from control galaxies lacking a black hole.

To facilitate referencing each specific sample through the text, we adopt the following notation: AGN samples are represented by the letter $\mathcal{S}$ and non-AGN control samples by the letter $\mathcal{C}$; subscripts indicate the variables controlled and superscripts represent the parent sample from which galaxies are drawn. For example, the control sample paired by stellar mass and drawn from the parent sample with no restriction (NR) is represented by $\mathcal{C}_{M_\ast}^{\rm NR}$, while its correspondent AGN sample is represented by $\mathcal{S}_{M_\ast}^{\rm NR}$. Analogously, the control sample paired by stellar and halo mass, which was drawn from the parent sample of galaxies with gas (WG), is represented by $\mathcal{C}_{M_\ast \& M_{\rm 200c} }^{\rm WG}$. To illustrate the structure of the galaxies analyzed here, in Appendix \ref{app:examples} we present the stellar and mass distributions of a few AGN-control pairs. 

To build each control sample, we apply the propensity score matching technique \citep{Rosembaum1983}, using the \textsc{MatchIt} package \citep{Ho2011}. This method allows us to create control samples in which the distributions of the control variables ($M_\ast$, SFR, $M_{\rm 200c}$) are as similar as possible between AGN and non-AGN samples, with each AGN host paired with a unique inactive galaxy. For the matching parameters, we adopted the Mahalanobis distance \citep{Mahalanobis1936} approach and the nearest-neighbor method. Additionally, we introduce a final cut ensuring that the maximum difference between individual AGN hosts and their paired control galaxy is 0.1, 0.1, and 0.2 dex in $\log(M_\ast)$, $\log(M_{\rm 200c})$, and $\log({\rm SFR})$, respectively. After the matching process, we perform Anderson-Darling two-sample tests \citep{AndersonDarling1952,Scholz1987} to determine whether the properties of control and AGN have similar distributions. This nonparametric statistical test has the null hypothesis that different samples are drawn from the same population, having identical distributions. We chose this specific test in our analysis because it is sensitive to differences in tails of the compared distributions, not only to differences in their central tendencies. When we found a significant difference - up to a significance level of 5\% - for a given physical property, we quantify it further through analysis of box plots, and median offsets between AGN and non-AGN samples (see Section \ref{sec:properties}).

Because of the intrinsic distributions of the control variables in TNG50-1 and the adopted criteria for selecting AGN here, not all dwarf AGN hosts can be paired with a unique non-AGN control galaxy. The direct implication of this is that we can robustly isolate the effect of AGN - with respect to $M_{\ast}$, $M_{\rm 200c}$, and SFR - only in a fraction of the dwarf hosts found in the simulation. However, we can guarantee that the effects seen in this subset of dwarf AGN are due to the recent black hole accretion. 

\section{Results}
\label{sec:results}
\subsection{Dwarf AGN fractions}
\label{sec:results_agn_fraction}

In the stellar mass regime adopted for this work, we found an overall black hole occupation fraction\footnote{The fraction of dwarf galaxies hosting a black hole.} of 44\% in TNG50-1 at $z=0$. When separating the dwarf galaxies by centrals and satellites, we found 53\% and 32\%, respectively. 
The overall occupation fraction mentioned above for dwarfs with $8 \leq \log(M_\ast/{\rm M_\odot}) \leq 9.5$ is roughly consistent with some theoretical works \citep{Ricarte-Natarajan2018, Askar2022} and observational constraints \citep{Trump2015, Miller2015}. However, it is worth mentioning that the predictions and constraints for occupation fraction can vary significantly \citep[see the review by][]{Greene2020}. An in-depth comparison of our results with observations requires a careful separation between satellites and central dwarf galaxies in observations, as the satellites in the simulation can have artificially lower fractions. We leave this analysis for future work. For an exhaustive and detailed analysis on the topic of BH occupation fractions in simulations, we refer the reader to the work conducted by \cite{Haidar2022}.

Regarding the population of active black holes, after adopting the AGN selection criteria of $\lambda_{\rm Edd} \geq 0.01$, we found 1017 AGN hosts from a total of 5909 dwarf galaxies, corresponding to an overall AGN fraction - number of galaxies with active black holes over the total number of galaxies - of 17\%. Separating for centrals and satellites, we found 24\% and 8\%, respectively. The lower AGN fraction on satellite galaxies can be explained mainly by two factors: the absence of bound gas in many satellites and the minimum halo mass for BH seeding. While almost all central galaxies have gas, only half of the satellite dwarf galaxies have a bound gaseous component. Since gas is required for AGN, this prevents many satellite dwarfs from hosting an AGN. Also, due to the black hole seeding model of TNG simulations, if the dwarf galaxy became a satellite before its former host halo reached the minimum halo seeding mass - $\log (M_{\rm FoF} / {\rm M_\odot}) = 10.868$, see Section \ref{sec:data_methods} -, then it will most likely remain without a black hole if it remains a satellite. Thus, a lower occupation fraction is found for the satellite dwarfs, contributing to a lower AGN fraction. For simplicity, in the following analysis and discussions of this work, we focus only on the population of central galaxies.

\begin{figure}
    \centering
    \includegraphics[width=0.48\textwidth]{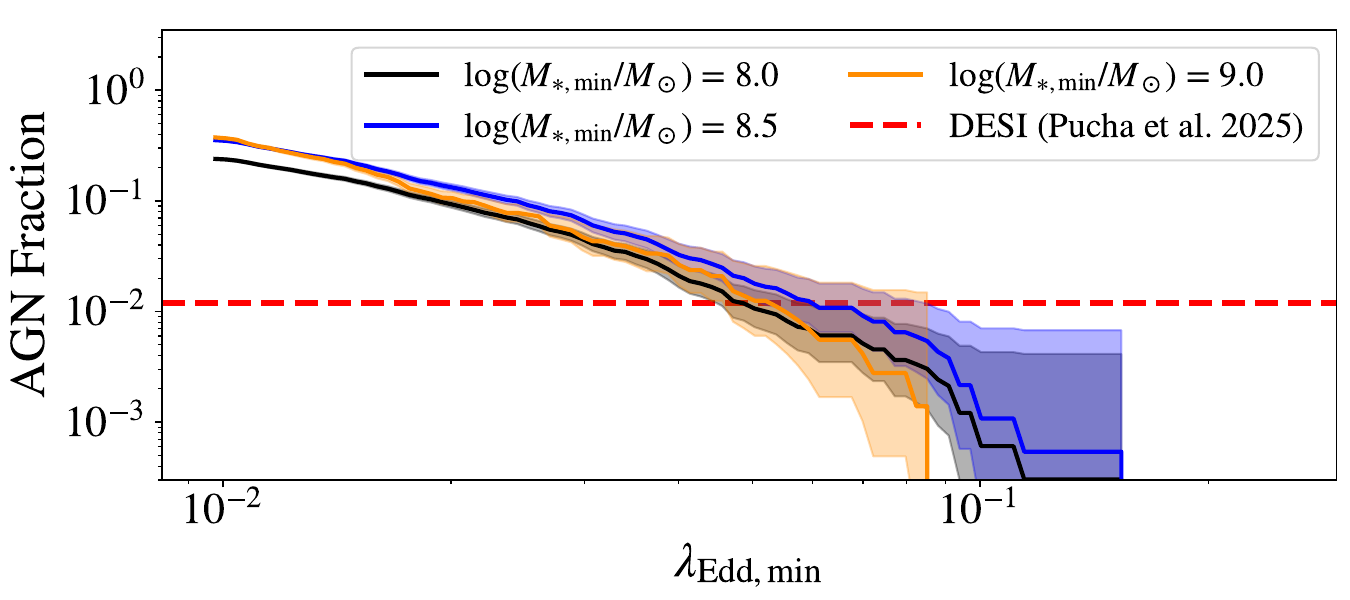}
    \includegraphics[width=0.48\textwidth]{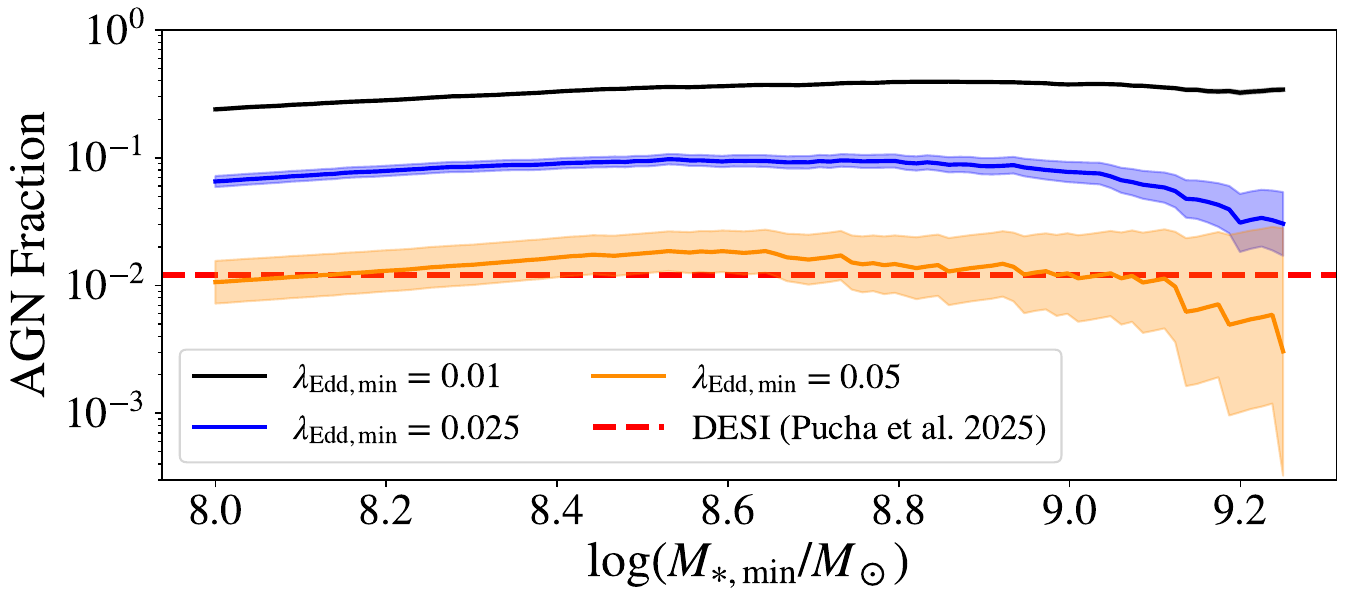}
    \caption{Fraction of central dwarf galaxies hosting AGN according to different minimum thresholds for variables used in the sample selection. Top panel: AGN fraction as a function of the minimum threshold for $\lambda_{\rm Edd}$. Bottom panel: AGN fraction as a function of the minimum threshold for stellar mass. In each panel, the solid lines show the fraction for different values of $\lambda_{\rm Edd, min}$ or $M_{\ast, \rm min}$, with the shaded regions representing the 95\% confidence intervals. The red dashed lines indicate the overall fraction of dwarf AGN candidates from the DESI survey \citep{Pucha2025}, for the whole mass range of $8 \lesssim \log (M_\ast/{\rm M_\odot}) \leq 9.5$. This value ($\sim$ 1.2\%) is derived from the values presented in their figure 7 (right panel), and is the fraction of dwarf galaxies identified as AGN/Composite in the [\ion{N}{ii}]-BPT diagram, over the total number of dwarf galaxies (including dwarf galaxies without emission lines).}
    \label{fig:agn_fraction}
\end{figure}

In Figure \ref{fig:agn_fraction} we explore how the AGN fraction in central dwarf galaxies changes as a function of the minimum threshold for $\lambda_{\rm Edd}$ and $M_\ast$ in our AGN selection. As the minimum threshold for $\lambda_{\rm Edd}$ increases, the AGN fraction quickly decreases, going from 24\% at $\lambda_{\rm Edd, min} = 0.01$ to 1\% at $\lambda_{\rm Edd, min} = 0.05$ - keeping $\log (M_{\ast, \rm min}/{\rm M_\odot}) = 8$ constant. The effect of changing the minimum stellar mass is much weaker, since the AGN fraction is mostly constant with respect to $M_{\ast, \rm min}$, only changing for more than a few percent at the $9 \leq \log(M_\ast/{\rm M_\odot}) \leq 9.5$ range. 

A recent work by \cite{Pucha2025} selected dwarf galaxies ($\log (M_\ast / {\rm M_\odot}) \leq 9.5$) with AGN signatures based on the  [\ion{N}{ii}]-BPT diagram \citep{BPT1981}, and found an AGN fraction of $\sim$2.1\% among the dwarf galaxies with emission lines. Using the data from their Figure 7 (right panel), we computed the AGN fraction over all dwarf galaxies, regardless of having emission lines, and found a value of $\sim$1.2\% (represented by the red dashed lines in Fig. \ref{fig:agn_fraction}). As it is clear from Fig. \ref{fig:agn_fraction}, the selection of dwarf AGN in the simulation based simply on $\lambda_{\rm Edd} \geq 0.01$ returns a fraction more than 10 times higher than observations. To reach an AGN fraction comparable to the values from \cite{Pucha2025}, the minimum threshold for the Eddington accretion would have to be $\lambda_{\rm Edd, min} \approx 0.05$, corresponding to a minimum AGN bolometric luminosity of $\log(L_{\rm bol}/{\rm ergs \ s^{-1}}) = 43.3$ for a host with $\log (M_{\rm BH}/\rm M_\odot) = 6.5$ - no obscuration considered. In Section \ref{sec:agn_fraction_obs}, we discuss the AGN fractions of TNG50-1 in the broader context of observational works that use different methods to select AGN in real dwarf galaxies.

\subsection{Internal properties and environment}
\label{sec:properties}

\begin{figure*}
    \hspace{0.2cm}
    \includegraphics[width=0.97\textwidth]{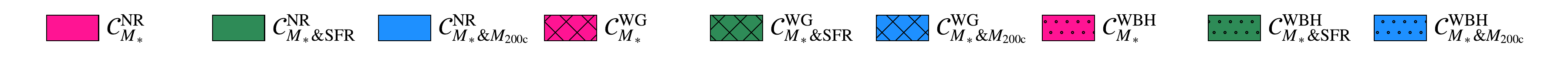}
    \includegraphics[width=0.99\textwidth]{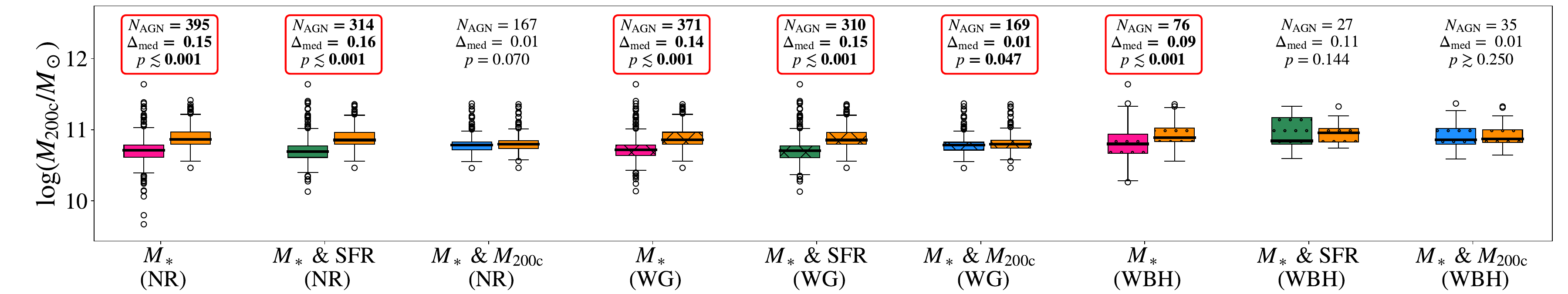}
    \includegraphics[width=0.99\textwidth]{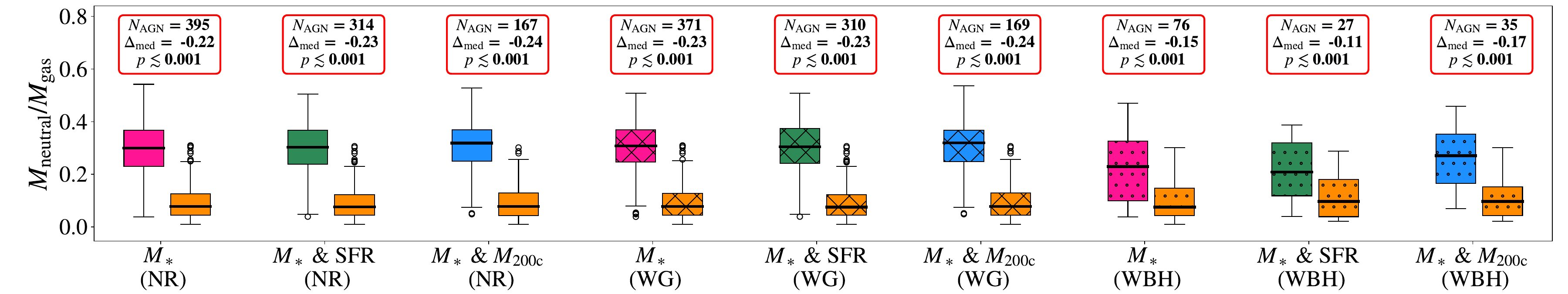}
    \includegraphics[width=0.99\textwidth]{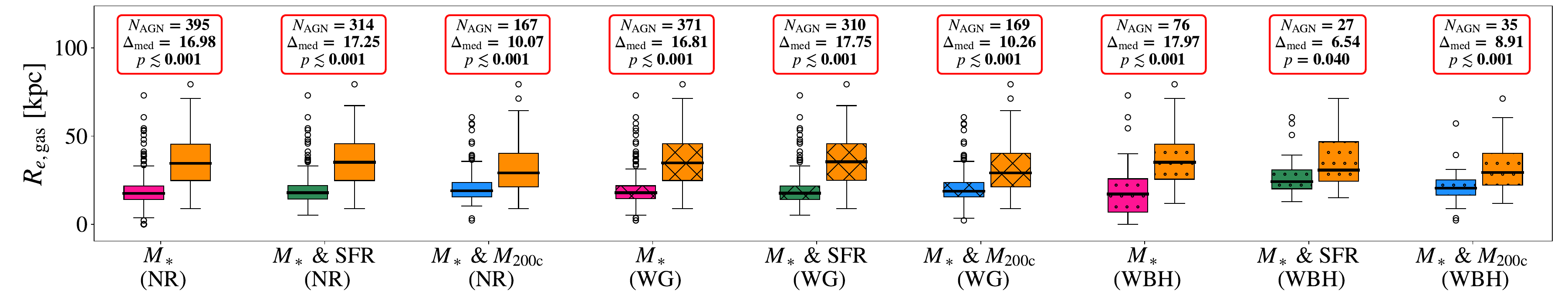}
    \caption{Box plots of AGN (orange) and non-AGN control (pink, green, blue) samples properties. Each box extends from the first to the third quartile of the data, with a line at the median. The whiskers extend from the box edges to the farthest data point lying within 1.5 times the inter-quartile range from the box. Data past the end of the whiskers are shown as empty black circles. AGN samples are represented by orange boxes, control samples paired by $M_\ast$, $M_\ast$ \& SFR or $M_\ast$ \& $M_{\rm 200c}$ are represented by pink, green, and blue boxes, respectively. Different hatch patterns indicate the different parent samples of non-AGN central galaxies from which the control samples were constructed: no pattern, "x" pattern, and dot pattern refer to the "No Restriction" (NR), "With Gas" (WG), and "With BH" (WBH) samples, respectively. The legend above the upper panel indicates the color and hatch pattern for each control sample shown in the figure. In each panel, above the boxes, we show the number of AGN hosts ($N_{\rm AGN}$), the difference between the median of the AGN and control ($\Delta_{\rm med}$), and the p-value of an Anderson-Darling two-sample test performed with the control and AGN paired samples. In the cases where the difference was considered statistically significant (p-value $< 0.05$), the upper text is highlighted as bold and enclosed in a red box. From top to bottom, the panels show box plots for host halo mass, neutral-to-total gas ratio, and gas half-mass radius.}
    \label{fig:boxplots1}
\end{figure*}

\begin{figure*}
    \hspace{0.2cm}
    \includegraphics[width=0.97\textwidth]{FIGURES/boxplots/boxplots_legend.pdf}
    \includegraphics[width=0.99\textwidth]{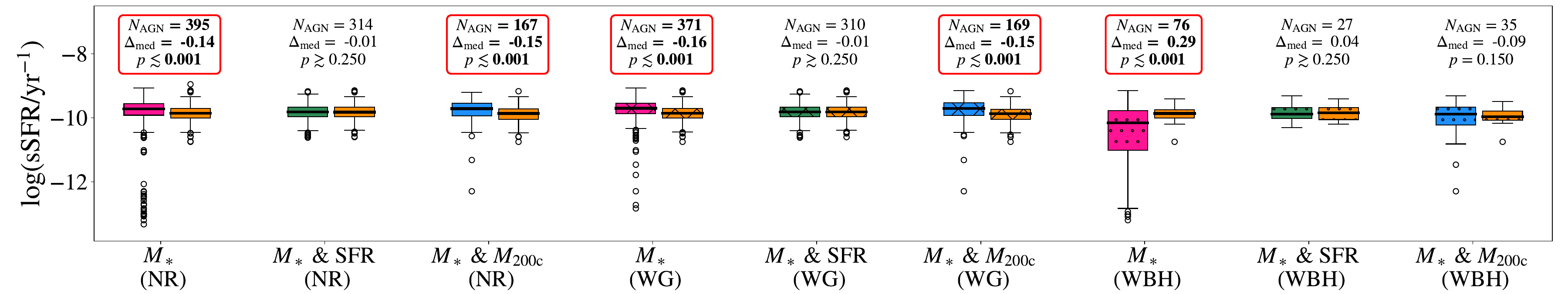}
    \includegraphics[width=0.99\textwidth]{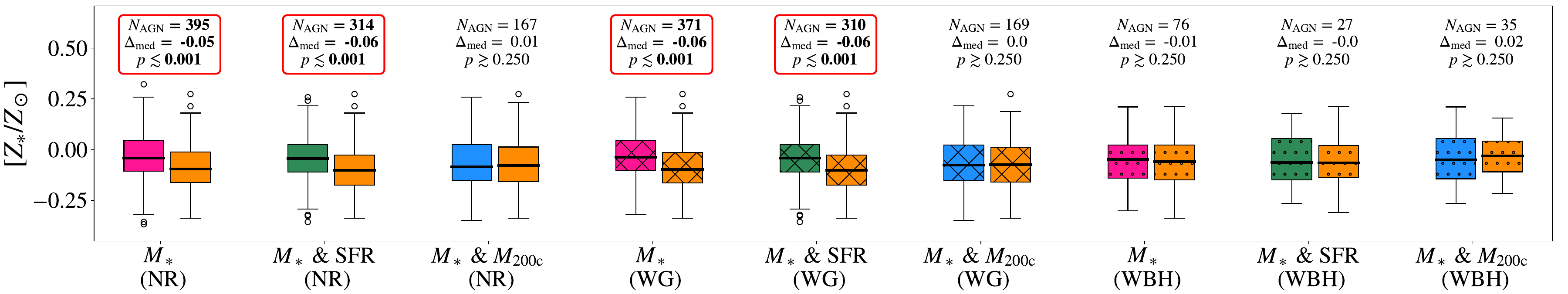}
    \includegraphics[width=0.99\textwidth]{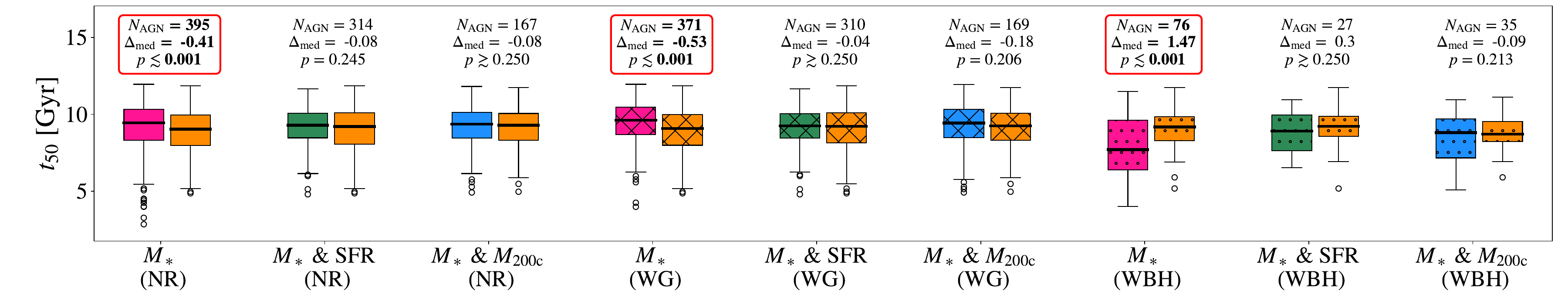}
    \includegraphics[width=0.99\textwidth]{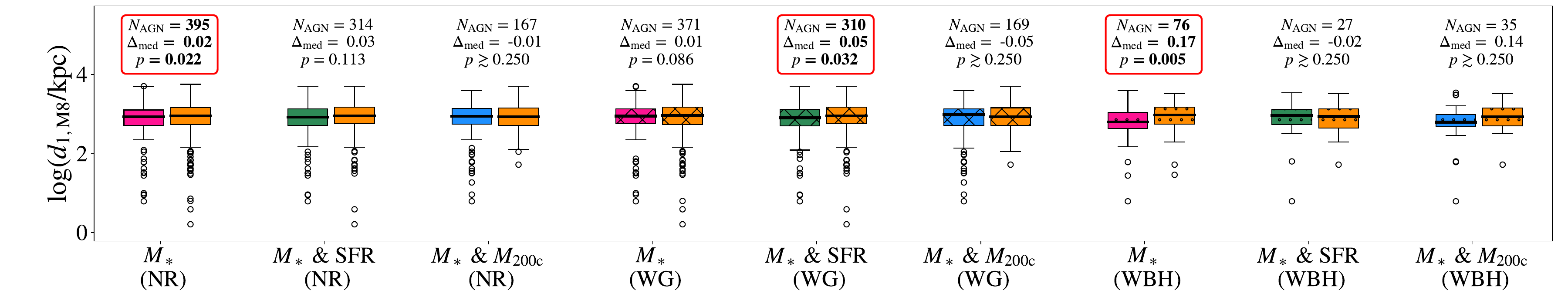}
    \caption{Same as Figure \ref{fig:boxplots1}. From top to bottom, the panels show box plots for: specific star formation rate, stellar metallicity, median stellar formation time, and distance to the 1st nearest neighbor (with $M_\ast \geq 10^8 \ \rm M_\odot$).}
    \label{fig:boxplots2}
\end{figure*}

Throughout this section, we illustrate the comparison of AGN and non-AGN properties through box plots, assessing the significance of the difference in their distributions with two-sample Anderson-Darling tests. We also use the difference between medians to quantify our statements, noted as $\Delta_{\rm med} = {\rm Med}(X_{\rm AGN}) - {\rm Med}(X_{\rm Control})$, where $X_{\rm sample}$ represents the set of values for a given physical quantity in a given sample. As is shown in the top panel of Figure \ref{fig:boxplots1}, AGN present higher halo masses when paired only by stellar mass and SFR. This is a direct result of the $M_\ast$-$M_{\rm 200c}$ joint distribution of the dwarf galaxies, as shown in Figure \ref{fig:whole_sample_Mstar_M200}, which is a consequence of the halo mass-based seeding model of TNG. This justifies the separate analysis of non-AGN control samples paired by $M_\ast$ and $M_{\rm 200c}$, to guarantee that the differences between AGN and non-AGN are not simply explained by an underlying distinction in their halo mass distributions. 

The most substantial differences that we found for dwarf AGN properties were in their gas component. In the lower panels of Figure \ref{fig:boxplots1}, we show the ratio of neutral-to-total gas mass ratio ($M_{\rm neutral}/M_{\rm gas}$) and the gas half-mass radius ($R_{e, \rm gas}$). It is evident that dwarf AGN are significantly different from their control in these properties, regardless of the adopted parent sample or control variables, although the differences are less pronounced in control samples with BH. The vast majority of control galaxies have $M_{\rm neutral}/M_{\rm gas}$ above 0.25, while the majority of AGN have $M_{\rm neutral}/M_{\rm gas} < 0.2$. Additionally, for almost all control samples, $\Delta_{\rm med}$ of $M_{\rm neutral}/M_{\rm gas}$ is close to 0.2, meaning that the dwarf AGN population has approximately 20\% less of its gas in the neutral form when compared to the non-AGN sample. If we also compare the extension of the gas radius in AGN and non-AGN, especially for $\mathcal{C}_{M_\ast \& M_{\rm 200c}}^{\rm WG}$, we find that $R_{e, \rm gas}$ in AGN is usually more than 10 kpc larger, with the majority of AGN having $R_{e, \rm gas} > 25$~kpc. Although some differences in $M_{\rm neutral}/M_{\rm gas}$ are systematically more minor for $\mathcal{C}_{M_\ast}^{\rm WBH}$, $\mathcal{C}_{M_\ast \& {\rm SFR}}^{\rm WBH}$ and $\mathcal{C}_{M_\ast \& M_{\rm 200c}}^{\rm WBH}$ control samples, they are still not negligible, implying that just the presence of black hole is not enough to explain the distinct gas properties in dwarf AGN. As discussed in Section \ref{sec:evolution}, the deficiency in neutral gas is not only connected with the black hole presence in these dwarf galaxies, but also to their recent activity in the radiative mode. 

We also analyzed other relevant properties of AGN hosts measured inside 2$R_{e,\ast}$, like their metallicity ($Z_{\ast}$), age ($t_{50}$), and specific SFR (sSFR), shown in Figure \ref{fig:boxplots2}. No significant large differences are found when controlling for halo and stellar mass, except for sSFR. AGN hosts tend to be less star-forming, with the median sSFR of control galaxies being 1.4 times larger than the median sSFR of AGN. Interestingly, observations of AGN in more massive hosts suggest opposite results \citep{Mountrichas2024,Riffel2024}. This difference in $\log ({\rm sSFR})$ is similar ($\Delta_{\rm med} \approx 0.1$ dex) for other non-AGN control samples where SFR is not controlled, and may be an indirect effect of the AGN hosts having less neutral gas available. 
In observational studies, it may be challenging to obtain the halo mass of dwarf galaxies, with non-AGN control samples only being able to be paired by stellar mass. When only $M_\ast$ is controlled in our analysis, in addition to having lower sSFR, AGN hosts also have lower stellar metallicities and lower $t_{50}$. Although statistically significant, the magnitude of these differences is small, as evidenced by the box plots and comparison of the medians in the figure. 

Furthermore, we analyzed the distances to nearest neighbors of the galaxies (Figure \ref{fig:boxplots2}) to explore possible relations between the environment and AGN activity. This analysis is also important to reinforce the results found for the neutral gas, excluding the possibility of AGN having different neutral gas fractions simply due to perturbations in their immediate vicinity. As shown in the lower panel of Figure \ref{fig:boxplots2}, for the 1st nearest neighbors ($d_{\rm 1,M8}$), most control samples show no significant difference in their distribution when compared to AGN samples, and those that have statistically significant differences (e.g. the sample $\mathcal{C}_{M_\ast}^{\rm NR}$), have it very slightly. For example, the sample $\mathcal{C}_{M_\ast}^{\rm WBH}$ shows the largest $\Delta_{\rm med}$ for $\log (d_{\rm 1,M8}/{\rm kpc})$, however this difference (0.17 dex) yet large in terms of physical size ($\sim 300$ kpc), is not physically relevant due to the typical values of $d_{\rm 1,M8}$. The distances to the first neighbors are larger than 300 kpc for 90\% of the galaxies, implying that the gravitational pull from the closest companions is already weak, and the difference found in the median of $d_{\rm 1,M8}$ should not be relevant for the AGN activity. Thus, since the difference in the distributions of distance to neighbors is even smaller for other control samples, it is unlikely that the presence of immediate neighbors at $d > 300$~kpc significantly impacts black hole activity of the simulated dwarf galaxies. The results remains the same if we consider other metrics to measure environment, such as the distance to the 1st massive nearest neighbor ($d_{\rm 1,M10}$), or distance to farthest neighbors ($d_{\rm 10,M8}$ and $d_{\rm 10,M10}$), which probe the environment in slightly larger scales (see Appendix \ref{app:extra_box}). In line with the result described above, observational works measuring the small-scale environment of AGN hosts also found that they have similar environments compared to non-AGN control galaxies matched in redshift, stellar mass, and morphology \citep{Rembold2024}.  
A previous work by \cite{Kristensen2021} analyzes the TNG100-1 simulation run and finds evidence of a non-negligible role of environment for AGN activity in dwarf galaxies with $9 \leq \log (M_\ast / {\rm M_\odot}) < 9.48$. We do not find the same trends when analyzing central dwarf galaxies in this work. The difference between the results may come from the absence of satellite galaxies in our analysis, the different stellar mass distribution of the AGN samples, and the difference in simulation volume or resolution. 

As described above, the strongest differences between AGN and non-AGN are in the gas half-mass radius and neutral gas content. However, the differences between $\mathcal{C}_{M_\ast}^{\rm WBH}$ and $\mathcal{S}_{M_\ast}^{\rm WBH}$ present a different behavior than other non-AGN/AGN paired samples. Although AGN hosts in $\mathcal{S}_{M_\ast}^{\rm WBH}$ are still more deficient than their control counterparts, $\mathcal{C}_{M_\ast}^{\rm WBH}$ shows more pronounced tails towards zero on the distributions of $M_{\rm neutral}/M_{\rm gas}$. When compared to other control samples, $\mathcal{C}_{M_\ast}^{\rm WBH}$ also shows more pronounced tails towards lower values of sSFR and $t_{50}$. When compared to its AGN counterpart, $\mathcal{C}_{M_\ast}^{\rm WBH}$ contains relatively older and more quiescent galaxies, indicating that if we impose the restriction of having a BH and only control for stellar mass, we may find that active dwarf galaxies formed more recently than inactive ones.
The distributions of $d_{\rm 10,M8}$, $d_{\rm 10,M10}$ and $d_{\rm 1,M10}$ for $\mathcal{C}_{M_\ast}^{\rm WBH}$ also present more pronounced tail towards smaller distances, suggesting slightly denser environments than their paired AGN. However, the difference in environment does not seem to be important given the absolute values of the distances to neighbors ($\gtrsim 300$ kpc).

\subsection{Differences in the gas component at $z=0$}
\label{sec:diff_gas_comp}
\begin{figure}
    \centering
    \includegraphics[width=0.47\textwidth]{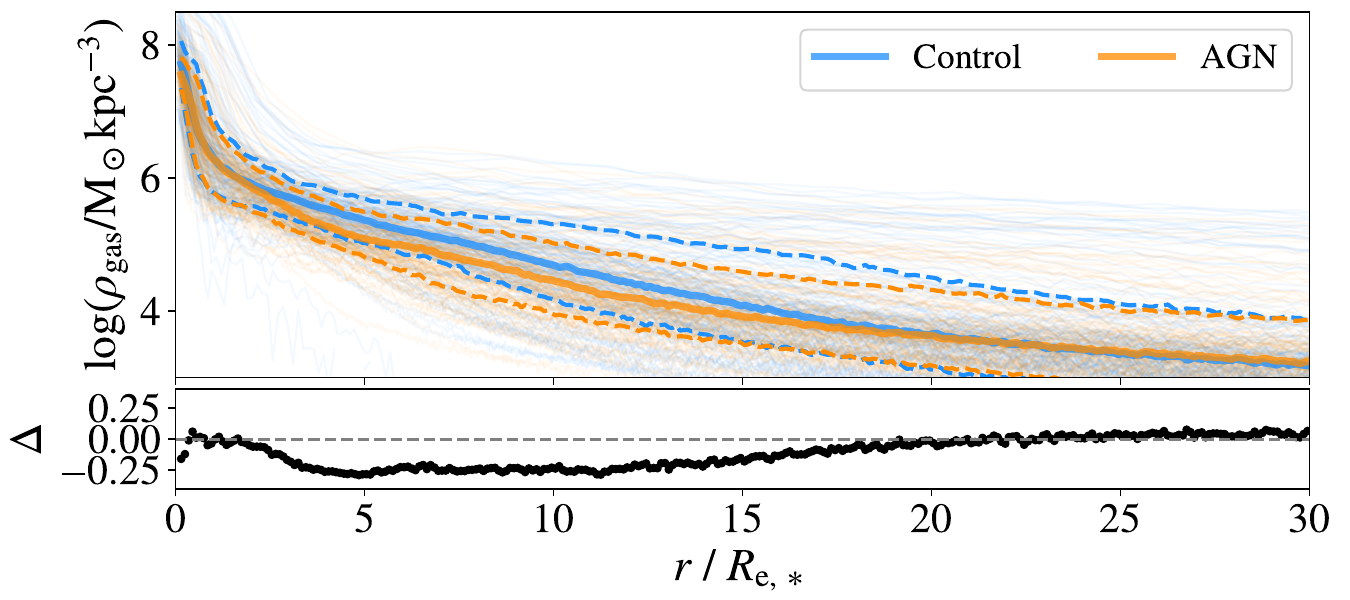}
    \includegraphics[width=0.47\textwidth]{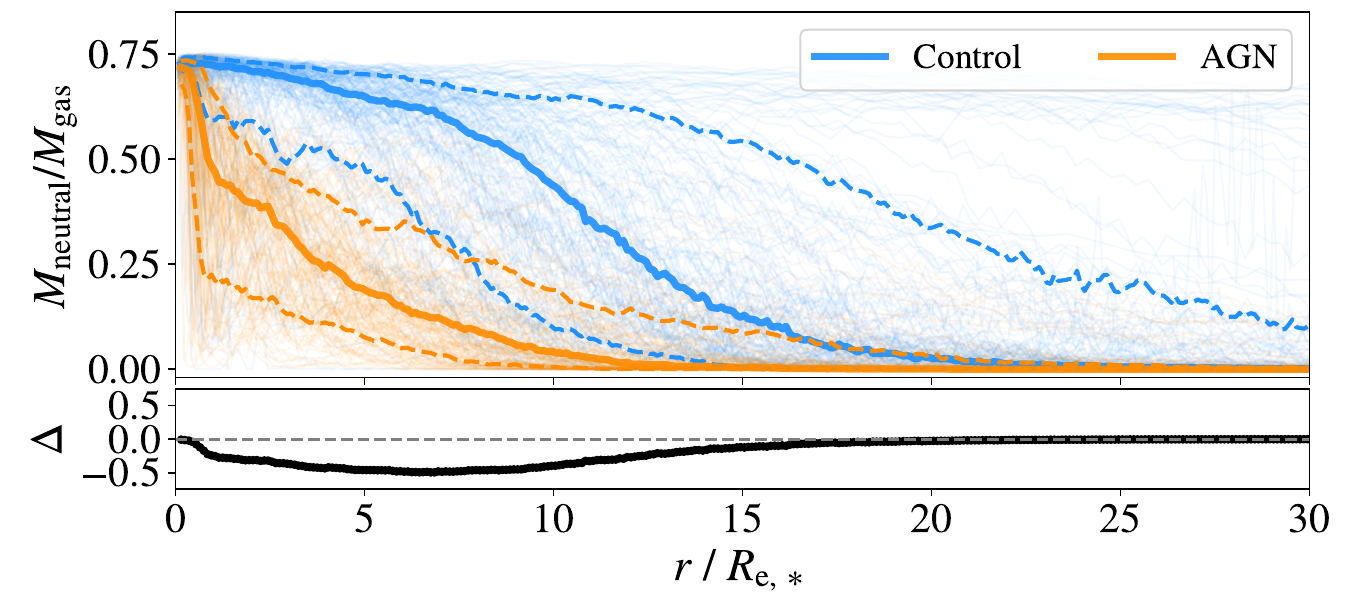}
    \includegraphics[width=0.47\textwidth]{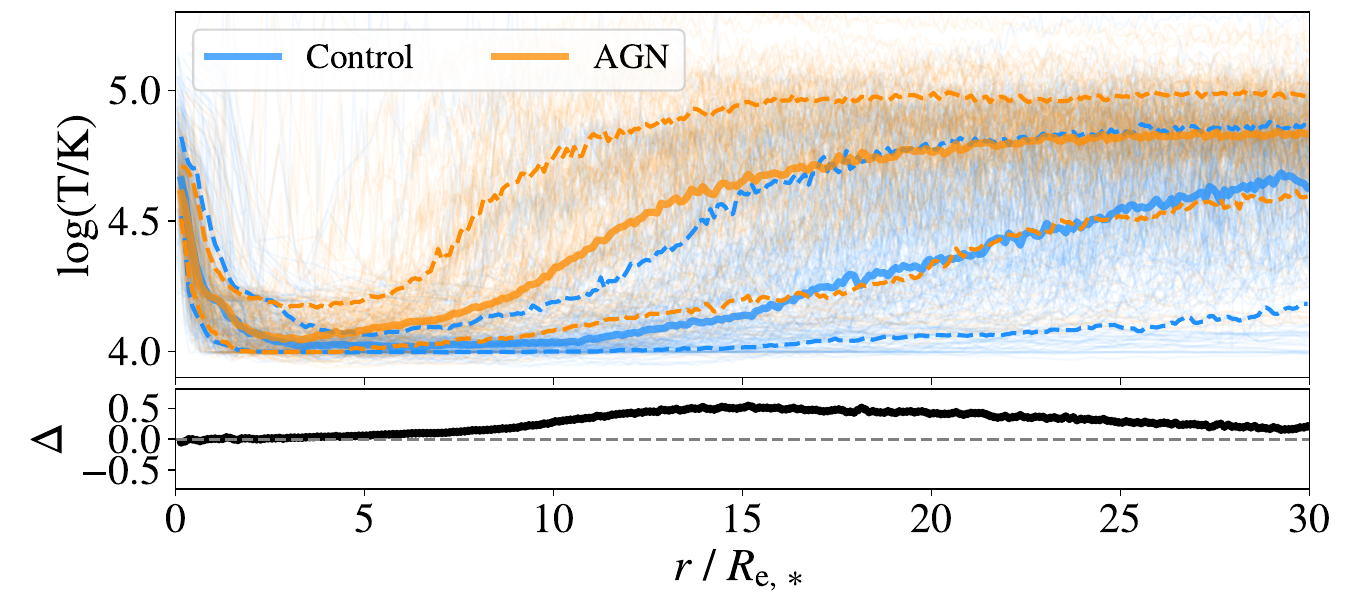}
    \caption{Spherical gas profiles of AGN (orange) and non-AGN control (blue) samples. From top to bottom, we show the radial profiles of the volumetric gas density, the neutral-to-total gas mass ratio, and the gas temperature. In both panels, the thin semitransparent lines represent the profiles of individual galaxies, the thick solid lines represent the median profile of the whole sample, and the dashed lines indicate the 16th and 84th percentiles. The radial distance to the galaxy center ($r$) is rescaled by $R_{\rm e,\ast}$. The difference between median AGN and non-AGN profiles for a given quantity ($\Delta$) is shown in black on the lower small sub-panels. The samples being compared here are $\mathcal{S}_{M_\ast \& M_{\rm 200c}}^{\rm WG}$ and $\mathcal{C}_{M_\ast \& M_{\rm 200c}}^{\rm WG}$.}
    \label{fig:profiles_wg}
\end{figure}

In this section we focus on the comparison of the specific AGN sample $\mathcal{S}_{M_\ast \& M_{\rm 200c}}^{\rm WG}$ with its respective control sample $\mathcal{C}_{M_\ast \& M_{\rm 200c}}^{\rm WG}$. The goal of focusing on this comparison is to ensure that differences in the neutral gas content are not consequences of a stellar or a halo mass bias, and also guarantee that all galaxies being compared have a gaseous component. We first analyze the difference in the gas radial profiles, more specifically, volumetric density ($\rho_{\rm gas}$), neutral-to-total mass ratio, and temperature profiles (Figure \ref{fig:profiles_wg}). 

From the comparison of density profiles, we see that AGN have slightly less dense gas from $\sim 2 R_{\rm e,\ast}$ to $\sim 20 R_{\rm e,\ast}$, however, the difference between the median profiles is well within the scatter of both samples. On the other hand, it is evident from the direct comparison of AGN and non-AGN median profiles that the neutral gas deficiency is strong already at the inner parts of the gas component ($\lesssim 2 R_{e,\ast}$), even considering the scatter of the profiles in each sample. The smaller values of $M_{\rm neutral}/M_{\rm gas}$ for the dwarf AGN are more significant in the $\sim 2 R_{\rm e,\ast}$ to $\sim 8 R_{\rm e,\ast}$ radial interval, where not even the scatter of AGN and non-AGN profiles overlap. Nonetheless, we can see that the AGN sample has relatively less neutral gas from $\sim 1 R_{\rm e,\ast}$ to $\sim 15 R_{\rm e,\ast}$. The temperature of the gas is related to the neutral hydrogen fraction in each gas cell - usually, hotter cells will have less neutral hydrogen -, but although we see clear differences in the neutral content for the inner part of the gas component, the same is not true for the gas temperature. The AGN have hotter gas in their outskirts ($r \gtrsim 10 R_{\rm e,\ast}$) and mostly similar temperatures for the gas within $\sim 5 R_{\rm e,\ast}$ - where gas cooling is stronger. 

\begin{figure}
    \centering
    \includegraphics[width=0.46\textwidth]{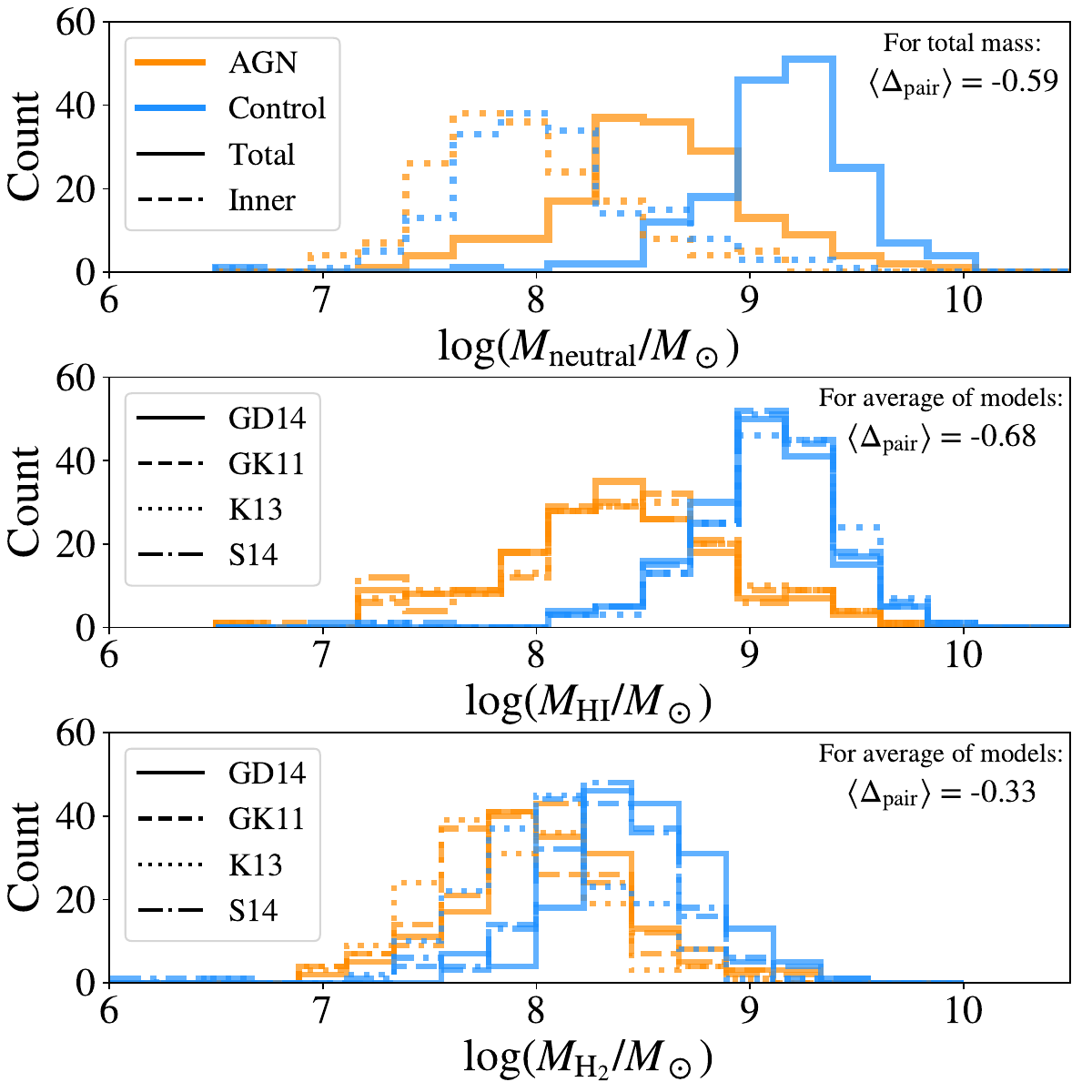}
    \caption{Histograms of neutral, neutral atomic (\ion{H}{i}), and molecular ($\rm H_2$) gas masses for AGN (orange) and non-AGN control (blue) samples. The upper panel shows the bound neutral gas masses, with the solid line indicating the total amount and the dashed line indicating only the mass within $2R_{e,\ast}$. The middle and lower panels show the total bound gas masses in the form of \ion{H}{i} and $\rm H_2$, taken from the TNG supplementary catalog \citep{Diemer2019}. Each line style of the histograms refers to a type of partition model considered by \cite{Diemer2019} in their post-processing: solid lines are from \cite{Gnedin_n_Draine2014}, dashed lines are from \cite{Gnedin_n_Draine2014}, dotted lines are from \cite{Krumholz2013}, and dotted-dashed lines are from \cite{Sternberg2014}. Here, we show different models to illustrate variations in the values of $M_{\rm \ion{H}{i}}$ and $M_{\rm H_2}$. In the upper right of each panel, we show the average pair-wise differences ($\Delta_{\rm pair}$) between the AGN and non-AGN logarithmic masses. The samples being compared here are $\mathcal{S}_{M_\ast \& M_{\rm 200c}}^{\rm WG}$ and $\mathcal{C}_{M_\ast \& M_{\rm 200c}}^{\rm WG}$.}
    \label{fig:HI_and_mol}
\end{figure}

In the upper panel of Figure \ref{fig:HI_and_mol}, we show the neutral gas masses in AGN and non-AGN at different apertures. It is clear that the total amount of neutral gas in the control galaxies is higher, with AGN hosts having, on average, 3.9 times less mass in this component than their matched non-AGN counterparts (from the $\mathcal{C}_{M_\star \& M_{\rm 200c}}^{\rm WG}$ sample). When only the stellar mass is controlled - more similar to what is commonly done in observational works \citep{Rembold2024, Alban2024, Gatto2025} - we found that the AGN hosts still have, on average, 2.5 times less neutral gas than non-AGN (see Appendix \ref{app:with_bh}).
To illustrate that the difference in $M_{\rm neutral}$ is stronger in the circumgalactic medium\footnote{Here defined as the gas outside $2 R_{\rm e,\ast}$ and still bound to the galaxy.} (CGM), we also show histograms of the masses within two stellar half-mass radii. As shown in the figure, the neutral gas mass confined to the inner parts of the galaxies is very similar between AGN and non-AGN samples. This result emphasizes that the strongest imprint the black hole accretion is leaving on dwarf galaxies is in the outskirts of their gaseous halo.

Using the supplementary catalog from \cite{Diemer2019}, we also analyze the difference in the individual phases of the neutral gas, that is, we study the mass distributions of \ion{H}{i} and molecular gas. As shown in the middle and lower panels of Fig. \ref{fig:HI_and_mol}, the deficit of neutral gas is also reflected in a deficit of \ion{H}{i} and molecular gas for the AGN, regardless of the partition model adopted. For the neutral atomic component, there is little variation on the distributions of mass with respect to models, and we find that $M_{\rm \ion{H}{i}}$ is, on average, 4.8 times lower in AGN when stellar and halo mass is matched in the control sample ($\mathcal{C}_{M_\ast \& M_{\rm 200c}}^{\rm WG}$), or 3 times lower when only stellar mass is matched ($\mathcal{C}_{M_\ast}^{\rm WG}$, see Appendix \ref{app:with_bh}). For the molecular component, the values of $M_{\rm H_2}$ show a stronger dependence on the adopted model, and on average, $M_{\rm H_2}$ is 2.1 times lower in the AGN relative to $\mathcal{C}_{M_\ast \& M_{\rm 200c}}^{\rm WG}$, and 1.6 times lower relative to $\mathcal{C}_{M_\ast}^{\rm WG}$.
There are a number of modeling choices underlying the values of $M_{\rm \ion{H}{i}}$ and $M_{\rm H_2}$ shown in Figure \ref{fig:HI_and_mol}, and the exploration of specific effects from each model is beyond the scope of this work - we refer the reader to the works of \cite{Diemer2018} and \cite{Diemer2019} for that. Nonetheless, the current results show that AGN presence in simulated dwarf galaxies of TNG50-1 is possibly associated with a decrease in their total \ion{H}{i} and molecular gas masses.

It is also important to note which of the two AGN feedback modes employed in TNG impacts the most. Active dwarf galaxies are selected here to have $\lambda_{\rm Edd} \geq 0.01$, and the transition for the low-accretion kinetic mode only happens if $\lambda_{\rm Edd} < \chi$, where $\chi = \min[0.002 \times (M_{\rm BH}/10^8 {\rm M_\odot})^2, 0.1]$ is a mass-dependent threshold \citep{Weinberger2017}. Since $\chi \lesssim 0.002$ for the dwarf galaxies hosting AGN at $z=0$, it is the high-accretion thermal feedback that dominates the liberated feedback energy.

\subsection{Evolution of gas properties}
\label{sec:evolution}

In the previous sections, we showed that the AGN hosts at $z=0$ have less neutral gas than non-AGN galaxies with similar stellar and halo masses. Since we can follow the complete history of individual galaxies in a simulation, we can explore the causal effects of black hole accretion on the gas phase of the simulated dwarf galaxies. As in the last section, here we focus on the $\mathcal{S}_{M_\ast \& M_{\rm 200c}}^{\rm WG}$ AGN sample and its $\mathcal{C}_{M_\ast \& M_{\rm 200c}}^{\rm WG}$ control sample for comparisons.

\begin{figure*}
    \centering
    \includegraphics[width=0.32\textwidth]{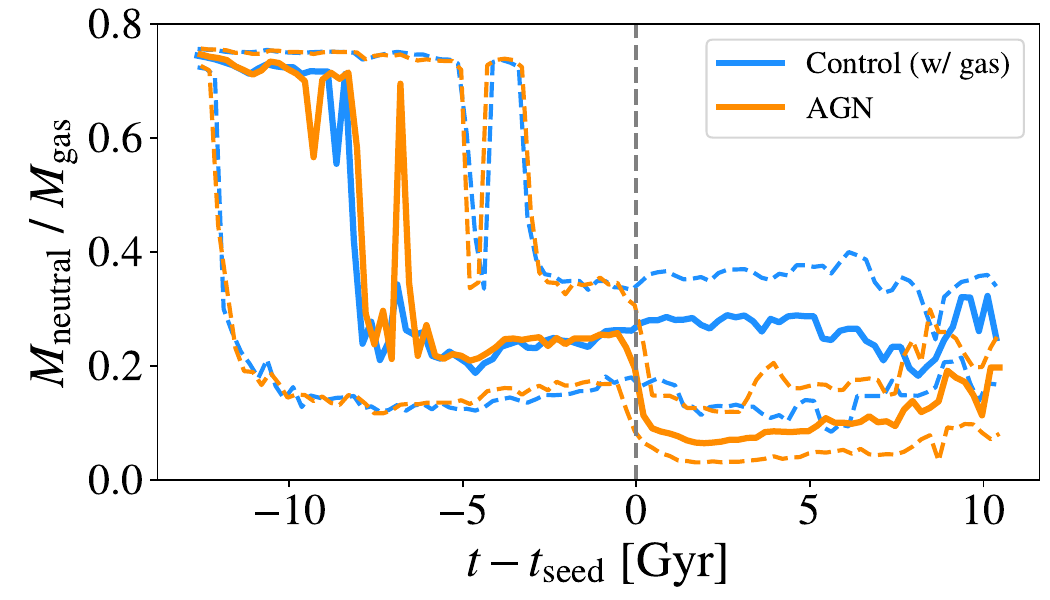}
    \includegraphics[width=0.32\textwidth]{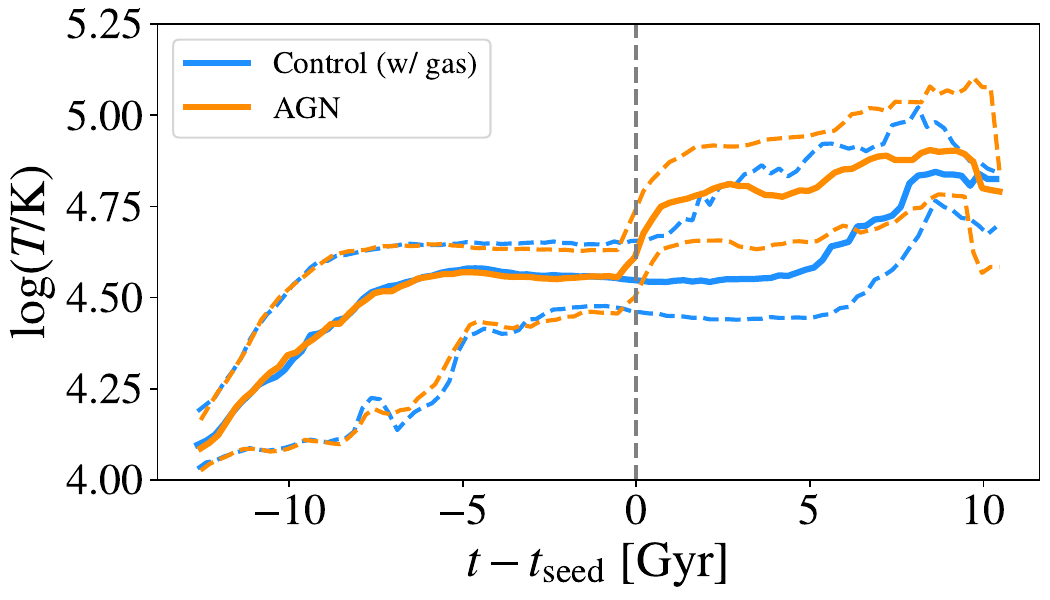}
    \includegraphics[width=0.32\textwidth]{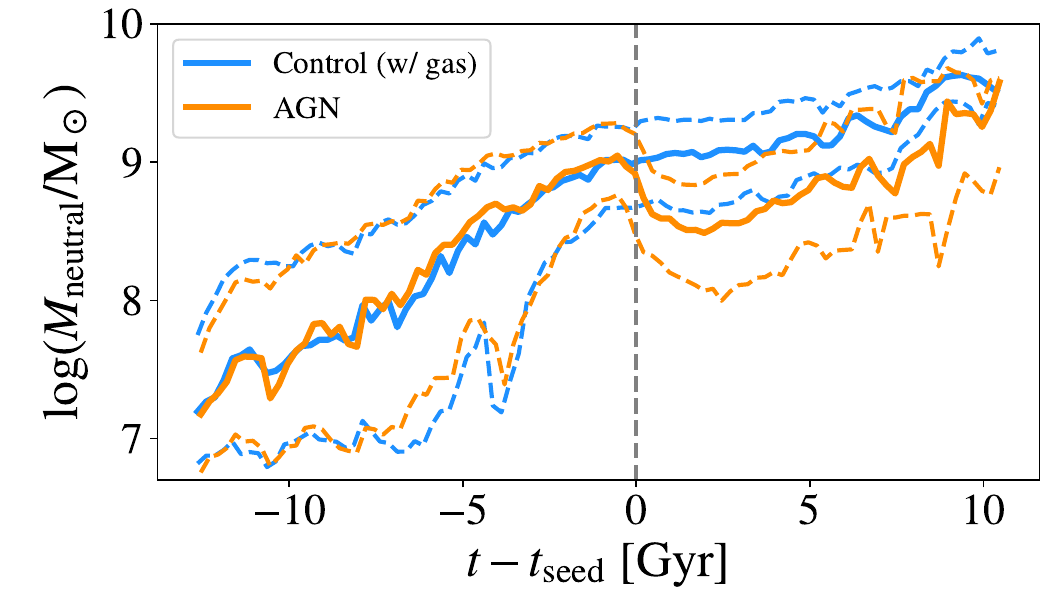}
    \includegraphics[width=0.32\textwidth]{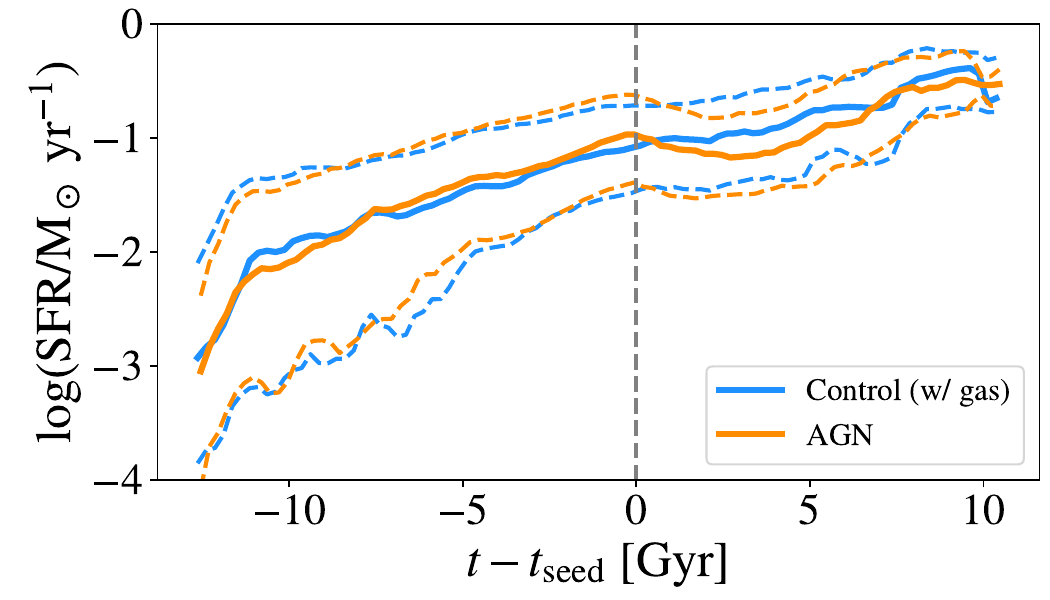}
    \includegraphics[width=0.32\textwidth]{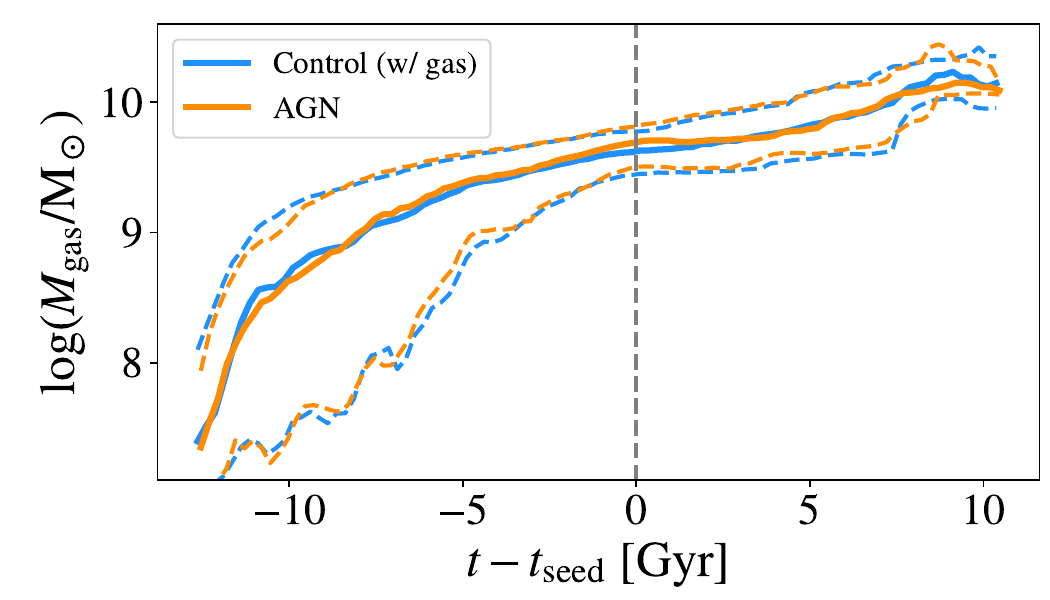}
    \includegraphics[width=0.32\textwidth]{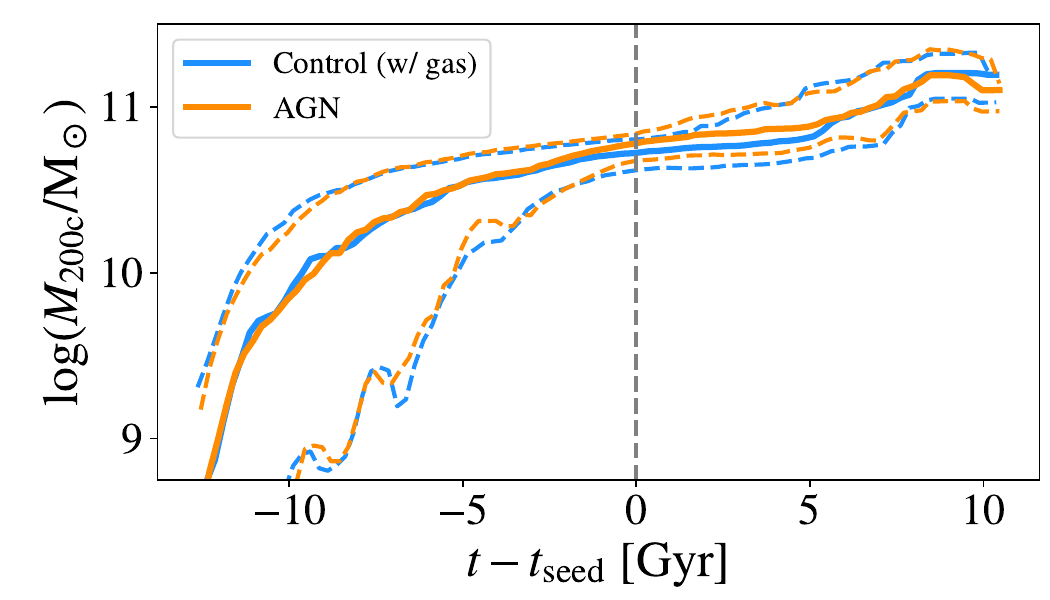}
    \caption{Evolution of properties in AGN (orange) and non-AGN control (blue) samples. The time variable on the horizontal axis is the cosmic time minus the seeding time ($t_{\rm seed}$) of the AGN host. where the median $t_{\rm seed}$ is 11 Gyr, and the 16th (84th) percentile is 7.3 Gyr (13 Gyr). From top to bottom and left to right, each panel represents the evolution of the neutral-to-total gas ratio,  gas temperature, neutral gas mass, star formation rate, total gas mass, and host halo mass. All bound gas particles are considered. The solid lines indicate the median evolution of a given quantity for the whole sample, while the dashed lines indicate the 16th and 84th percentiles. Different from Figure \ref{fig:profiles_wg}, here we choose not to plot the data of each galaxy for better visualization. The samples being compared here are $\mathcal{S}_{M_\ast \& M_{\rm 200c}}^{\rm WG}$ and $\mathcal{C}_{M_\ast \& M_{\rm 200c}}^{\rm WG}$.}
    \label{fig:history_wg}
\end{figure*}

In Figure \ref{fig:history_wg}, we show the evolution of different properties of AGN and non-AGN control samples. It is clear from the evolution of the $M_{\rm neutral}/M_{\rm gas}$ that the AGN hosts suffer a sudden decline in their neutral content after the seeding of the black holes, which are actively accreting as soon as they are seeded\footnote{In the IllustrisTNG BH model, as long as there is gas available, the black holes are accreting at the pure Bondi–Hoyle–Lyttleton rate, limited by the Eddington rate \citep{Weinberger2017}.}. The evolution of the gas temperature also reflects this effect since the AGN also present higher temperatures compared to the control galaxies at $t>t_{\rm seed}$. The AGN essentially stay in the high-accretion regime during their entire evolution. Thus, the direct effect of AGN is the injection of thermal energy into the gas near the black hole \citep{Weinberger2017}. It could be argued that the decrease in the neutral gas content is related to an increase in the SFR of the galaxies, which in turn would cause more intense stellar feedback. However, as is shown in the bottom left panel of Figure \ref{fig:history_wg}, non-AGN and AGN have similar SFR before $t_{\rm seed}$, making it unlikely that star formation processes are solely responsible for the decrease of $M_{\rm neutral}$. Indeed, the median SFR of the AGN hosts immediately after $t_{\rm seed}$ is even slightly lower, although the difference is within the scatter of both samples. Furthermore, there is no strong difference in the evolution of the total gas mass or the halo mass, excluding the possibility of the decrease in the neutral gas fraction being a consequence of different halo assembly histories in AGN and non-AGN samples.  

On average, for $\mathcal{C}_{M_\ast \& M_{\rm 200c}}^{\rm WG}$, the difference in neutral gas mass is set in place quickly after $t_{\rm seed}$, being present for several Gyr after. However, by analyzing galaxies in $\mathcal{C}_{M_\ast \& M_{\rm 200c}}^{\rm WBH}$ - control sample with restriction of having a black hole (see Appendix \ref{app:with_bh}) - we found a different picture. The difference in neutral gas does not appear immediately after black hole seeding, but $\sim 4$~Gyr later, when the accretion rates in control galaxies start to decrease (see Figure \ref{fig:history_edd_ratio}). Thus, the neutral gas deficiency in dwarf AGN is not only directly linked to the black hole seeding, but also to its black hole matter accretion history in the last few Gyr. 

\begin{figure*}
    \centering
    \includegraphics[width=0.245\textwidth]{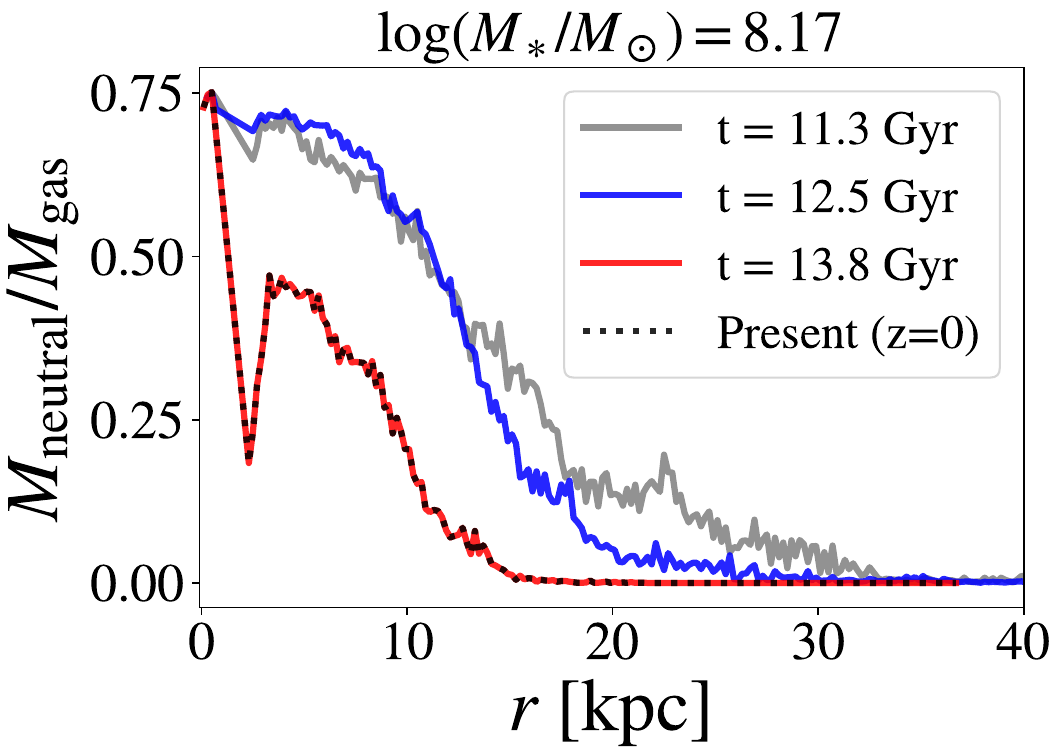}
    \includegraphics[width=0.245\textwidth]{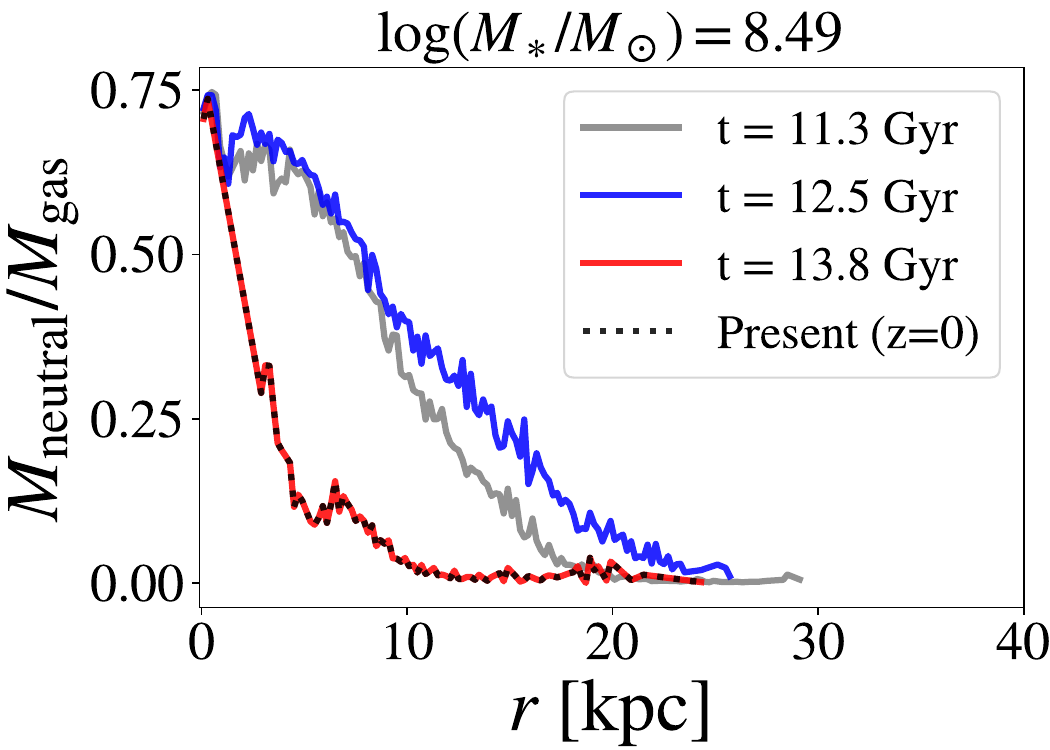}
    \includegraphics[width=0.245\textwidth]{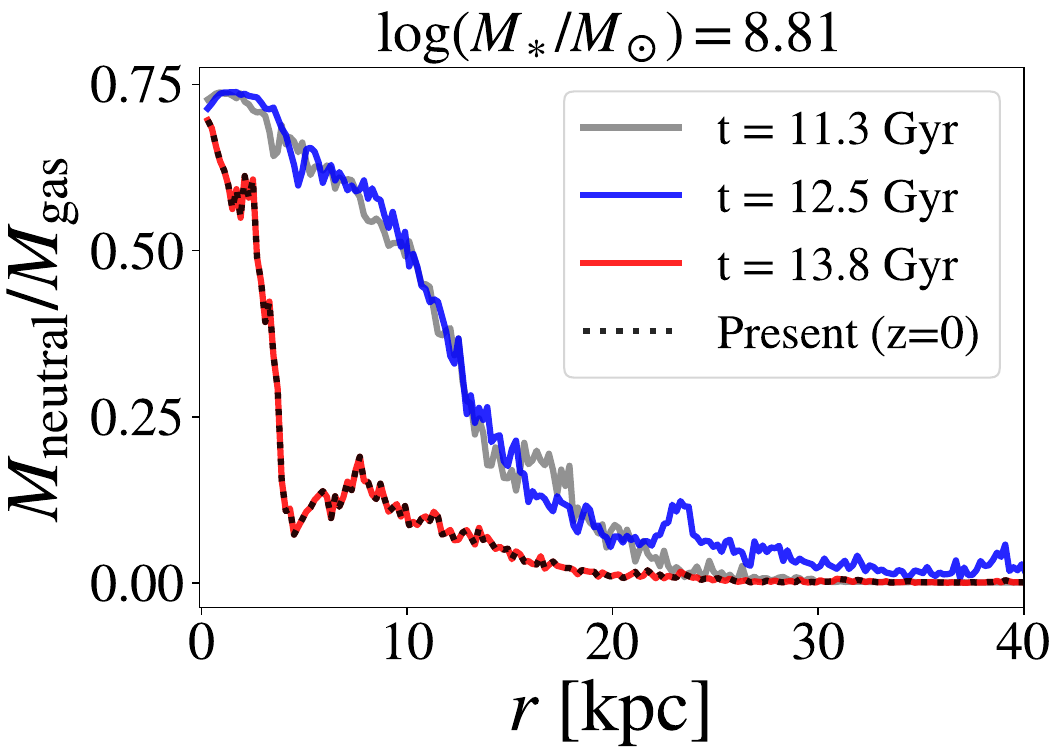}
    \includegraphics[width=0.245\textwidth]{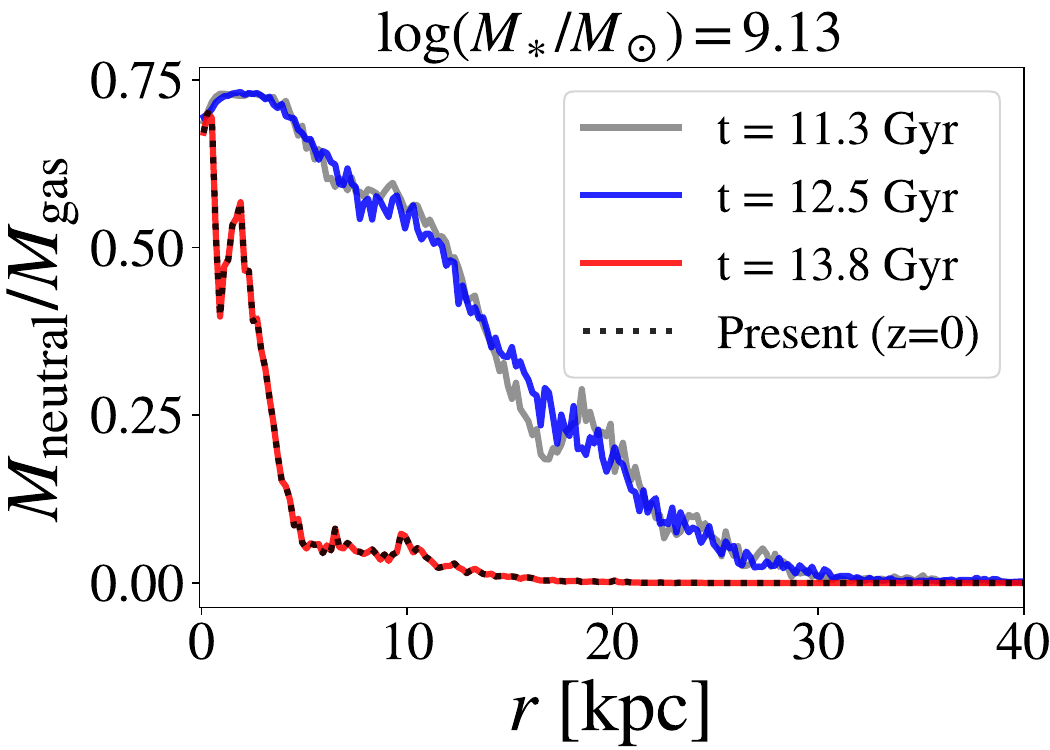}

    \includegraphics[width=0.245\textwidth]{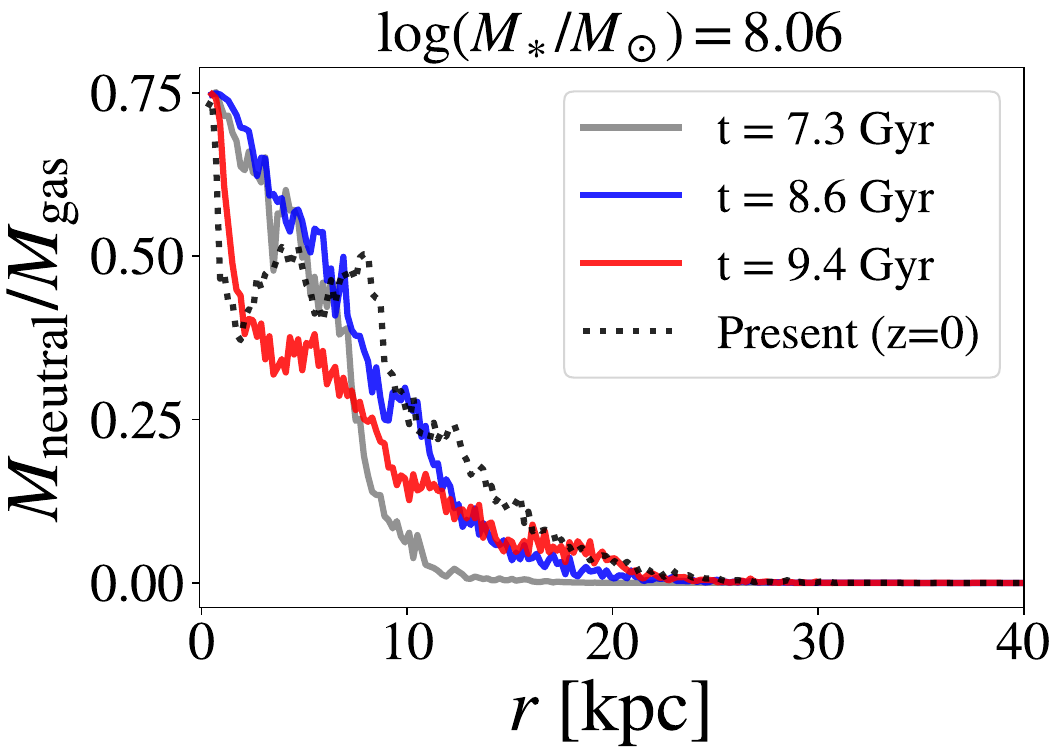}
    \includegraphics[width=0.245\textwidth]{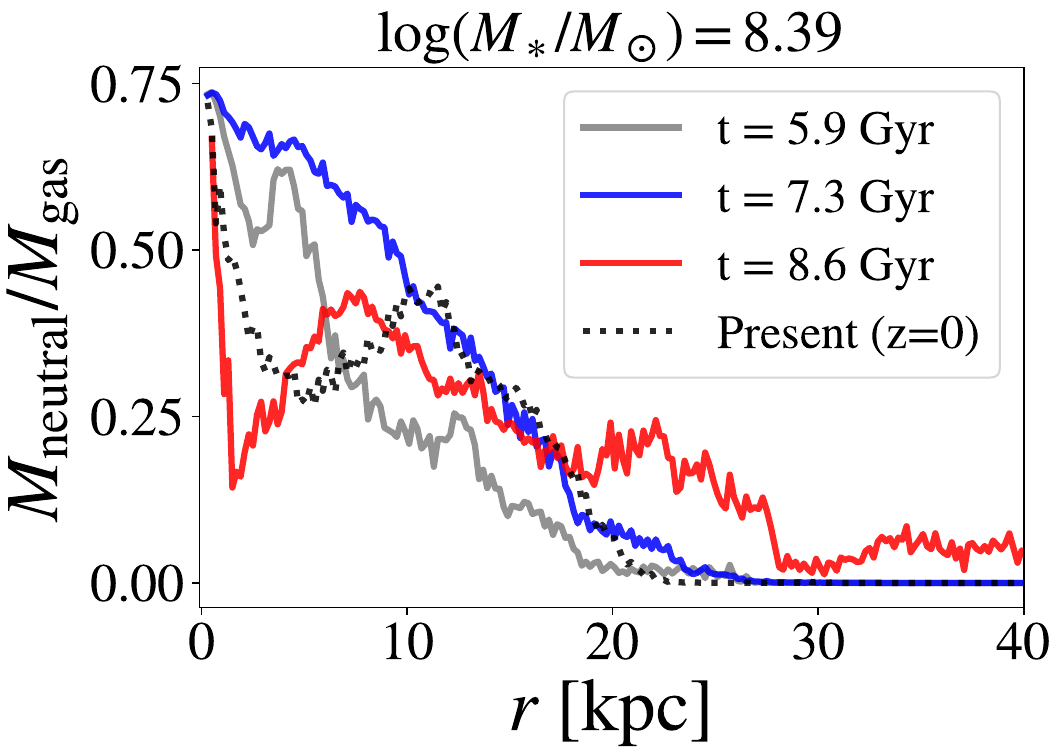}
    \includegraphics[width=0.245\textwidth]{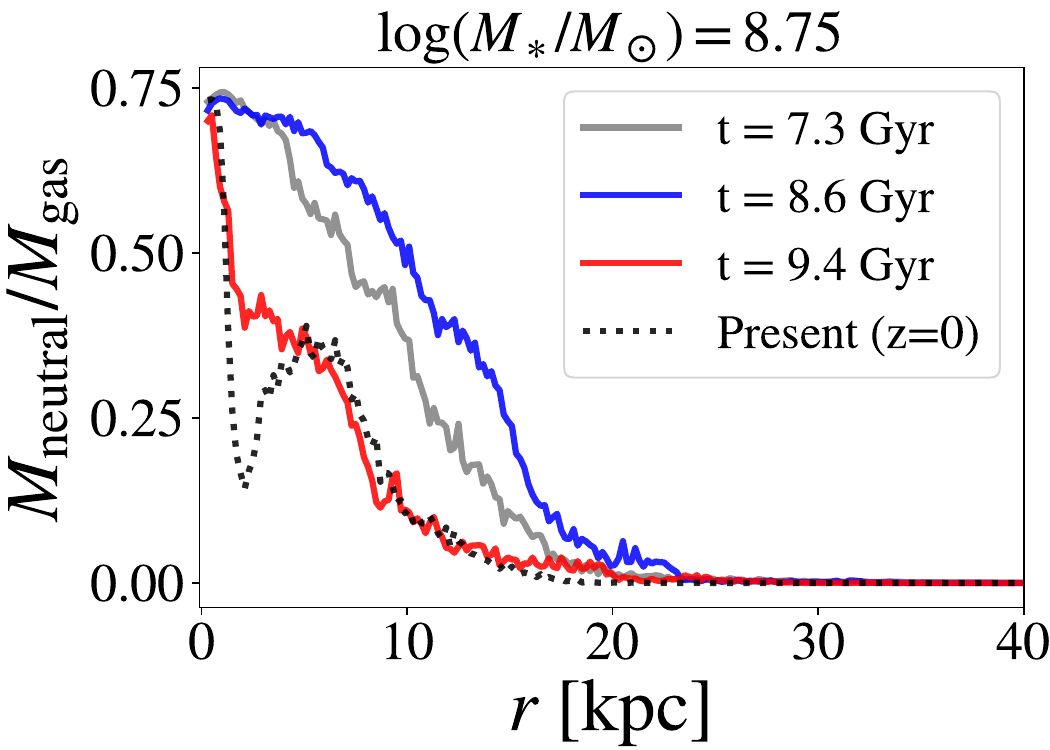}
    \includegraphics[width=0.245\textwidth]{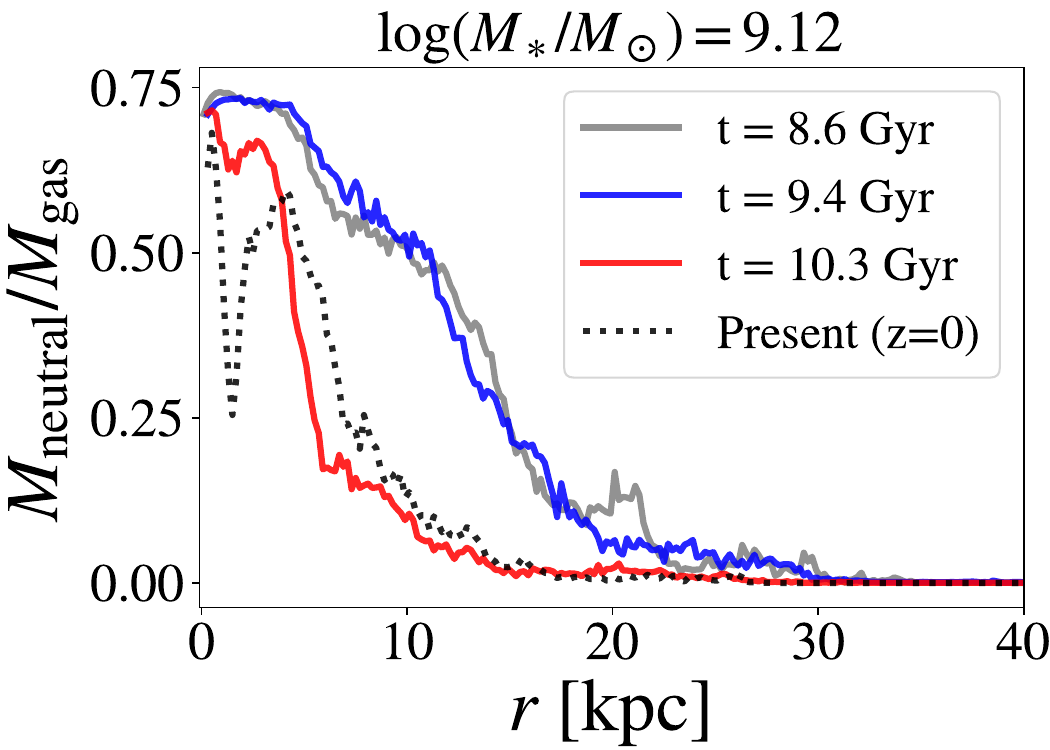}    
    
    \caption{Examples of neutral-to-total gas profile evolution for a few AGN selected at $z=0$. In each panel, we show the profile of a specific galaxy at different times, before and after the black hole seeding. The two latest available profiles before $t_{\rm seed}$ are shown as blue and gray solid lines, the earliest available profile after $t_{\rm seed}$ is shown as a red solid line, and the present profile is shown as a black dotted line. The availability of the profiles depends on the type of simulation snapshot; see Section \ref{sec:method_evolution} for details. Each row is dedicated to a bin of black hole seeding times, in the top row $t_{\rm seed} < 1$ Gyr, and in the bottom row $t_{\rm seed} > 3$ Gyr. From left to right, the stellar mass of the example galaxies increases in bins of roughly 0.25 dex.}
    \label{fig:gas_fraction_evolution}
\end{figure*}

We can also see the direct effect of AGN feedback in the neutral-to-total gas profiles of individual galaxies, as illustrated with a few examples in Figure \ref{fig:gas_fraction_evolution}. We show the profiles at three different times, for randomly selected AGN hosts with different stellar masses and different $t_{\rm seed}$. In general, all the $M_{\rm neutral} / M_{\rm gas}$ profiles suffer a strong decrease in their values after the AGN starts, with the most noticeable difference appearing on the galaxies with recent BH seeding ($t_{\rm seed} < 1$ Gyr), regardless of stellar mass. Due to the limited size of the sample, a binning of the galaxies in the space of $M_\ast$ and $t_{\rm seed}$ results in only a few galaxies per bin, hindering a robust statistical analysis. Thus, we restrict our analysis of the effect of these variables on the profiles to a qualitative presentation of illustrative random examples. Since the impact of AGN on $M_{\rm neutral} / M_{\rm gas}$ profiles is more evident on the hosts with more recent $t_{\rm seed}$, it is possible that the AGN does not permanently decrease the neutral gas content on dwarf galaxies.

\section{Discussion}
\label{sec:discussion}

\subsection{Dwarf AGN demographics}
\label{sec:agn_fraction_obs}
The demographics of AGN in dwarf galaxies are essential for understanding galaxy evolution \citep{Mezcua2020, Arjona-Galvez2024} and may inform studies on massive black hole formation \citep{Greene2020, Reines2022}. Thus, checking the BH occupation and AGN fractions on state-of-the-art cosmological simulations is important. In observations, the selection of dwarf AGN is challenging, and the determination of the AGN fraction is subject to several biases \citep{Hainline2016, Agostino2019, Latimer2021, Sturm2025}, which can result in significant variations of reported fractions \citep{Haidar2022}. For example, sources identified by X-rays and infrared diagnostics are often not classified as AGN through optical spectroscopy \citep{Wasleske2024}. This lack of overlap in the samples selected by different techniques can be partly explained by a few factors, such as dilution of the AGN signatures within star formation \citep{Moran2002, Trump2015}, varying ionizing spectrum with decreasing BH mass \citep{Cann2019}, off-nuclear emission \citep{Thygesen2023}, and changes in the emission line ratios due to low metallicities \citep{Groves2006}.

As shown in Section \ref{sec:results_agn_fraction}, our analysis of the small volume of TNG50 yields an overall BH occupation fraction on dwarf galaxies of 44\%, which is broadly consistent with observational constraints for $\log (M_\ast / {\rm M_\odot}) \leq 9.5$ \citep{Miller2015, Greene2020, Cho2024}. Furthermore, we found AGN fractions that span more than one order of magnitude - $\sim 1$\% ($\lambda_{\rm Edd, min} = 0.05$) to $\sim 24$\% ($\lambda_{\rm Edd, min} = 0.01$) - depending on the minimum accretion rate adopted for AGN selection. 
In contrast, different observational estimates, based on X-ray \citep{Schram2013,Lemons2015,Miller2015,Mezcua2018,Birchall2020} and optical \citep{Reines2013,Sartori2015,Cho2024} selection methods, found AGN fractions usually below 2\% for galaxies with $\log (M_\ast / {\rm M_\odot}) \leq 9.5$. Only a few exceptions in the literature report fractions of the same order of magnitude as we find in TNG50. For example, \cite{Kaviraj2019} selected AGN in dwarf galaxies at $0.1 \leq z \leq 0.3$ using WISE infrared photometry, and found the AGN fraction to be in the range of 10\% to 30\% for $8 \leq \log (M_\ast / {\rm M_\odot}) \leq 9.5$. A more recent work, by \cite{Mezcua2024} analyzed data from MaNGA DR17 and selected AGN using a combination of [\ion{N}{ii}]–BPT, [\ion{S}{ii}]–BPT and [\ion{O}{i}]–BPT diagrams and the WHAN diagram \citep{CidFernandes2010}. Their initial estimate for AGN fraction in MaNGA was of $\sim 20$~\%, but after a correction in the stellar masses and SFRs used in their paper \citep{Mezcua2025correction}, considering only galaxies with $\log (M_\ast/{\rm M_\odot}) \leq 9.5$ in their sample, the updated fraction\footnote{Their sample have galaxies with stellar masses up to $\log (M_\ast/{\rm M_\odot}) = 10$, thus using their whole sample to compute the fraction may result in different values than discussed here.} falls to $\sim 14$~\%. On the other hand, a more recent work by \cite{Pucha2025} presents an analysis of DESI survey and selects AGN also using BPT diagrams, finding an AGN fraction of $\sim$~1.2\% (computed from their figure 7) over all dwarfs with $\log (M_{\rm \ast}/{\rm M_\odot}) \leq {9.5}$. For a comprehensive list of AGN fractions estimated for galaxies with $\log(M_\ast/{\rm M_\odot}) \leq 10$ at low redshift, we refer the reader to Appendix A of \cite{Haidar2022}, which exhaustively lists the reported AGN fractions in the literature. 

Except for a few works, most of the AGN fraction estimates in the literature are at least one order of magnitude below the value of 24\% that we found in our fiducial selection sample here ($\lambda_{\rm Edd,min} = 0.01$). This discrepancy between the simulation and observations may arise from a heavy BH seed implemented in TNG \citep{Habouzit2021, Haidar2022}, but also from an intrinsic difficulty in detecting all active black holes in observed dwarf galaxies, as discussed above. Nonetheless, it is important to be careful when comparing AGN fractions from studies that use different AGN detection techniques, since they can trace different AGN populations \citep{Riffel2023, Wasleske2024,Alban2024, Tian2025}, and also possibly introduce their own specific biases on the estimates of AGN fraction in dwarfs. 

\begin{figure}
    \centering
    \includegraphics[width=0.45\textwidth]{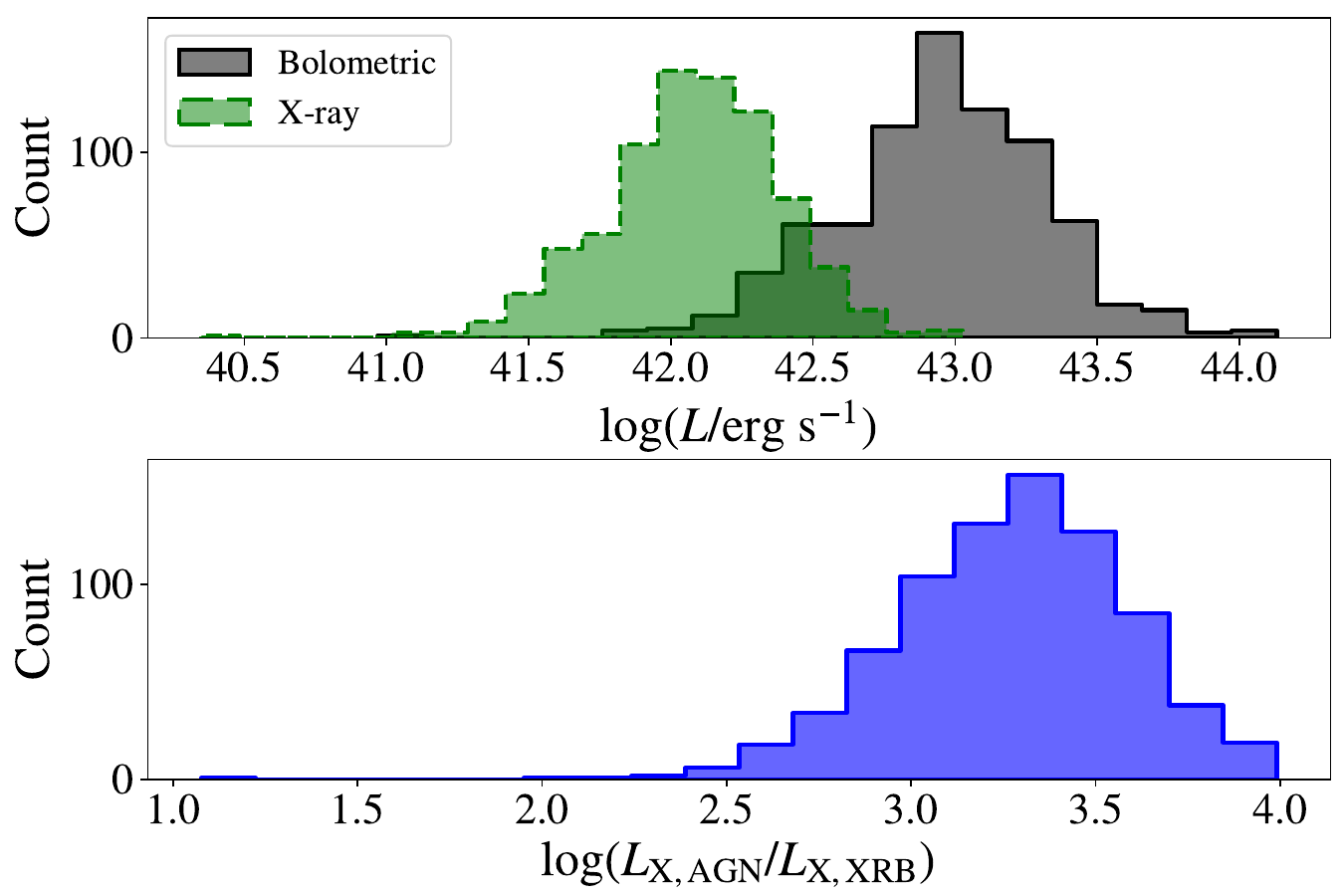}
    \caption{Top: histogram of AGN luminosities, bolometric (black), and total (0.5-10 keV) X-ray (green). Bottom: histogram of the ratio between AGN and X-ray binaries (XRB) luminosities. No extinction or obscuration is considered (see Section \ref{sec:data_methods} for details on calculation).}
    \label{fig:luminosities}
\end{figure}

As we show in the top panel of Figure \ref{fig:luminosities}, the bulk of the dwarf AGN identified in TNG50 have bolometric luminosities ($L_{\rm bol}$) mostly between $10^{42.5}$ and $10^{43.5}$ erg $\rm s^{-1}$, with X-ray luminosities mostly in the range of $10^{41.5}$ to $10^{42.5}$ erg $\rm s^{-1}$. It has been argued that the X-ray detection of AGN in dwarf galaxies could be contaminated by X-ray binaries \citep{Mezcua2018,Thygesen2023}; thus, it is valid to compare the luminosities of both types of sources in order to understand which dominates the X-ray emission. We can also see in the lower panel of Figure \ref{fig:luminosities} that XRB luminosities are expected to be more than 100 times fainter than AGN luminosities. Similarly, \cite{Schirra2021} analyzed TNG100 and found that the AGN usually dominate the X-ray luminosity for $\log (M_\ast/{\rm M_\odot}) < 10.5$. Thus, according to this simple estimate, AGN should dominate the X-ray luminosities in active dwarf galaxies with similar $M_\ast$ and $L_{\rm bol}$ as those studied here. 

In our analysis, the minimum stellar mass chosen for the selection of dwarf AGN galaxies does not strongly affect the AGN fraction, at least in the $8 \lesssim \log (M_\ast / {\rm M_\odot}) \lesssim 9$ mass interval. However, as we increase the minimum threshold for the Eddington ratio, the AGN fraction rapidly decreases, and since previous works found a relation between BH accretion rate and \ion{H}{i} gas \citep{Li2025}, it is relevant to verify if our results change with $\lambda_{\rm Edd}$. As shown in Figure \ref{fig:agn_fraction}, the AGN fraction is most similar to observational estimates only when $\lambda_{\rm Edd,min} \gtrsim 0.05$. If we consider AGN only dwarf hosts with $\lambda_{\rm Edd} \geq 0.05$, the results for neutral gas deficiency remain the same (see Appendix \ref{app:selection}).

\subsection{Neutral gas deficiency}
As shown by the evolution of the global gas properties and the evolution of individual profiles, the decrease in the neutral gas mass is a direct effect of AGN feedback on dwarf galaxies. The reduction of the neutral component is not only caused by the initial triggering of the feedback but is also maintained by the cumulative effect of multiple accretion peaks over a few billion years. Previous theoretical works also find that the cumulative energy injected by the AGN onto the gas can decrease the atomic-to-stellar mass ratio \citep{Weinberger2017, Ma2022, Li2025}. While it has been demonstrated that the kinetic feedback mode holds an important role in the quenching of more massive simulated galaxies \citep{Weinberger2017,Terrazas2020}, this type of feedback is not present in the dwarf AGN analyzed here. The black hole feedback in action is the high-accretion thermal feedback mode employed in the IllustrisTNG model. Similar results were found by \cite{Li2025}, which analyzed larger - but with lower resolution - simulations (TNG100 and EAGLE) and found that \ion{H}{i} gas is regulated mainly by thermal-mode AGN feedback in TNG. \cite{Sharma2020} also finds lower \ion{H}{i} content in isolated dwarf galaxies ($9 \leq \log (M_\ast / {\rm M_\odot}) \leq 10 $), with over-massive black holes in the ROMULUS25 simulation. Furthermore, using the AURIGA galaxy formation physics model, \cite{Arjona-Galvez2024} compared simulation runs with and without AGN feedback in dwarf galaxies. They found that the feedback in these systems leads to a decrease in the amount of \ion{H}{i} gas. However, the reduction in \ion{H}{i} is not due to the gas being expelled from the galaxy, but rather due to heating of the gas in the vicinity of the AGN, which lowers the fraction of gas in the cold, neutral phase. However, \cite{Koudmani2022} finds different results regarding the gas content in dwarf AGN. They performed a large suite of zoom-in simulations varying seeding times, seeding masses, and the length of the AGN duty cycle, and founds that the \ion{H}{i} gas masses of dwarfs are either completely suppressed or within the scatter of the observed $M_{\rm \ion{H}{i}} - M_{\ast}$ relations for field dwarfs.

On the observational domain, different results are reported on the \ion{H}{i} content of AGN hosts. While some studies found that AGN hosts and inactive galaxies have similar $M_{\rm \ion{H}{i}}/M_{\ast}$ for $\log(M_\ast/{\rm M_\odot}) \geq 10$ \citep{Fabello2011}, others found that AGN can be even more gas-rich than their non-AGN counterparts \citep{Ho2008}. Furthermore, a few works predict mixed results for the stellar mass range of $9 \leq \log(M_\ast/{\rm M_\odot}) \lesssim 10$ \citep{Bradford2018, Elisson2019}. \cite{Bradford2018} investigates the global \ion{H}{i} content of isolated galaxies selected from the SDSS spectroscopic survey with optical evidence of AGN, identifying a set of galaxies at large distances from the BPT star-forming sequence having lower than expected \ion{H}{i} masses. The galaxies in which this \ion{H}{i} gas deficit was found are in the stellar mass range of $9.2 < \log (M_\ast/{\rm M_\odot}) < 9.5$. In another work, \cite{Elisson2019} investigates the \ion{H}{i} gas fractions of AGN hosts in the extended GALEX Arecibo SDSS Survey (xGASS), finding that at fixed stellar mass, AGN hosts with $9 < \log (M_\ast/{\rm M_\odot}) < 9.6$ show an \ion{H}{i} deficit that reaches a factor of $\sim 2$ when the control sample is matched by stellar mass and SFR. 

\begin{figure}
    \centering
    \includegraphics[width=0.49\textwidth]{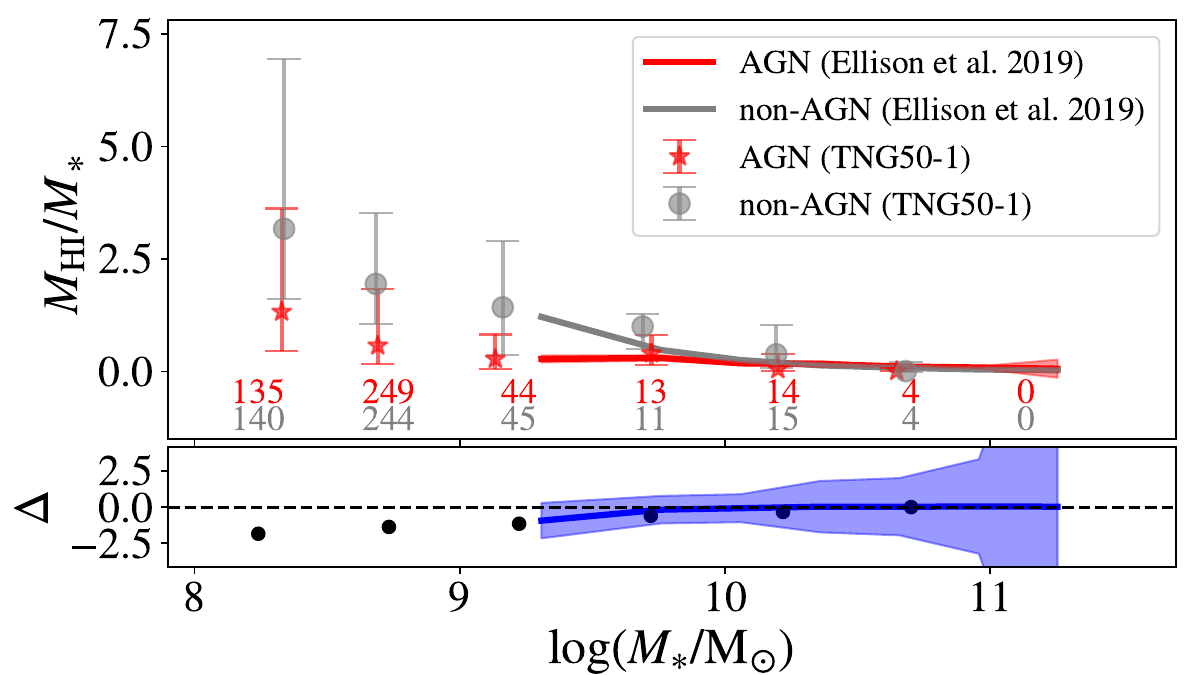}
    \caption{Median atomic-to-stellar mass ratio ($M_{\rm \ion{H}{i}}/M_\ast$) for AGN (red stars) and non-AGN (gray circles) in different stellar mass bins (upper panel) of TNG50-1. Error bars mark the 16th and 84th percentiles for each bin. The small lower panel indicates the difference between the medians of the AGN ($\lambda_{\rm Edd} \geq 0.01$) and non-AGN samples. The position of the markers on the x-axis corresponds to the median stellar mass of the galaxies in the bin. The number of objects in each mass bin is indicated by their corresponding colors. The samples AGN and non-AGN are paired by stellar mass and contain only central galaxies. We also plot the $M_{\rm \ion{H}{i}}/M_\ast$ values (red and gray solid lines) and uncertainties (shaded regions) reported on \cite{Elisson2019} (their figure 8), which compares AGN and non-AGN in fixed stellar-mass bins. In the bottom panel, we show as a blue line the corresponding AGN to non-AGN difference from \cite{Elisson2019}.}
    \label{fig:HI_Mstar_frac}
\end{figure}

To investigate the trend of \ion{H}{i} gas deficiency as a function of stellar mass, we selected all central galaxies in TNG50-1 containing neutral gas and computed $M_{\rm \ion{H}{i}}/M_{\ast}$ in logarithmic stellar mass bins of 0.5 dex from $\log(M_\ast/{\rm M_\odot}) = 8$ to $\log(M_\ast/{\rm M_\odot}) = 11.5$. We then create a sample of AGN ($\lambda_{\rm Edd} \geq 0.01$) and a control sample of non-AGN matched only by stellar mass\footnote{We performed testes pairing also by SFR, and our conclusions remain the same.} , similar to what is commonly done in other observational works \citep{Rembold2024, Alban2024, Gatto2025}. The comparison of the atomic-to-stellar mass ratio of AGN versus non-AGN is shown in Figure \ref{fig:HI_Mstar_frac} - with $M_{\rm \ion{H}{i}}$ being the average of the four volumetric estimates from \cite{Diemer2019}. We find a qualitatively similar trend as the observational results from \cite{Elisson2019}, that is, the deficiency of \ion{H}{i} gas in AGN hosts becomes stronger for lower stellar masses, reaching the maximum difference in $M_{\rm \ion{H}{i}}/M_{\ast}$ at the minimum $M_\ast$ considered here. 

Our findings could be further tested through a detailed comparison with radio observations. For example, one could combine catalogs of dwarf AGN candidates from SDSS and DESI \citep{Reines2013,Pucha2025} with the single-dish mass estimates from the Arecibo Legacy Fast ALFA (ALFALFA) survey \citep{Haynes2018} or the more recent FAST All Sky \ion{H}{i} Survey (FASHI) \citep{Zhang2024}, both with large coverage of the north sky. More than 80\% of the dwarf AGN selected in this work have $\log (M_{\rm \ion{H}{i}}/{\rm M_\odot}) > 8$, which are measurable up to $z \approx 0.015$ combining ALFALFA and FASHI catalogs, or $z \approx 0.01$ assuming only half of the total \ion{H}{i} mass is detected. However, for a robust comparison, with a large sample (number of objects $>$ 500) of dwarf AGN candidates, deeper radio observations would be needed to cover redshifts up to $z=0.05$. 

Regarding molecular gas, studies on local luminous AGN found that they have similar molecular gas fractions to inactive galaxies with similar infrared luminosity (proxy to stellar mass) \citep{Rosario2018}, similar sSFR \citep{Saintonge2017}, or compared to normal star-forming disc-dominated galaxies \citep{Husemann2017}. However, these works address the molecular gas content of AGN in hosts mostly with $\log (M_\ast / {\rm M_\odot}) \gtrsim 9.5$, and a similar analysis focusing on a sample of confirmed dwarf AGN would be useful to test our results.

As shown in the examples of Figure \ref{fig:example_pair} and in our results in Section \ref{sec:properties}, the dwarf galaxies hosting AGN can have similar stellar structures and similar stellar properties as other non-AGN galaxies of comparable stellar and halo masses. However, the distribution and global properties of their CGM can be radically different, especially in terms of the gas thermodynamic properties. A hotter gas reservoir, with less neutral gas available, may entail a decline of future star formation on the central dwarf galaxies that suffered from the black hole feedback. Furthermore, given the frailty of the CGM of dwarf galaxies during interaction with more massive galaxies \citep{Pearson2016,Zhu2024}, the AGN feedback could aggravate the situation, facilitating further gas removal in case they become satellites in groups or clusters. 
In this sense, the analysis of the CGM in the low-mass regime may be a powerful approach to understanding the baryon cycle of dwarf galaxies \citep{Piacitelli2025} and also to test the predictions of cosmological simulations employing completely different subgrid physics \citep{Zinger2020,Liu2025CGM,Medlock2025}. Thus, our results contribute to ongoing efforts by providing useful insights into the interaction between AGN and the gaseous component of central dwarf galaxies, especially in light of upcoming surveys from new facilities (e.g., Vera C. Rubin Observatory, Square Kilometer Array), which will expand our view of galaxies down to lower masses.

\section{Conclusions}
\label{sec:conclusions}
In this work, we searched for dwarf galaxies hosting AGN in the TNG50-1 cosmological simulation and investigated their demographics. We also studied the effects of AGN feedback by comparing these active dwarf galaxies with inactive ones using multiple control samples, paired by stellar and halo mass. From our analysis of central dwarf galaxies ($8 \leq \log M_\ast/{\rm M_\odot} \leq 9.5$) at $z=0$, we can summarize our conclusions as follows:

\begin{itemize}
    \item The fraction of dwarf galaxies hosting AGN at $z=0$ varies from 1\% ($\lambda_{\rm Edd,min} = 0.05$) to 24\% ($\lambda_{\rm Edd,min} = 0.01$), depending on the adopted threshold for Eddington ratio (Figure \ref{fig:agn_fraction}). The simulation may overestimate the fraction of active BHs in dwarf galaxies compared to observations, though different Eddington ratio cuts and AGN detection methods can yield varying reported fractions;
    \item AGN hosts are deficient in neutral gas when compared to non-AGN of similar stellar mass, SFR, and halo mass. They also have more extended gas halos, having gas radii at least $\sim 10$~kpc larger (Figure \ref{fig:boxplots1});
    \item The sample of AGN hosts have $\log ({\rm sSFR})$ lower by $\gtrsim 0.1$~dex compared to non-AGN of similar stellar and halo mass. On the other hand, we found no differences in stellar metallicity and age (Figure \ref{fig:boxplots2});
    \item We also found no distinguishable difference in the environment of dwarf AGN hosts and control galaxies as measured by the distance to their neighbors (Figure \ref{fig:boxplots2});
    \item The neutral gas deficiency in AGN hosts is stronger beyond two stellar half-mass radii ($2R_{\rm e,\ast} \approx 2.8$~kpc, typically). The onset of this dearth in neutral gas in the CGM of dwarf galaxies is caused by the BH thermal feedback (Figures \ref{fig:profiles_wg} and \ref{fig:history_wg}); 
    \item The neutral gas component in the AGN sample is, on average, $\sim 3.9$ times less massive than in the non-AGN control sample (matched by stellar and halo mass). Similarly, the \ion{H}{i} mass in AGN hosts is $\sim 4.8$ times lower (Figure \ref{fig:HI_and_mol}). If only stellar mass is matched, dwarf AGN still have, on average, $\sim 2.5$ times less neutral mass and $\sim 3$ times less \ion{H}{i} mass;
    \item The $M_{\rm \ion{H}{i}}/M_\ast$ fraction of dwarf AGN and non-AGN becomes increasingly different towards lower stellar masses, with most dwarf AGN having $M_{\rm \ion{H}{i}}/M_\ast < 2$ for $\log (M_\ast/{\rm M_\odot}) < 10$ (Figure \ref{fig:HI_Mstar_frac}); 
    
\end{itemize}

The high AGN fraction found in this work relative to observations - at $\log (M_\ast / {\rm M_\odot}) \leq 9.5$ - highlights the importance of precise black hole seeding models in cosmological simulations, especially for the low-mass regime. Additionally, it motivates a future detailed investigation on AGN selection methods and how each method relates to different cuts in the Eddington ratio. On the other hand, the prediction of lower neutral gas fractions in the gas reservoirs of dwarf galaxies is qualitatively consistent with observational results \citep{Elisson2019}, and it is important because it represents a direct imprint of the subgrid models employed in IllustrisTNG. Further testing this prediction with future observational works can be valuable to the development of more accurate and effective models in future cosmological simulations. Current and future observational facilities (e.g., JWST, Vera C. Rubin Observatory, SKA, ATHENA) will likely provide a rich and large amount of data on dwarf galaxies across different wavelengths, allowing the study of these galaxies with unprecedented detail. Finally, investigating AGN activity in dwarf galaxies is important for understanding the exact impact of black holes on the low-mass end of the galaxy stellar mass function.

\begin{acknowledgements}
The authors thank the referee for the comments that led to
an improved version of the manuscript. RFF thanks the support of Conselho Nacional de Desenvolvimento Científico e Tecnológico (CNPq) and the support from Coordenação de Aperfeiçoamento de Pessoal de Nível Superior (CAPES). 

MT thanks the support of CNPq (process \#312541/2021-0).  

RAR acknowledges the support from Conselho Nacional de Desenvolvimento Cient\'ifico e Tecnol\'ogico (CNPq; Proj. 303450/2022-3, 403398/2023-1, \& 441722/2023-7) and Coordena\c c\~ao de Aperfei\c coamento de Pessoal de N\'ivel Superior (CAPES;  Proj. 88887.894973/2023-00).

RR acknowledges support from  Conselho Nacional de Desenvolvimento Cient\'{i}fico e Tecnol\'ogico  ( CNPq, Proj. CNPq-445231/2024-6,311223/2020-6, 404238/2021-1, and 310413/2025-7), Funda\c{c}\~ao de amparo \`{a} pesquisa do Rio Grande do Sul (FAPERGS, Proj. 19/1750-2 and 24/2551-0001282-6) and Coordena\c{c}\~ao de Aperfei\c{c}oamento de Pessoal de N\'{i}vel Superior (CAPES, 88881.109987/2025-01).

BDO acknowledges the support from the Coordena{\c c}{\~a}o de Aperfei{\c c}oamento de Pessoal de N\'ivel Superior (CAPES-Brasil, 88887.985730/2024-00).

This work was conducted during a scholarship supported by the International Cooperation Program PROBRAL at the Ruprecht Karl University of Heidelberg. Financed by CAPES – Brazilian Federal Agency for Support and Evaluation of Graduate Education within the Ministry of Education of Brazil.

The IllustrisTNG simulations were undertaken with compute time awarded by the Gauss Centre for Supercomputing (GCS) under GCS Large-Scale Projects GCS-ILLU and GCS-DWAR on the GCS share of the supercomputer Hazel Hen at the High Performance Computing Center Stuttgart (HLRS), as well as on the machines of the Max Planck Computing and Data Facility (MPCDF) in Garching, Germany.

\end{acknowledgements}

\bibliographystyle{aa}
\bibliography{main}

\begin{appendix}

\FloatBarrier
\section{Persistence of neutral gas deficiency for different control samples}
\label{app:with_bh}
In this appendix, we show the results for the neutral gas deficiency in dwarf AGN compared with non-AGN control samples other than $\mathcal{C}_{M_\ast \& M_{\rm 200c}}^{\rm WG}$, which was the main focus in Sections \ref{sec:diff_gas_comp} and \ref{sec:evolution}. The following subsections show that our conclusions on the neutral gas deficiency are qualitatively the same as those presented in Section \ref{sec:conclusions}, regardless of the control galaxies having black holes or being paired only by stellar mass.

\FloatBarrier
\subsection{Pairing only by stellar mass}
\begin{figure}
    \centering
    \includegraphics[width=0.45\textwidth]{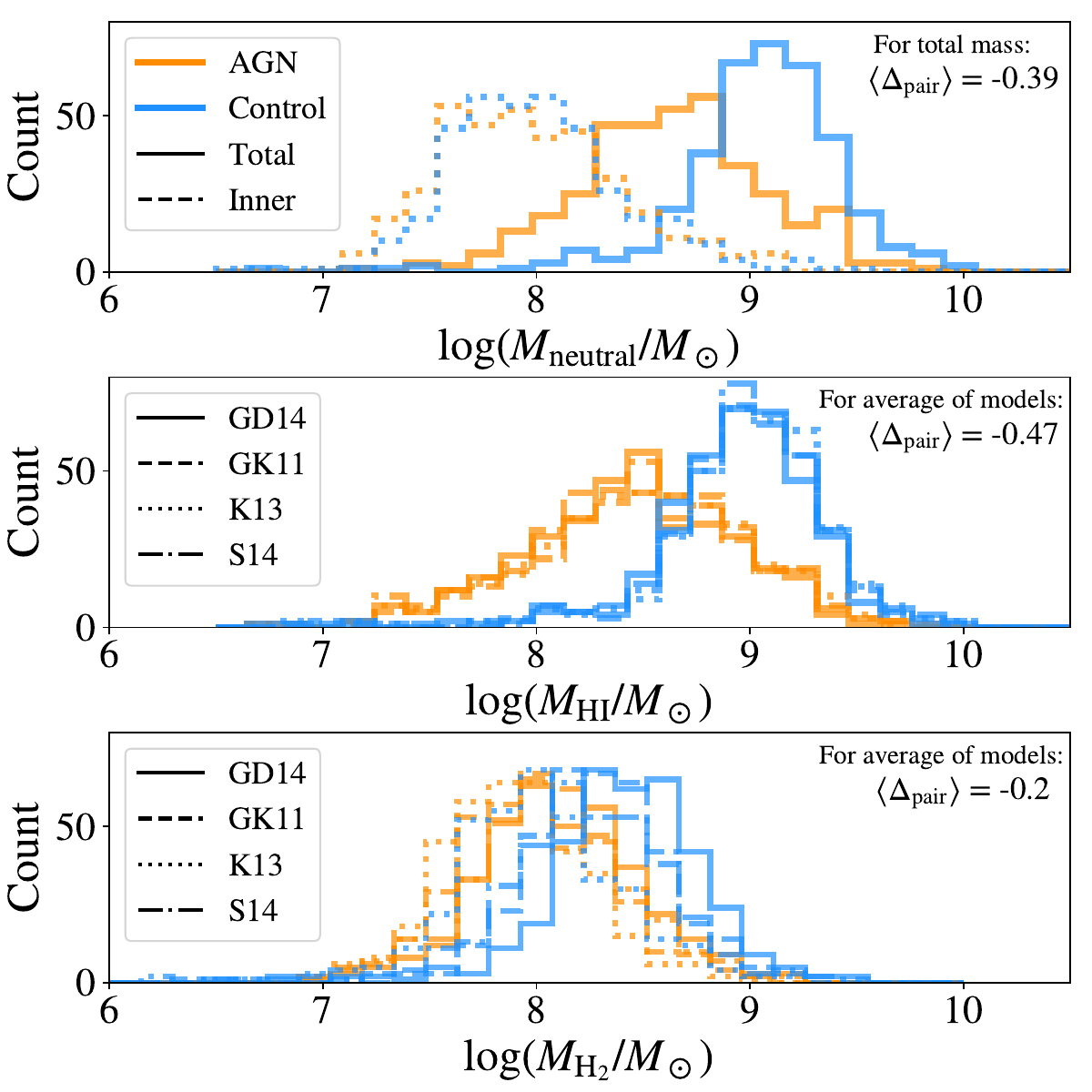}
    \caption{Same as in Figure \ref{fig:HI_and_mol}, but for the samples $\mathcal{S}_{M_\ast}^{\rm WG}$ and $\mathcal{C}_{M_\ast}^{\rm WG}$.}
    \label{fig:HI_and_mol_wg}
\end{figure}

In Figure \ref{fig:HI_and_mol_wg} we show the mass distribution for different components of the neutral gas for the samples $\mathcal{C}_{M_\ast}^{\rm WG}$ and $\mathcal{S}_{M_\ast}^{\rm WG}$ - where only stellar mass is used for pairing. This is an important test, because in observational studies, it can be challenging to create control samples of inactive galaxies paired also by halo mass. As it is evident from comparing all the panels in Figures \ref{fig:HI_and_mol} and \ref{fig:HI_and_mol_wg}, our conclusions on active dwarf galaxies having less neutral gas mass remain qualitatively the same. The main difference is that the deficiency is weaker when we only control $M_\ast$, with the AGN hosts having, on average, 2.5 (3) times less neutral (HI) gas than non-AGN galaxies.

\FloatBarrier
\subsection{Control sample of inactive galaxies with BH}

\begin{figure}
    \centering
    \includegraphics[width=0.45\textwidth]{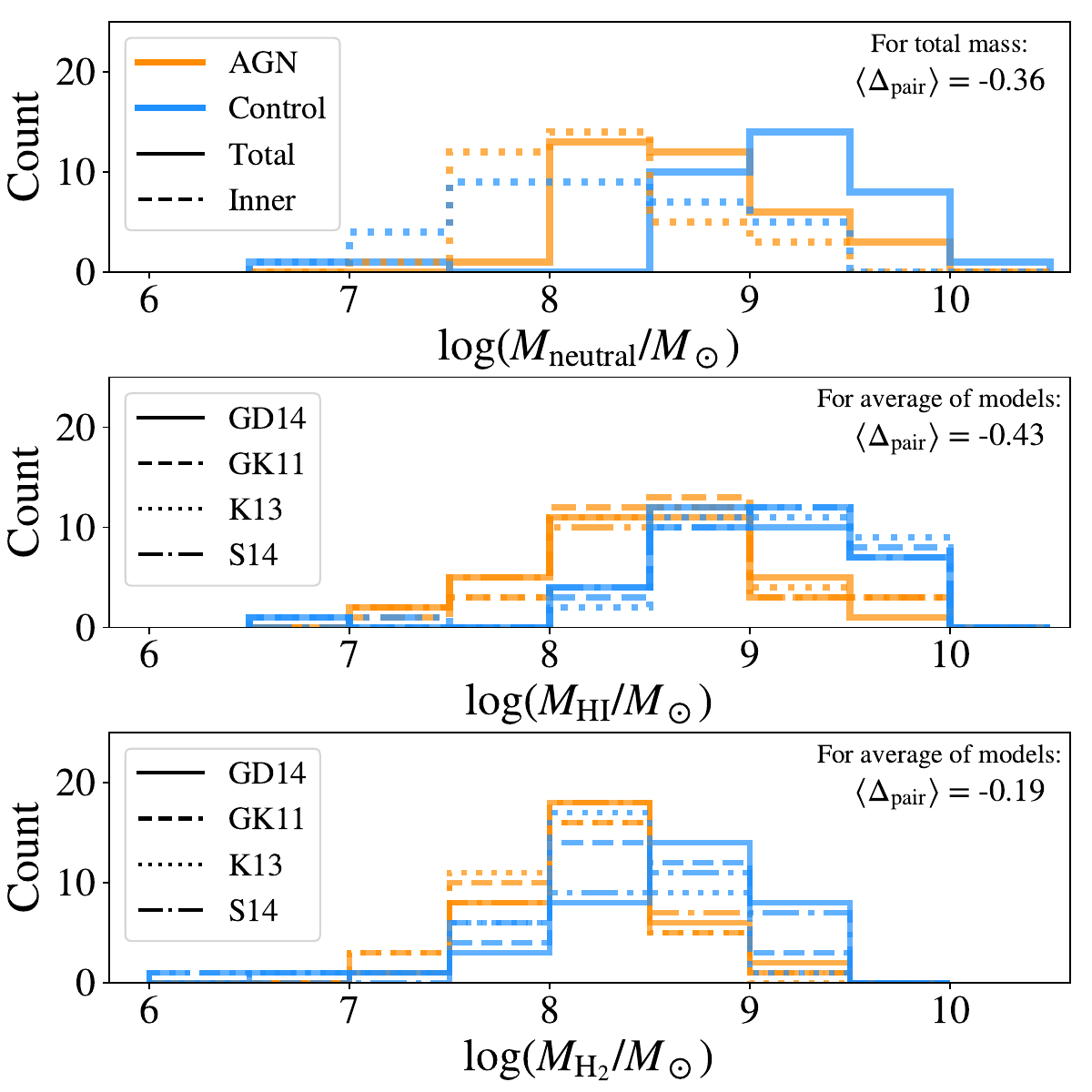}
    \caption{Same as in Figure \ref{fig:HI_and_mol}, but for the samples $\mathcal{S}_{M_\ast \& M_{\rm 200c}}^{\rm WBH}$ and $\mathcal{C}_{M_\ast \& M_{\rm 200c}}^{\rm WBH}$.}
    \label{fig:HI_and_mol_wbh}
\end{figure}

In this subsection, we show that if we compare the samples $\mathcal{S}_{M_\ast \& M_{\rm 200c}}^{\rm WBH}$ and $\mathcal{C}_{M_\ast \& M_{\rm 200c}}^{\rm WBH}$ - in which all galaxies have a BH - our conclusions on the neutral gas deficiency remains qualitatively the same. As it is evident from comparing all the panels in Figures \ref{fig:HI_and_mol} and \ref{fig:HI_and_mol_wbh}, the main difference is that the deficiency is weaker, with the AGN hosts having, on average, 2.3 (2.7) times less neutral (HI) gas than non-AGN galaxies.

\begin{figure*}
    \centering
    \includegraphics[width=0.33\textwidth]{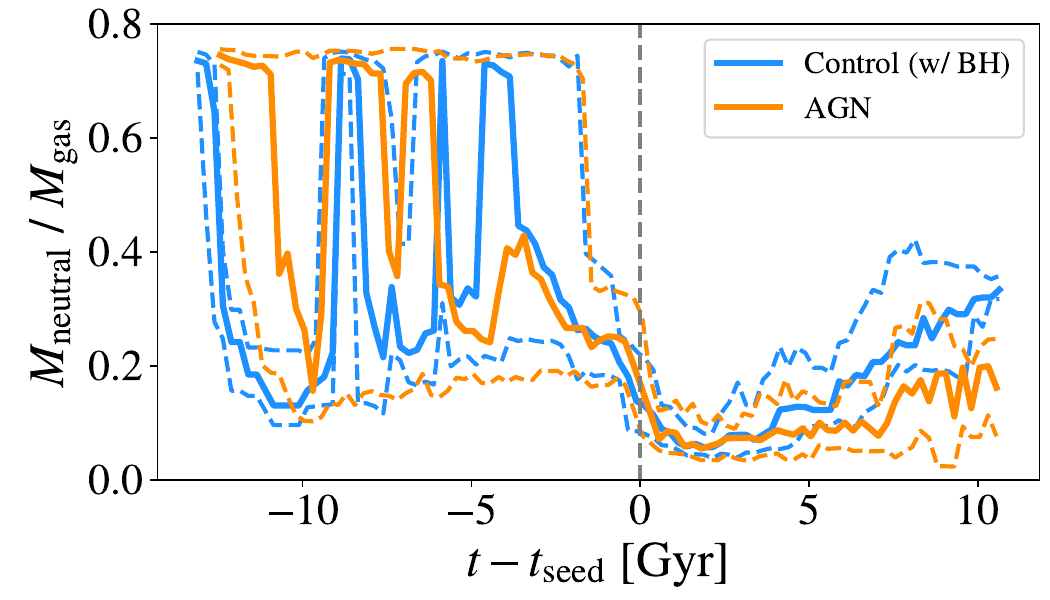}
    \includegraphics[width=0.33\textwidth]{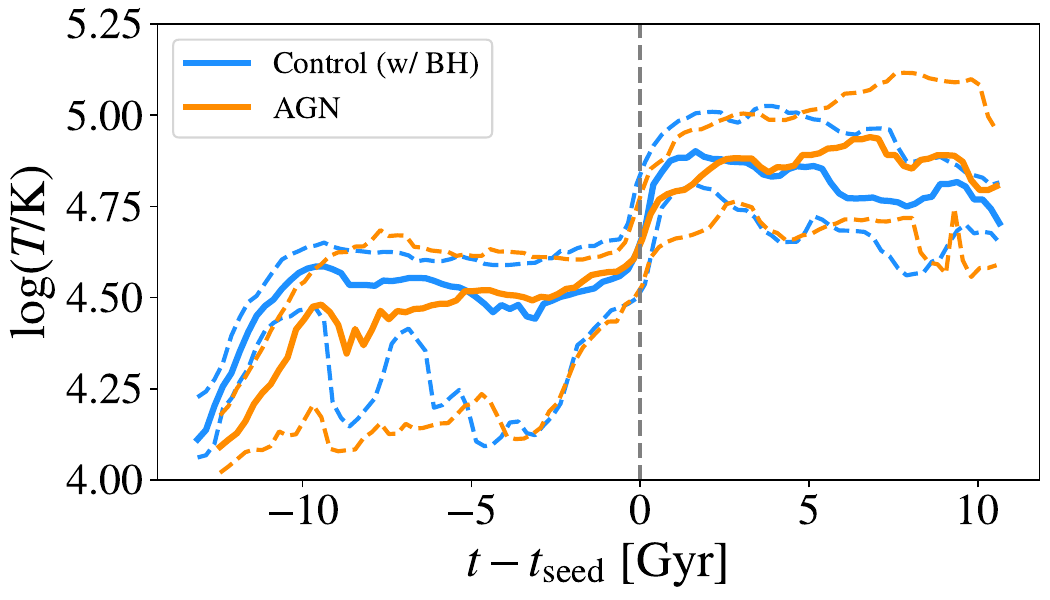}
    \includegraphics[width=0.33\textwidth]{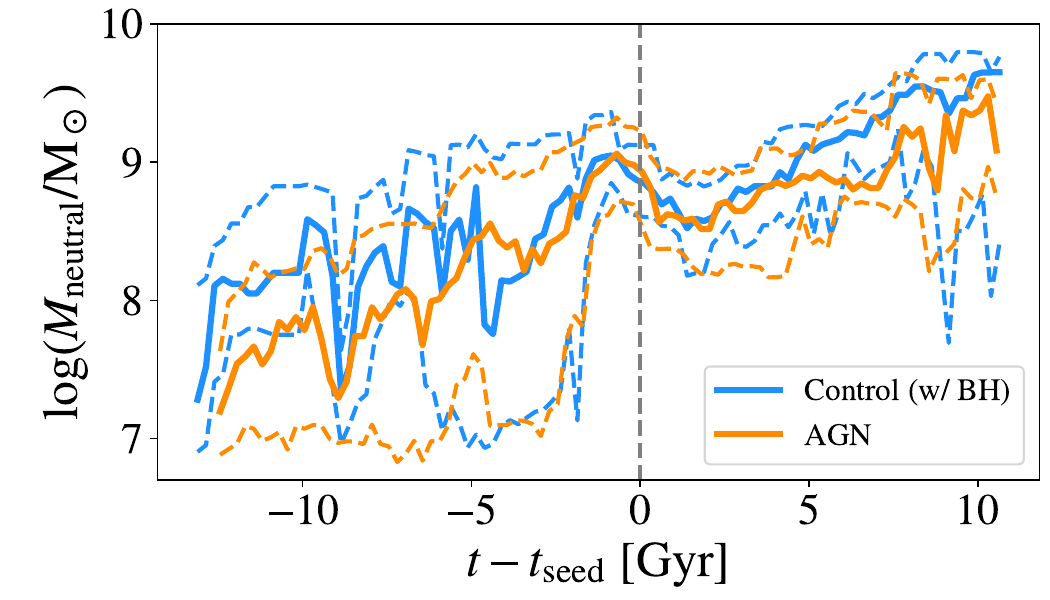}
    \includegraphics[width=0.33\textwidth]{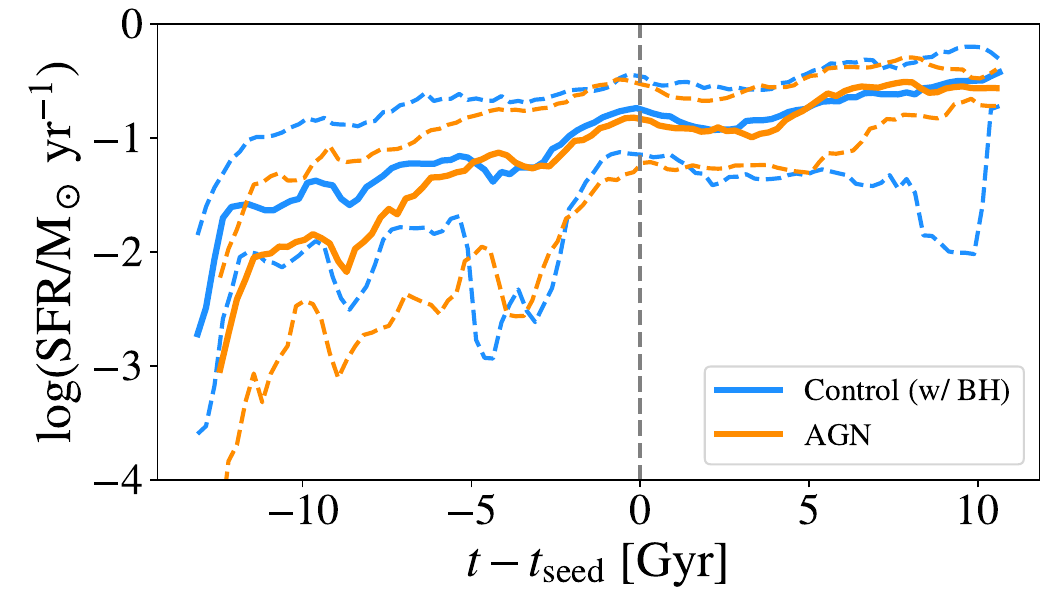}
    \includegraphics[width=0.33\textwidth]{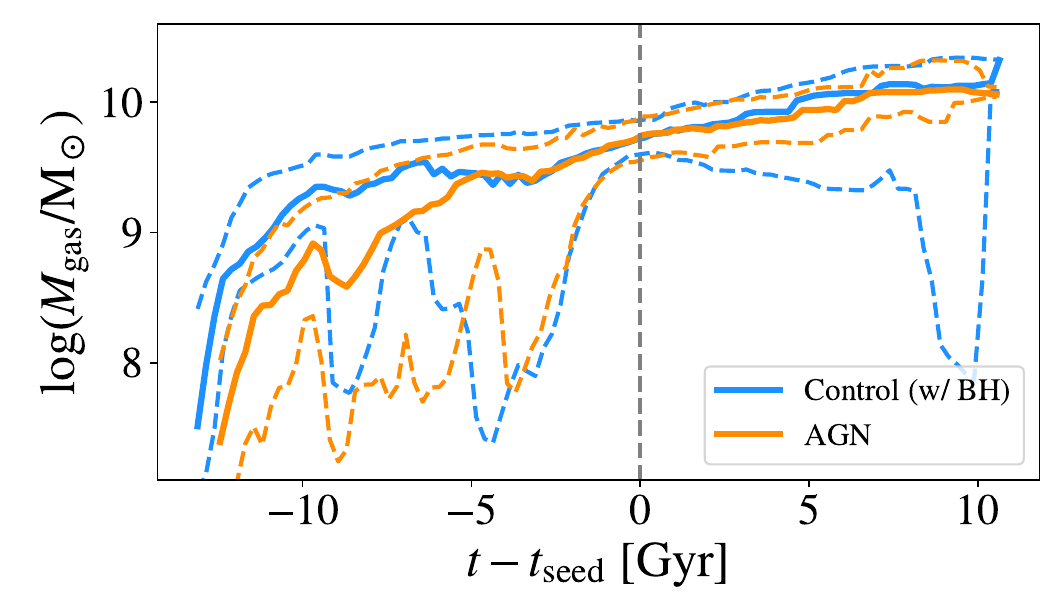}
    \includegraphics[width=0.33\textwidth]{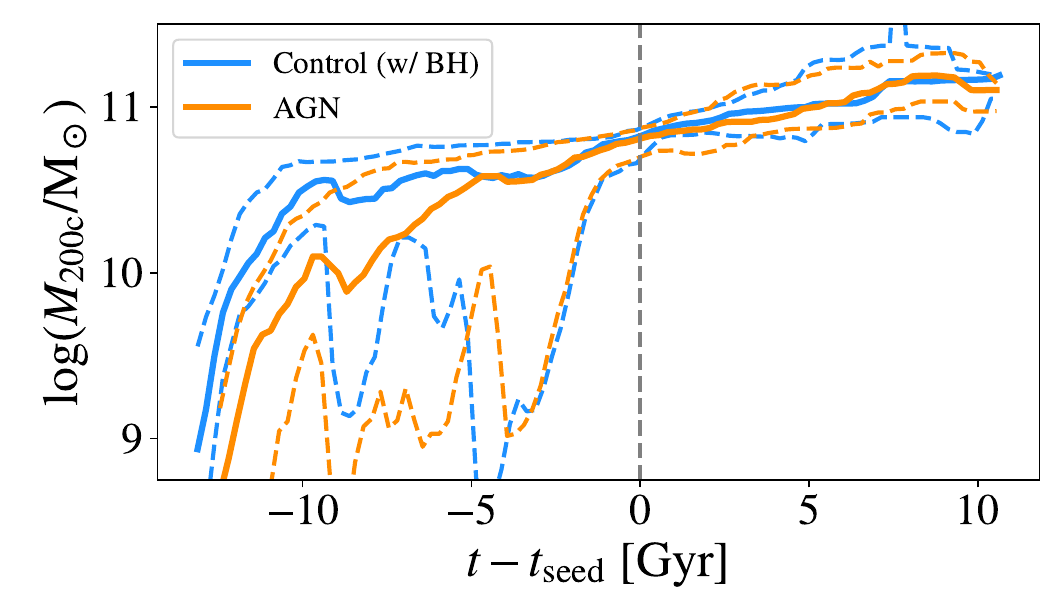}
    \caption{Same as in Figure \ref{fig:history_wg}, but for the samples $\mathcal{S}_{M_\ast \& M_{\rm 200c}}^{\rm WBH}$ and $\mathcal{C}_{M_\ast \& M_{\rm 200c}}^{\rm WBH}$.}
    \label{fig:history_wbh}
\end{figure*}

\begin{figure}
    \centering
    \includegraphics[width=0.49\textwidth]{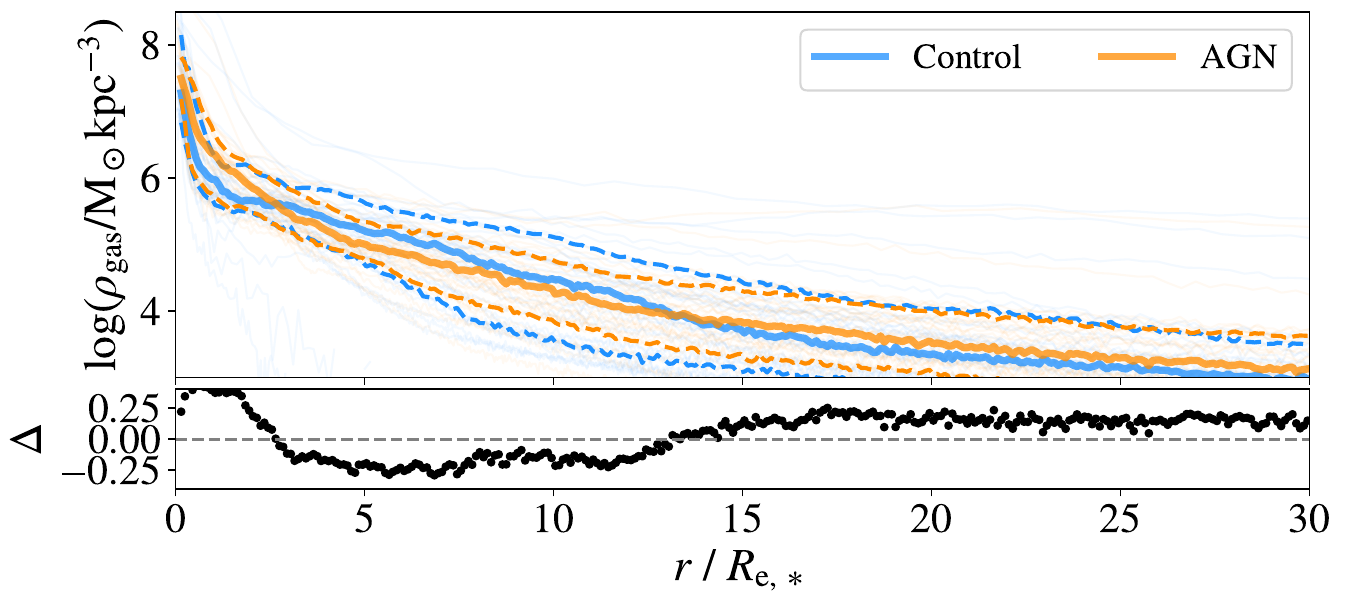}
    \includegraphics[width=0.49\textwidth]{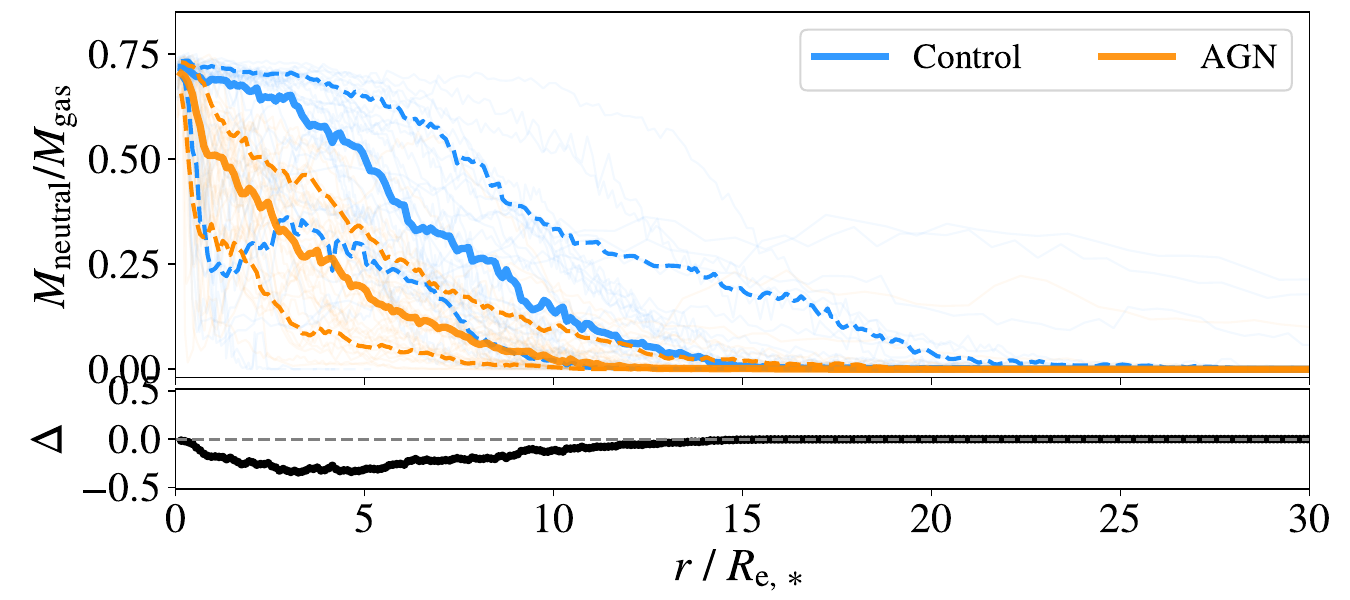}
    \includegraphics[width=0.49\textwidth]{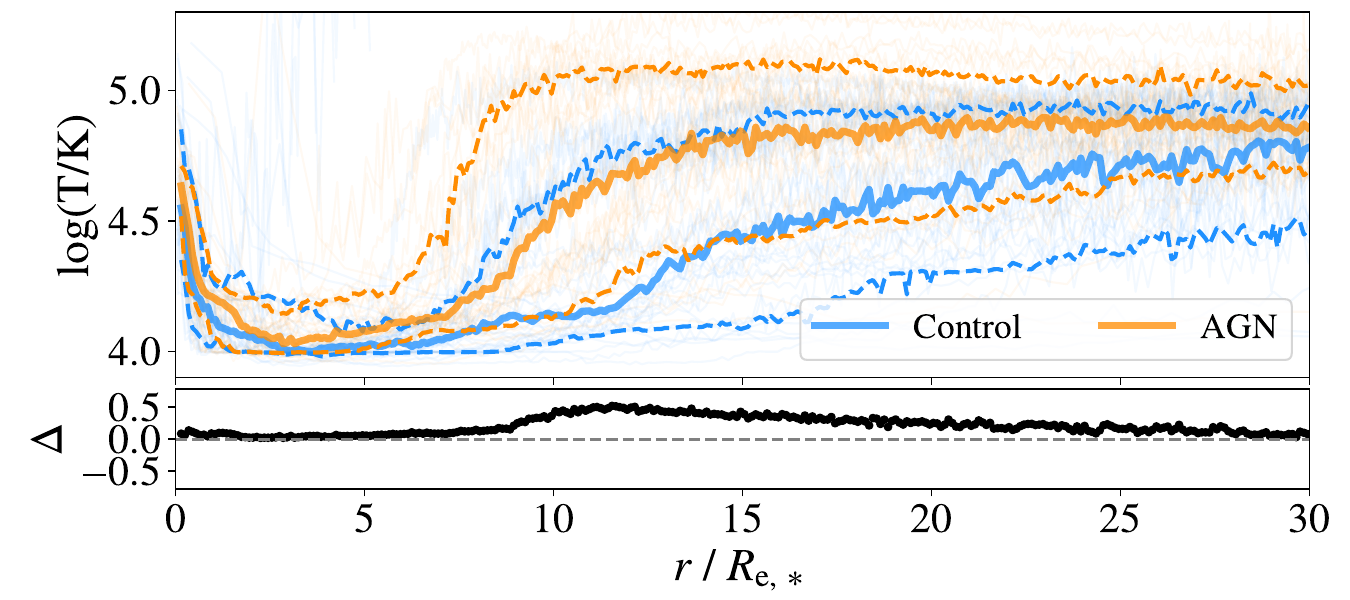}
    \caption{Same as in Figure \ref{fig:profiles_wg}, but for the samples $\mathcal{S}_{M_\ast \& M_{\rm 200c}}^{\rm WBH}$ and $\mathcal{C}_{M_\ast \& M_{\rm 200c}}^{\rm WBH}$.}
    \label{fig:profiles_wbh}
\end{figure}

Additionally, we checked if the results for the profiles and evolution of AGN hosts change when considering control galaxies that necessarily have a black hole. The differences between the median profiles of AGN and non-AGN samples shown in Figure \ref{fig:profiles_wbh} present similar trends as those observed in Figure \ref{fig:profiles_wg}. However, the evolution of the neutral-to-total gas mass ratio is different, as we can see by comparing the upper left panels from Figures \ref{fig:history_wg} and \ref{fig:history_wbh}. Differently from $\mathcal{C}_{M_\ast \& M_{\rm 200c}}^{\rm WG}$, the $\mathcal{C}_{M_\ast \& M_{\rm 200c}}^{\rm WBH}$ sample presents a similar decrease in neutral gas as their paired AGN in $\mathcal{S}_{M_\ast \& M_{\rm 200c}}^{\rm WBH}$. However, as time passes, the galaxies in $\mathcal{C}_{M_\ast \& M_{\rm 200c}}^{\rm WBH}$ accrete more neutral gas (upper right panel of Figure \ref{fig:history_wbh}) and their $M_{\rm neutral}/M_{\rm gas}$ increases.

\begin{figure}
    \centering
    \includegraphics[width=0.48\textwidth]{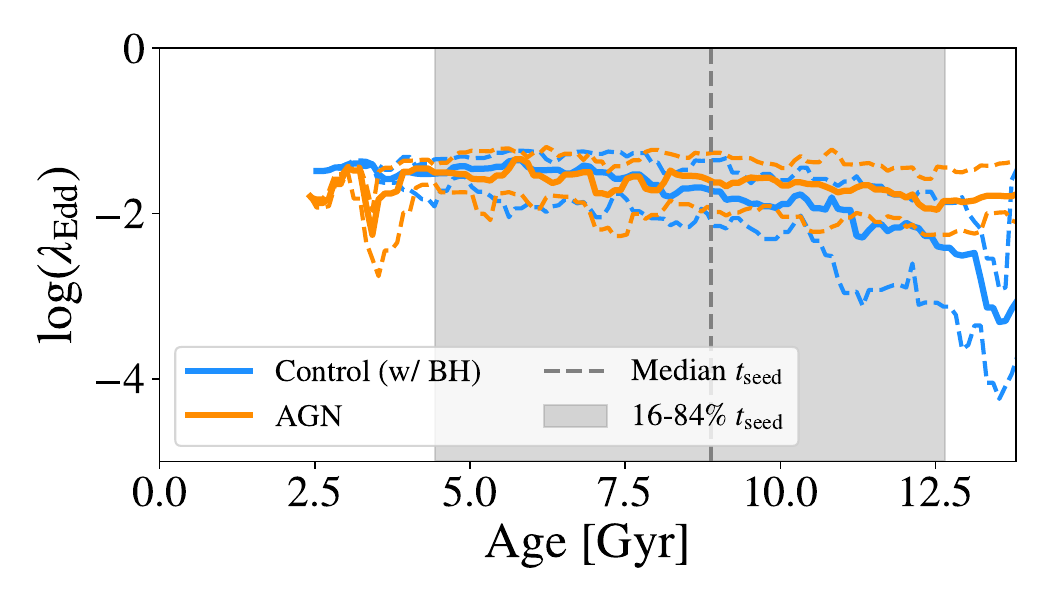}
    \caption{Eddington ratio AGN (orange) and non-AGN control (blue) samples as a function of the age of the Universe. The solid lines indicate the median evolution for the whole sample, while the dashed lines indicate the 16th and 84th percentiles. The vertical dashed line indicates the median black hole seeding time (AGN and non-AGN), and the shaded region encompasses the values between the 16th and 84th percentiles. The samples being compared here are $\mathcal{S}_{M_\ast \& M_{\rm 200c}}^{\rm WBH}$ and $\mathcal{C}_{M_\ast \& M_{\rm 200c}}^{\rm WBH}$.}
    \label{fig:history_edd_ratio}
\end{figure}

\begin{figure}
    \centering
    \includegraphics[width=0.48\textwidth]{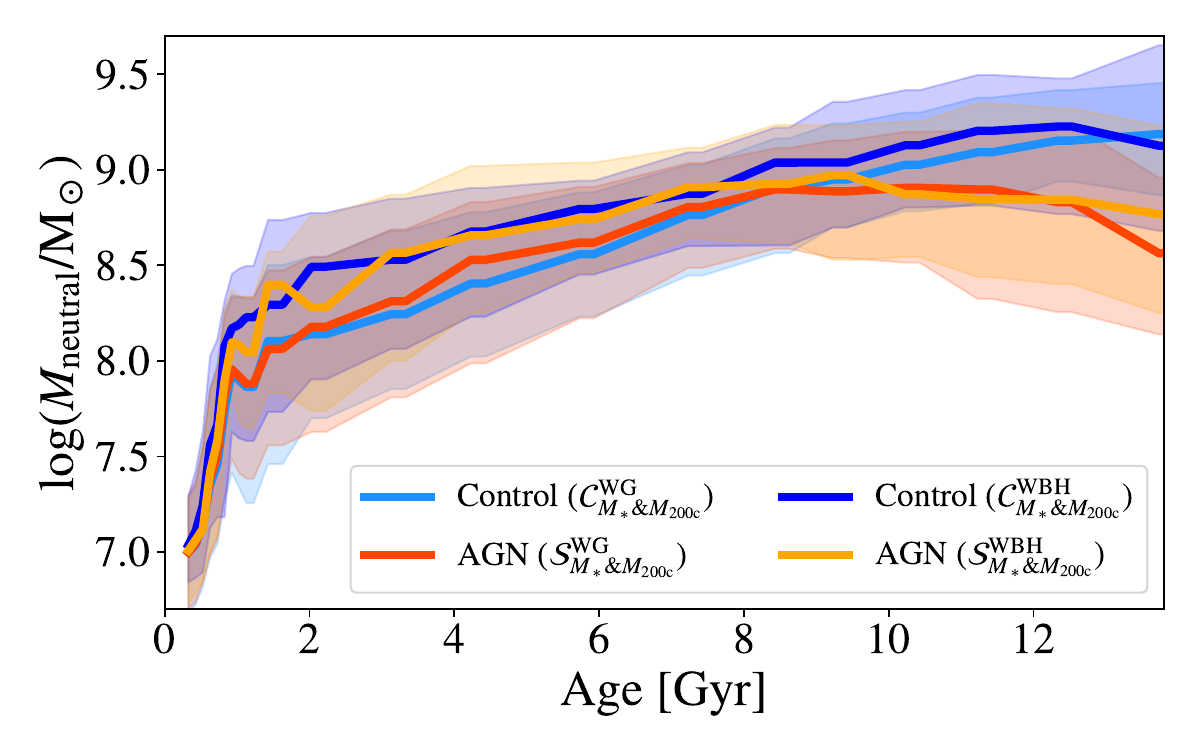}
    \caption{Comparison of neutral gas mass evolution between different AGN (shades of orange) and non-AGN control (shades of blue) samples. The horizontal axis shows the age of the Universe. The thick solid lines indicate the median evolution for the whole sample, while the shaded regions encompass the values between the 16th and 84th percentiles.}
    \label{fig:history_extra}
\end{figure}

As we show in Figure \ref{fig:history_edd_ratio}, the median Eddington ratio of the non-AGN control galaxies decreases in the following Gyr after seeding, while the AGN hosts do not, also explaining their difference in neutral gas content at $z=0$. Additionally, in Figure \ref{fig:history_extra} we compare the neutral gas mass evolution of the $\mathcal{C}_{M_\ast \& M_{\rm 200c}}^{\rm WG}$ and $\mathcal{C}_{M_\ast \& M_{\rm 200c}}^{\rm WBH}$ samples. It is clear that both control samples (blue lines) show a roughly similar history in the last $\sim 5$~Gyr, which is the period that their AGN counterparts (orange lines) have a decrease in neutral gas mass.

\FloatBarrier
\section{Different AGN selection criteria}
\label{app:selection}
In this appendix, we explore how our main results, regarding the neutral gas component, are affected by changes in our AGN selection in the simulation. More specifically, changes involving the Eddington ratio.

\begin{figure}
    \centering
    \includegraphics[width=0.45\textwidth]{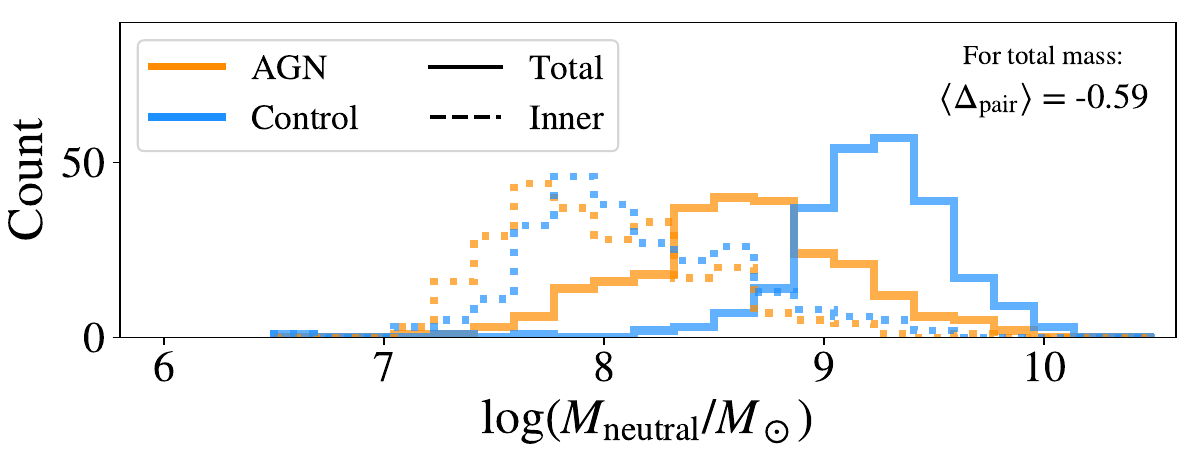}
    \includegraphics[width=0.45\textwidth]{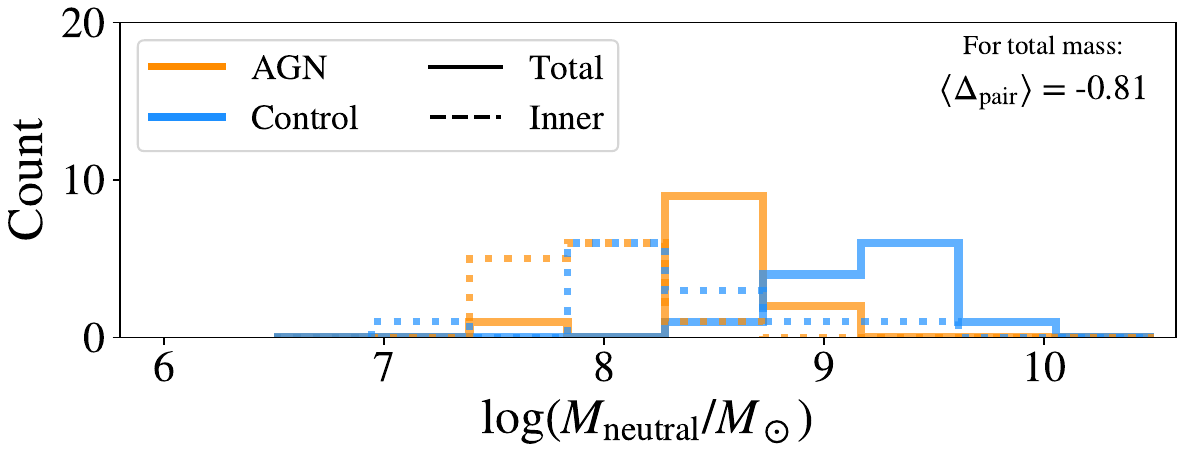}
    \caption{Histograms of neutral gas mass in AGN and non-AGN control galaxies. Colors and line styles are as in the upper panel of Figure \ref{fig:HI_and_mol}, but changing the selection criteria of AGN. The upper panel shows the mass distributions for AGN selected using instantaneous $\lambda_{\rm Edd}$, and the lower panel shows the mass distributions for AGN selected using the $\lambda_{\rm Edd,min}=0.05$. The samples being compared here are $\mathcal{S}_{M_\ast \& M_{\rm 200c}}^{\rm WBH}$ and $\mathcal{C}_{M_\ast \& M_{\rm 200c}}^{\rm WBH}$.}
    \label{fig:extra_neutral_hists}
\end{figure}

As described in our Methods section, to select AGN in TNG50-1, we used the Eddington ratio, more specifically, an average of this quantity over the last three snapshots in the simulation. However, we could also use the instantaneous value of the Eddington ratio to select the AGN. Thus, in Figure \ref{fig:extra_neutral_hists} we show the histograms of the neutral gas mass for AGN selected using instantaneous $\lambda_{\rm Edd}$ and its corresponding control sample of inactive dwarf galaxies (paired by stellar and halo mass). As it is clear from the figure, the AGN still have lower masses than non-AGN, regardless of the change in the selection method.

\begin{figure}
    \centering
    \includegraphics[width=0.48\textwidth]{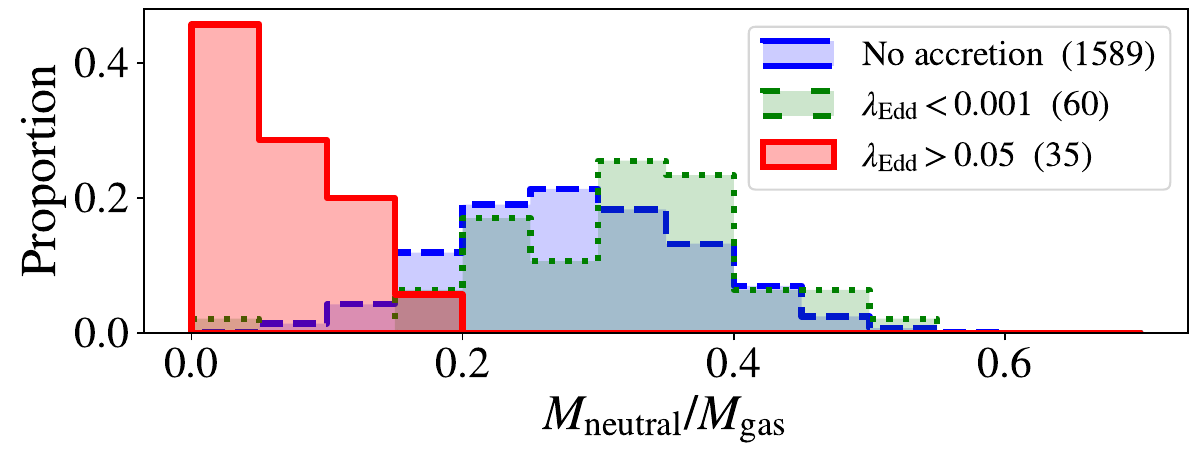}
    \caption{Histograms of neutral-to-total gas mass fraction for all central dwarf galaxies in TNG50-1 with stellar masses in the range $8 \leq \log(M_\ast / {\rm M_\odot}) \leq 9.5$. }
    \label{fig:extra_highacc}
\end{figure}

To test whether our main results on the neutral gas deficiency remain even if only the strongest AGN are selected, similarly to Figure \ref{fig:HI_and_mol}, we show in Figure \ref{fig:extra_neutral_hists} the neutral gas masses of AGN and non-AGN. Additionally, in Figure \ref{fig:extra_highacc} we show the neutral-to-total gas mass fraction for dwarf galaxies in TNG50-1 with different values of $\lambda_{\rm Edd}$. As it is clear from the figures, the general trend of our results remains the same if only dwarf AGN with $\lambda_{\rm Edd} \geq 0.05$ are selected.


\FloatBarrier
\section{Other properties and environment measures}

\begin{figure*}
    \hfill
    \includegraphics[width=0.92\textwidth]{FIGURES/boxplots/boxplots_legend.pdf}
    \includegraphics[width=0.97\textwidth]{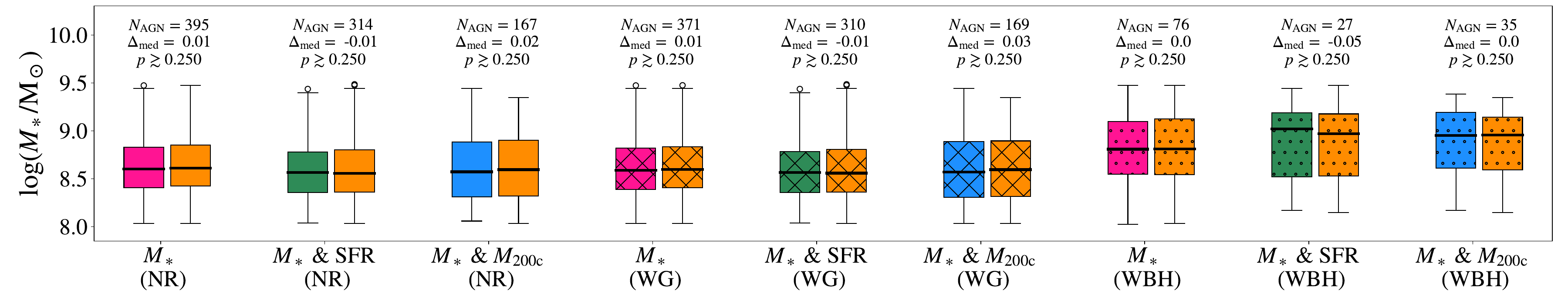}
    \includegraphics[width=0.97\textwidth]{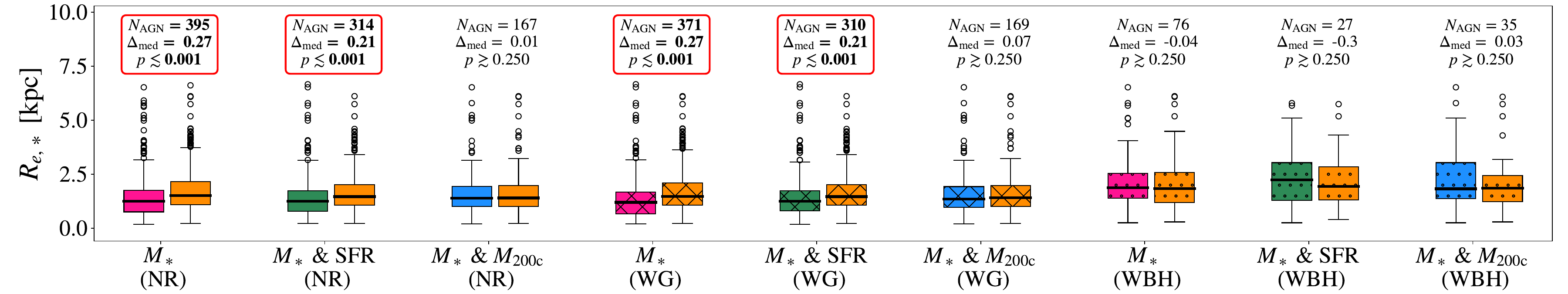}
    \includegraphics[width=0.97\textwidth]{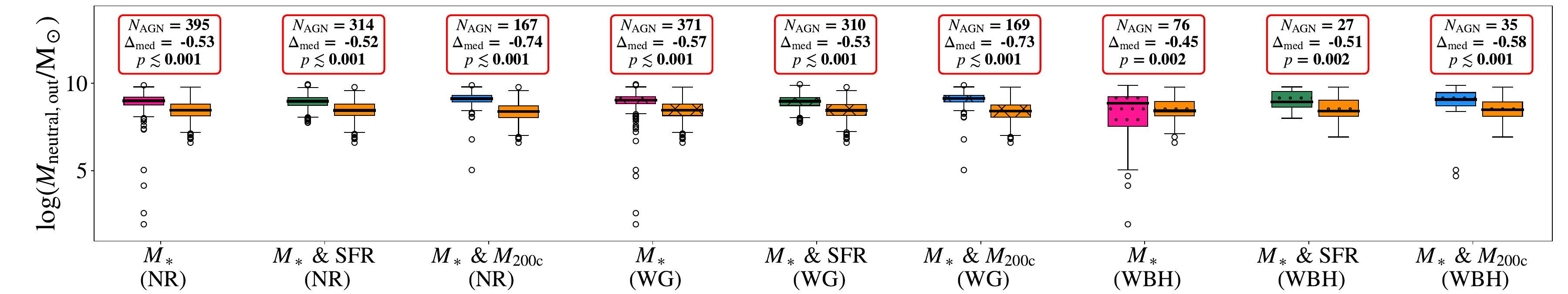}
    \includegraphics[width=0.97\textwidth]{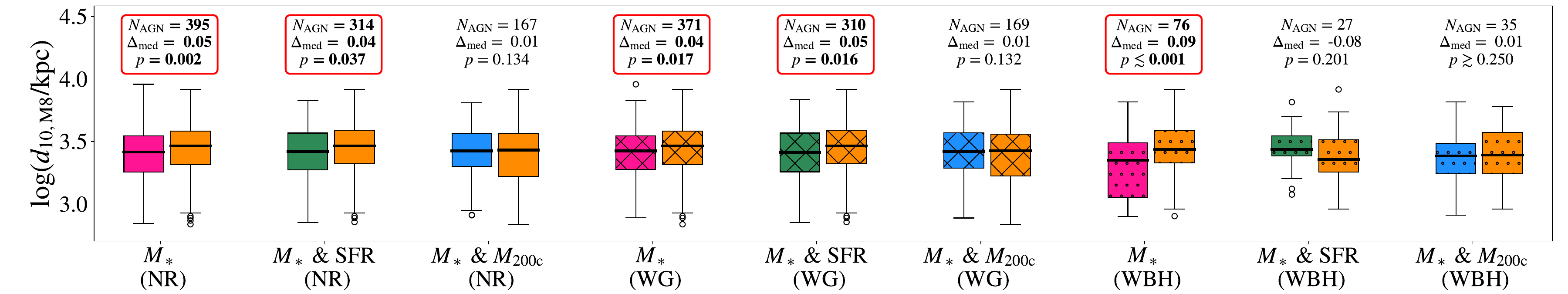}
    \includegraphics[width=0.97\textwidth]{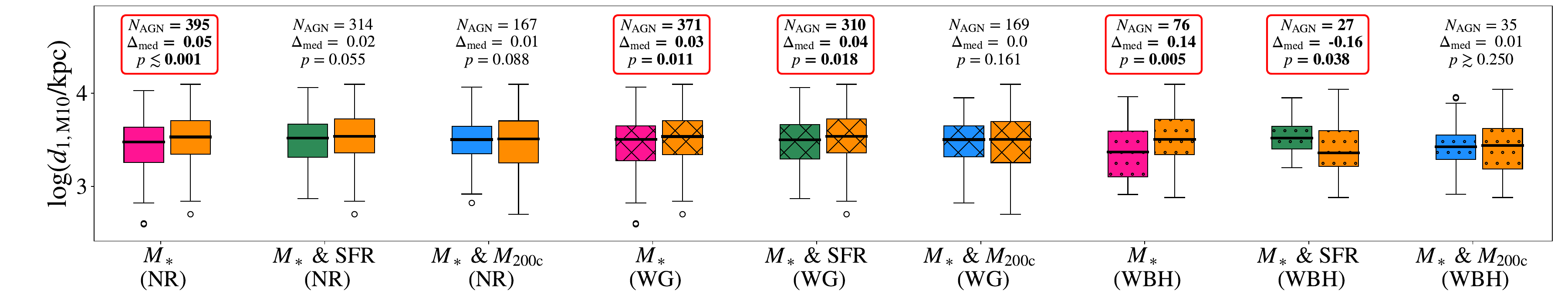}
    \includegraphics[width=0.97\textwidth]{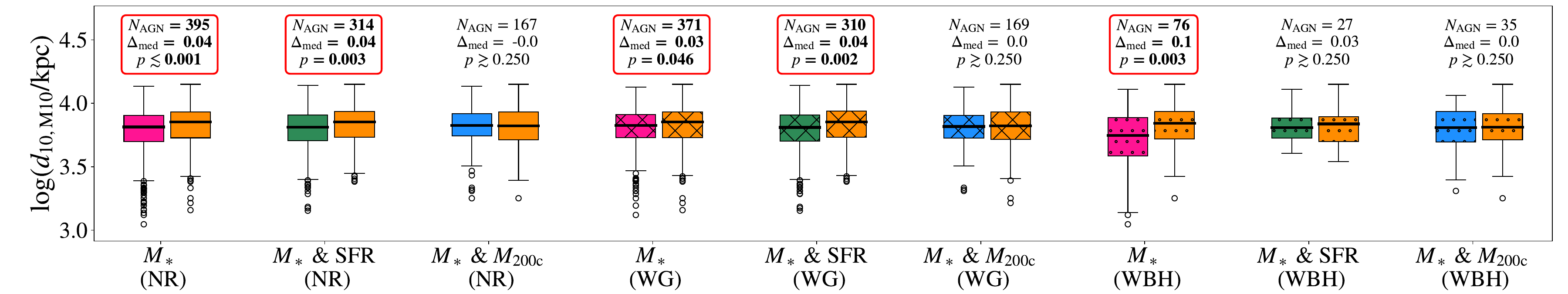}
    \caption{Same as Figure \ref{fig:boxplots1}. From top to bottom the panels show box plots for: stellar mass, stellar half-mass radius, neutral gas mass outside 2$R_{e,\ast}$, distance to the 10th nearest neighbor with $M_\ast \geq 10^8 \ \rm M_\odot$, distance to the 1st nearest neighbor with $M_\ast \geq 10^{10} \ \rm M_\odot$, distance to the 10th nearest neighbor with $M_\ast \geq 10^{10} \ \rm M_\odot$.}
    \label{fig:boxplots3_extra}
\end{figure*}

In this appendix, we present some extra physical quantities that were not presented in Section \ref{sec:properties}. The upper panel of Figure \ref{fig:boxplots3_extra} shows that all AGN samples are appropriately paired to their respective non-AGN control samples in stellar mass. Continuing from top to bottom in the panels of Figure \ref{fig:boxplots3_extra}, we also show the stellar half-mass radius. This stellar radius is similar for AGN host and non-AGN galaxies once the halo mass is controlled. In the third panel from the top, we also present the neutral gas mass outside 2$R_{e,\ast}$, which complements the results we found in Figure \ref{fig:profiles_wg}, with the neutral gas deficiency extending to the outskirts of active dwarf galaxies. In the last three lower panels of the figure, we show different distances to neighbors used to measure the environment. As it is clear from the information shown, no significant difference in environment is found once the halo mass is controlled.

\label{app:extra_box}

\FloatBarrier
\section{Examples of AGN \& non-AGN pairs} 
\label{app:examples}
\begin{figure*}
    \centering
    \includegraphics[width=0.18\linewidth]{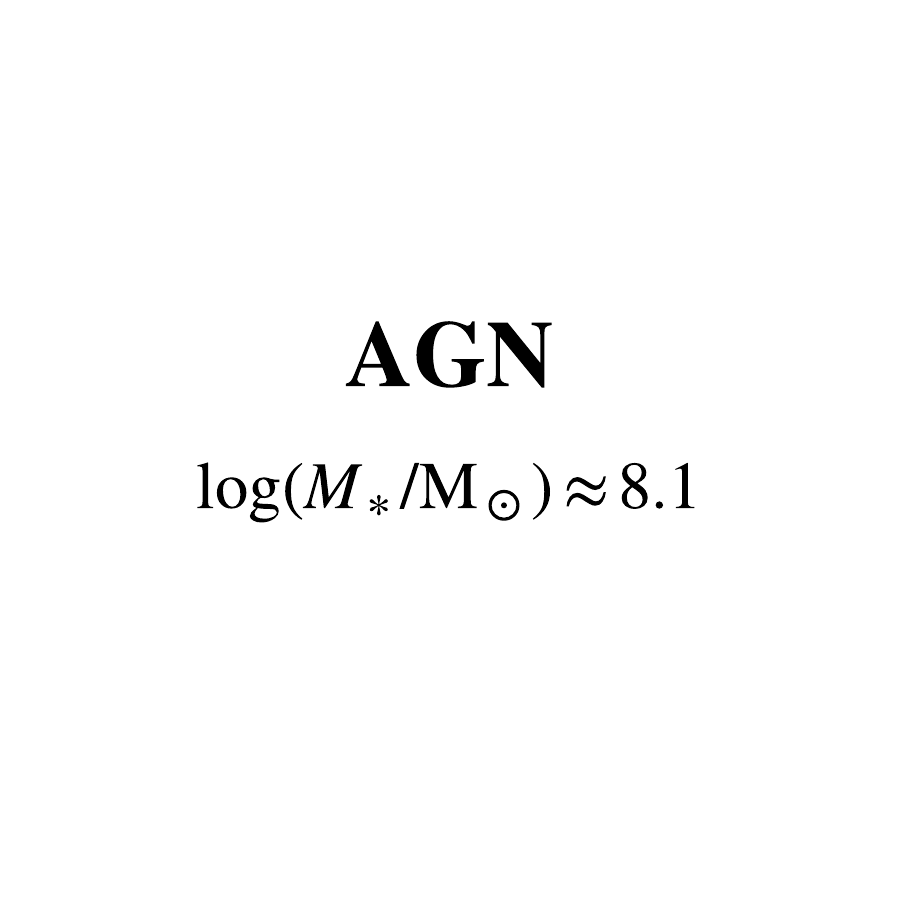}
    \includegraphics[width=0.81\linewidth]{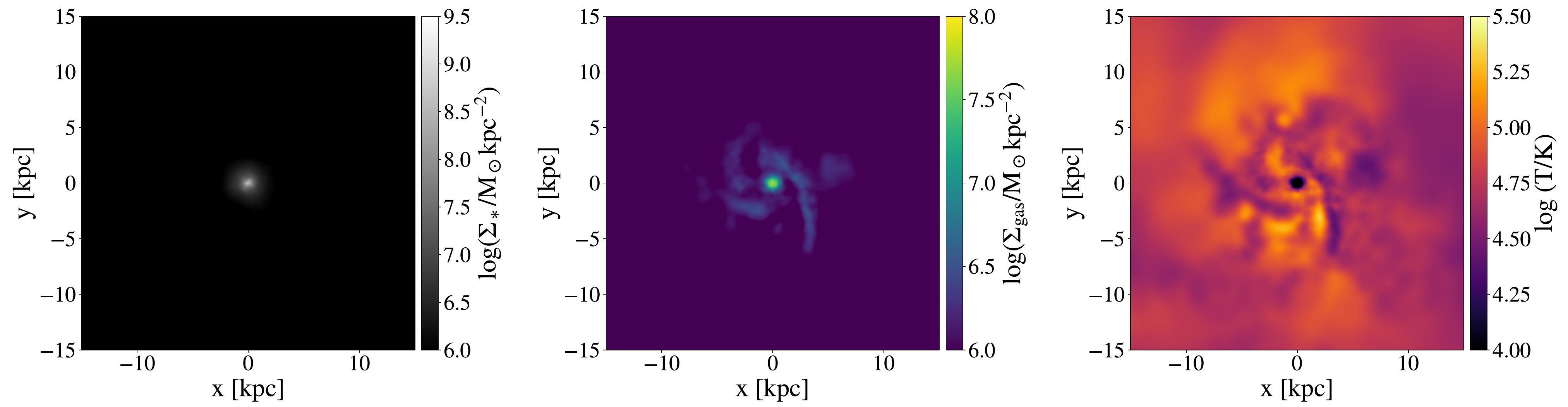}
    \includegraphics[width=0.18\linewidth]{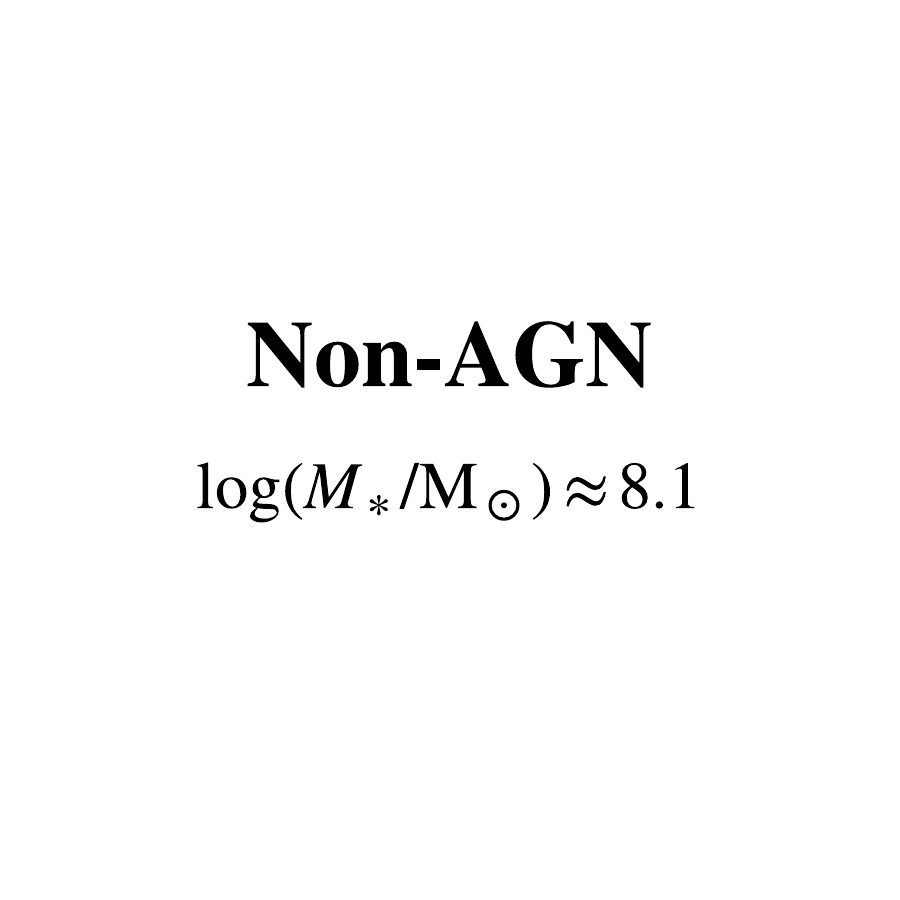}
    \includegraphics[width=0.81\linewidth]{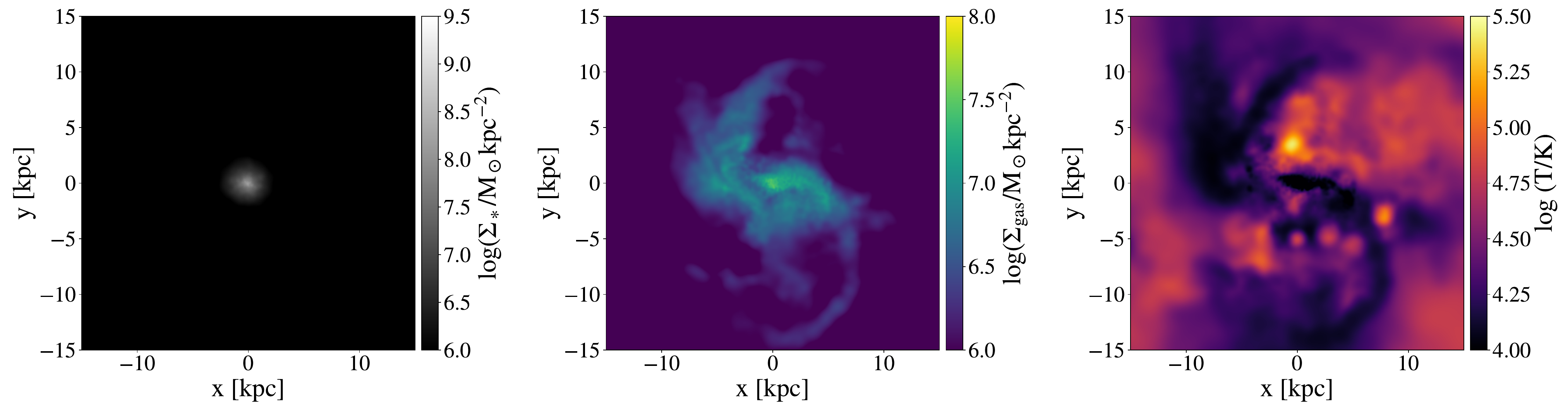}

    \includegraphics[width=0.18\linewidth]{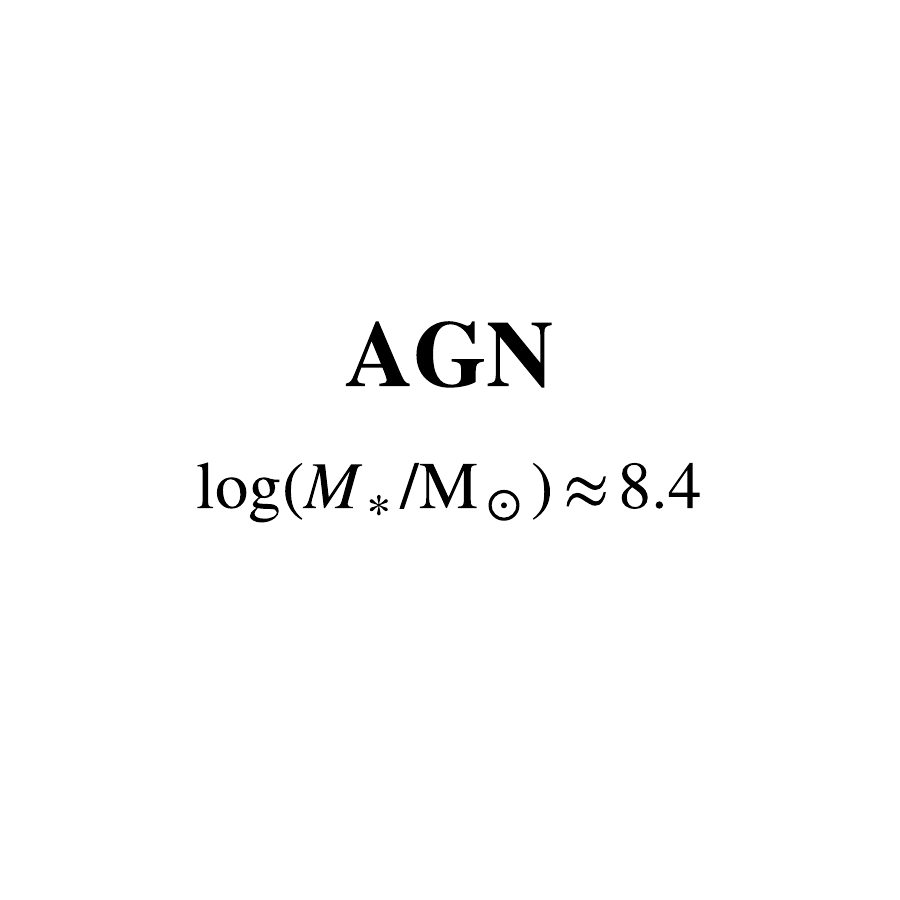}
    \includegraphics[width=0.81\linewidth]{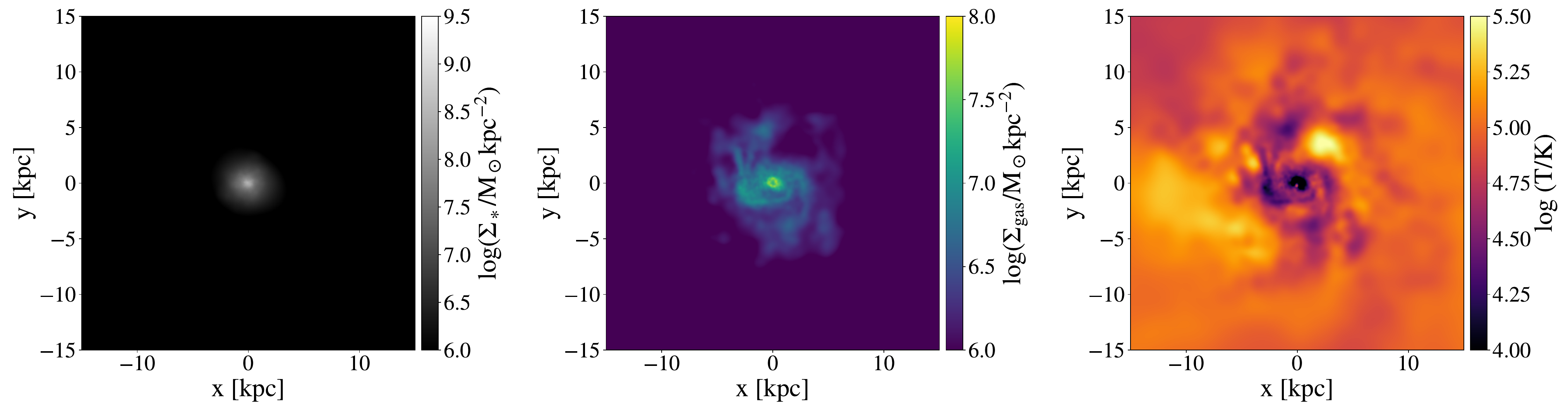}
    \includegraphics[width=0.18\linewidth]{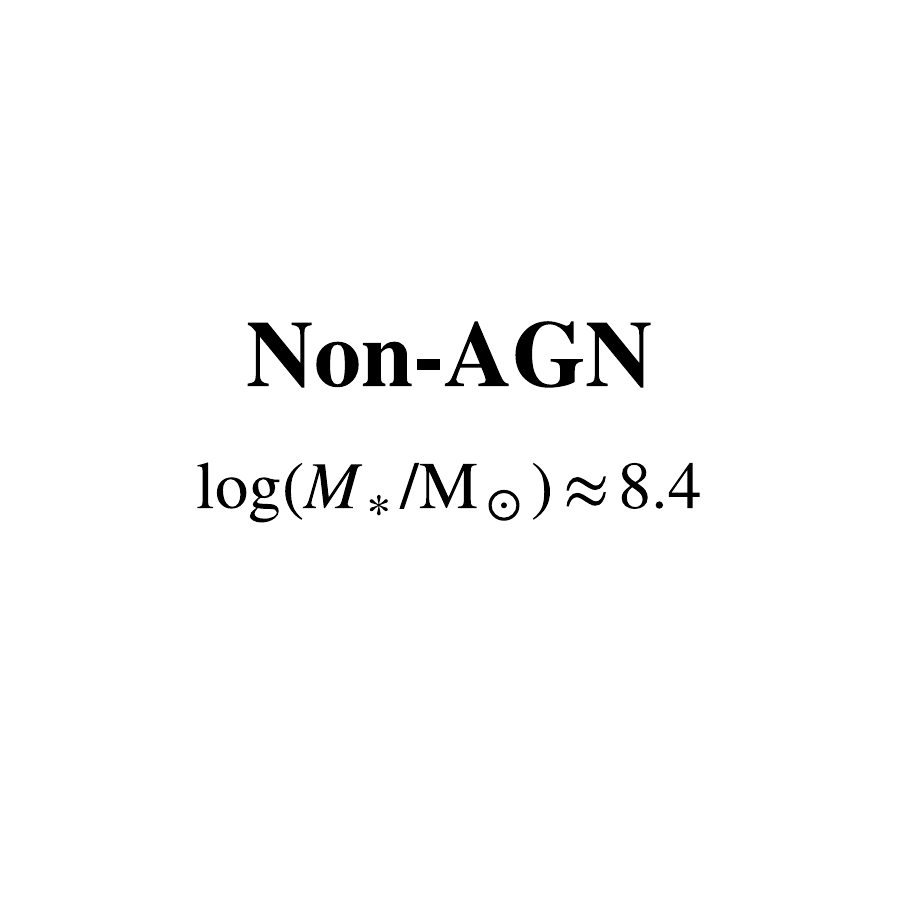}
    \includegraphics[width=0.81\linewidth]{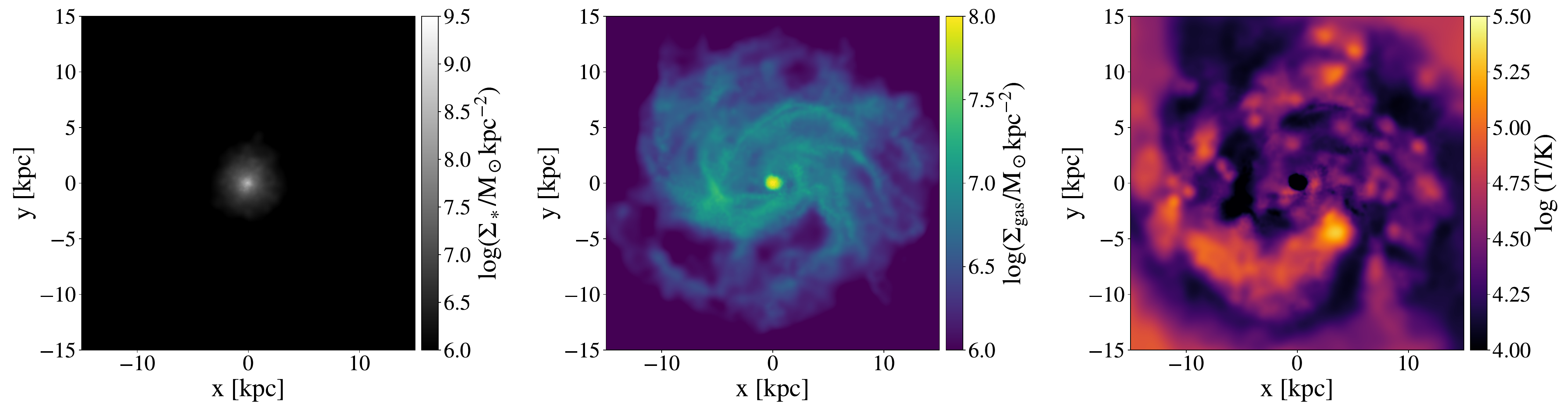}

    \includegraphics[width=0.18\linewidth]{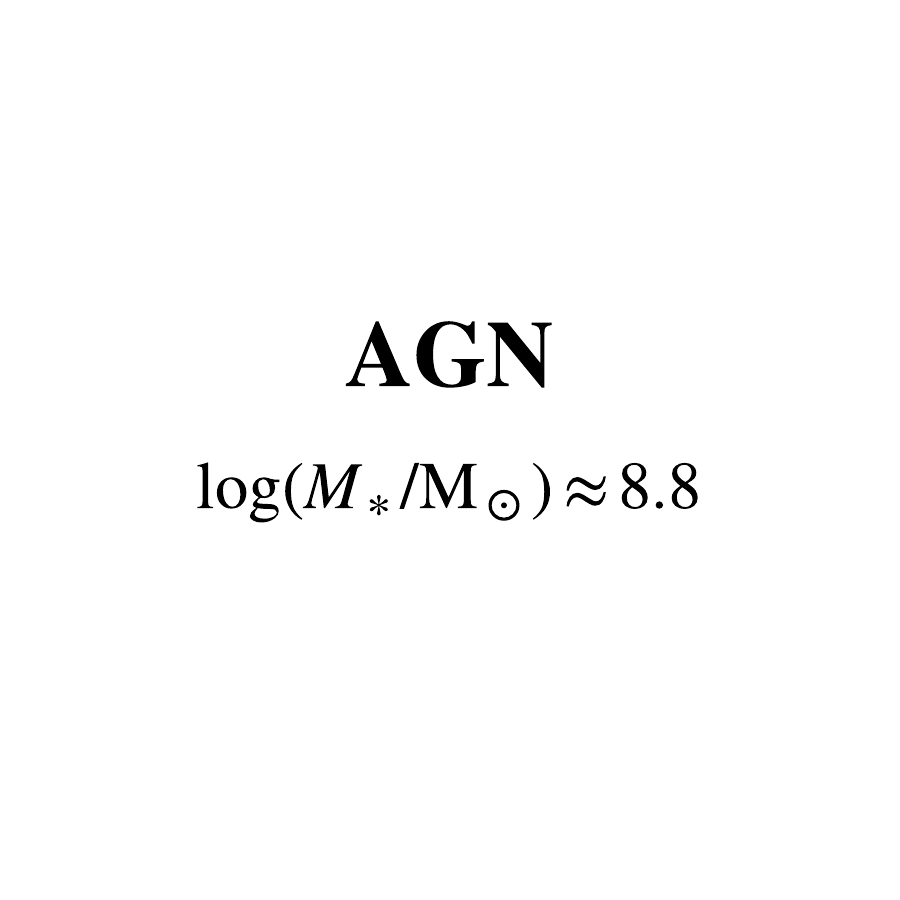}
    \includegraphics[width=0.81\linewidth]{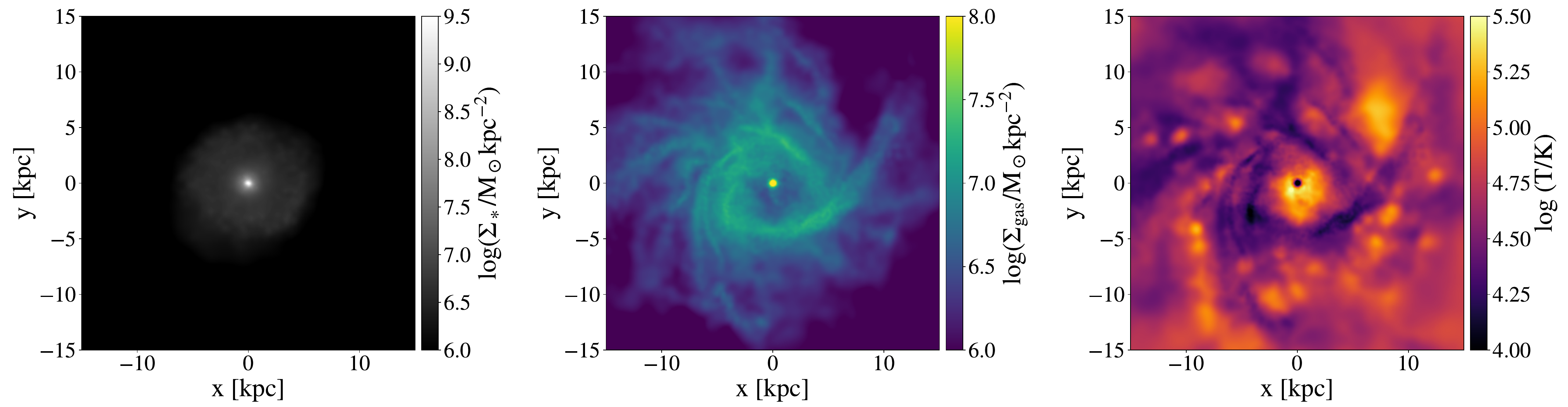}
    \includegraphics[width=0.18\linewidth]{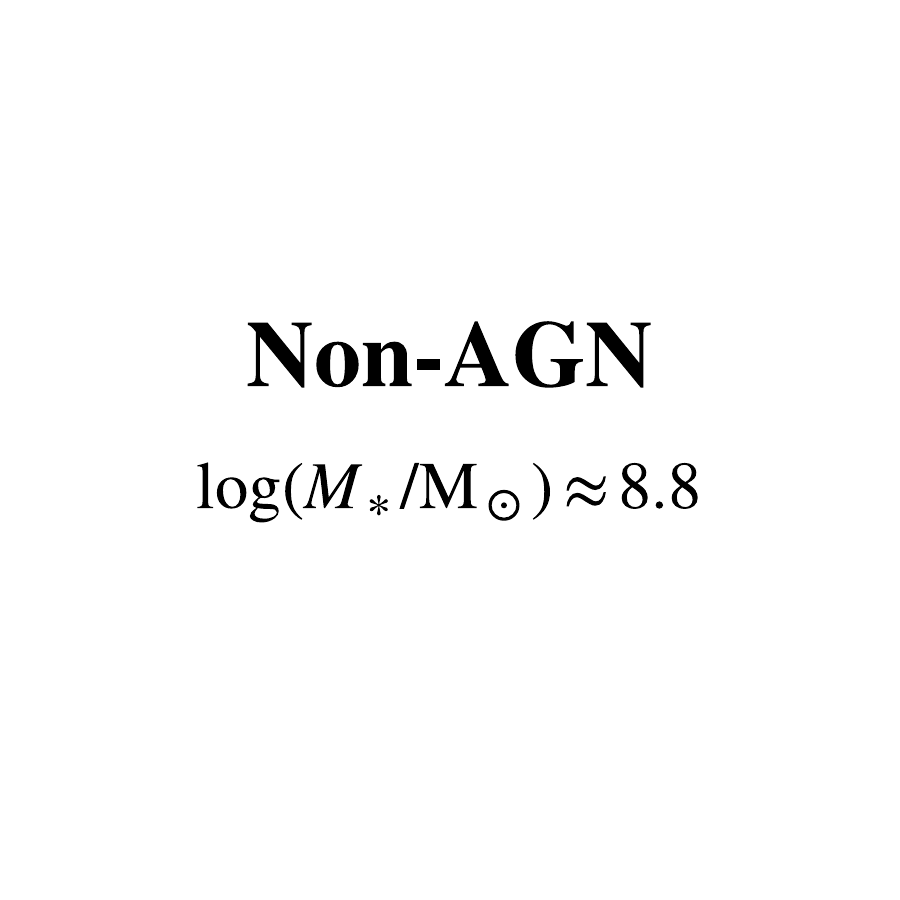}
    \includegraphics[width=0.81\linewidth]{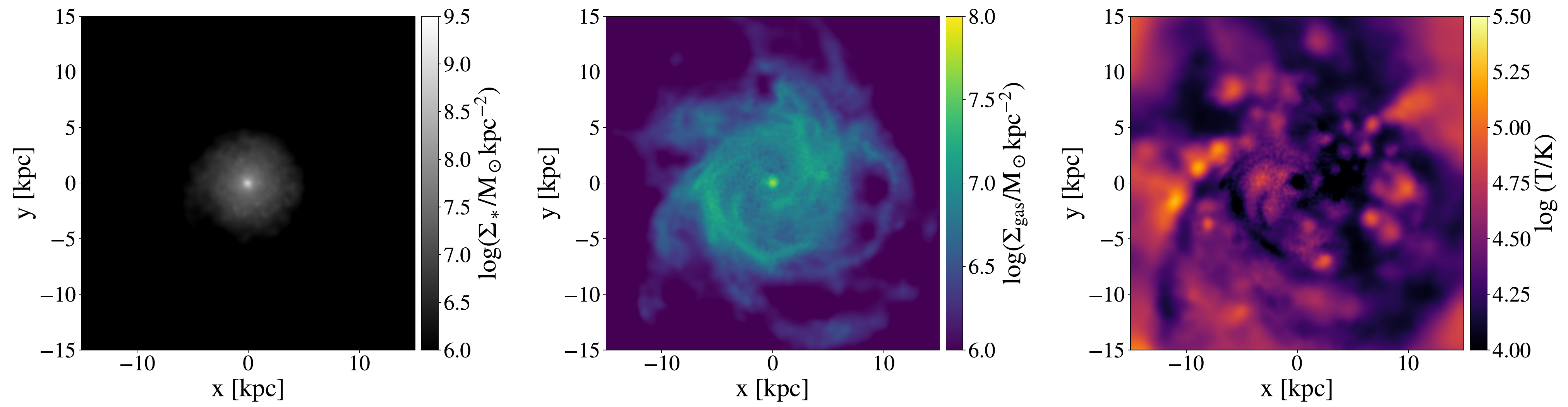}
    
    \caption{Examples of three AGN-control pairs. Each row shows the different maps for a given galaxy in the face-on orientation, from left to right, the quantities are: stellar surface density ($\Sigma_\ast$), gas surface density ($\Sigma_{\rm gas}$), and gas temperature ($T$). Each two rows represent an AGN-control pair, with AGN in the upper row and non-AGN in the lower row for each pair. Stellar masses are indicated on the left of the maps.}

    \label{fig:example_pair}
\end{figure*}

To illustrate the baryonic structures of the dwarf galaxies hosting AGN and their matched control galaxies, we present a few maps of stellar and gas particles in Figure \ref{fig:example_pair}. The few examples shown here reveal that the two-dimensional distribution of gas density and temperature can be very different between the AGN and non-AGN pairs of the simulations.

\end{appendix}

\end{document}